\documentclass{cwruthesis}
\usepackage[hyphens]{url}
\usepackage{lipsum}
\usepackage{graphicx}
\usepackage{hyperref}
\hypersetup{colorlinks=true, linkcolor=black, filecolor=black, urlcolor=black, citecolor=black}
\usepackage[acronym,toc]{glossaries}
\usepackage[intoc]{nomencl}

\newcommand{\la}{\langle}
\newcommand{\ra}{\rangle}
\newcommand{\w}{\omega}

\newcommand{\be}{\begin{equation}}
\newcommand{\ee}{\end{equation}}
\newcommand{\bea}{\begin{eqnarray}}
\newcommand{\eea}{\end{eqnarray}}
\newcommand{\bes}{\begin{subequations}}
\newcommand{\ees}{\end{subequations}}

\usepackage{mathptmx}
\usepackage[T1]{fontenc}
\usepackage[utf8]{inputenc}
\usepackage{amsfonts, amsmath, amsthm, amssymb}
\usepackage{newtxtext,newtxmath}

\theoremstyle{definition}

\usepackage[hyphens]{url}
\usepackage{lipsum}
\usepackage{graphicx}
\usepackage{physics}
\usepackage{relsize}
\usepackage{amsmath,latexsym}
\usepackage{braket}
\usepackage{hyperref}
\hypersetup{colorlinks=true, linkcolor=black, filecolor=black, urlcolor=black, citecolor=black}
\usepackage[acronym,toc]{glossaries}
\usepackage{layout}
\usepackage[intoc]{nomencl}

\usepackage{mathptmx}
\usepackage[T1]{fontenc}
  
\begin{document}
\pagenumbering{roman}

\title{\begin{large}  Solutions to the Mode Equation for a Quantized Massless Scalar Field Outside a Black Hole that Forms from the Collapse of a Null Shell: Late-time Behaviors and Computation of the Stress-energy Tensor\end{large}} 
\author{\begin{large}Shohreh Gholizadeh Siahmazgi\end{large}} 
\maketitle
\begin{acknowledgements}
I would like to acknowledge and give my warmest thanks to my supervisor, Dr. Paul Anderson, who made this work possible. His guidance and support carried me through all the stages of this research work.
 
I would also like to give my special thanks to my committee members, Dr. Eric Carlson, Dr. Gregory Cook, Dr. John Gemmer, and Dr. Freddie Salsbury for their useful comments and suggestions, and my collaborator, Alessandro Fabbri for his guidance during my research.

I would like to thank Raymond Clark who made some power series expansions for solutions to the radial mode equation that were used in the Fortran program that was used to solve the mode equation. 

The research of Paul Anderson was supported in part by the National Science Foundation under Grants No. PHY-1308325, No. PHY-1505875, No. PHY-1912584, and No. PHY- 230918.  Some of the numerical work was done using the WFU DEAC cluster; we thank the WFU Provost's Office and Information Systems for their generous support. 

The research of Alessandro Fabbri was supported by Spanish grants FIS2017-84440-C2-1-P funded by
MCIN/AEI/10.13039/501100011033
”ERDF A way of making Europe”, Grant PID2020-116567GB-C21 funded by
MCIN/AEI/10.13039/501100011033, and the project PROMETEO/2020/079
(Generalitat Valenciana).
\end{acknowledgements}
\begin{KeepFromToc}
\end{KeepFromToc}
\begin{KeepFromToc}
  \tableofcontents
\end{KeepFromToc}

\listoffigures

\begin{abbreviations}
\textbf{Abbreviations}\quad \quad \quad  \textbf{Definitions}\\[10pt]
QFTCS \quad \quad \quad \quad \quad \; Quantum Field Theory in Curved Space \\[6pt]
RSET \quad \quad \quad \quad \quad \; \; \;Renormalized Stress-energy Tensor \\[6pt]
SdS \quad \quad \quad \quad \quad \quad  \quad \;Schwarzschild-de Sitter  \\[6pt]
\end{abbreviations}
\begin{abstract}
The late-time behaviors of the modes in the Unruh state for various eternal two-dimensional black holes are discussed. The Unruh state is designed
to be a state that mimics the late-time behaviors of quantized fields that Hawking predicted for the black holes that form from collapse. Evidence is provided that the late-time behaviors of some of the modes of the quantum fields and the symmetric two-point function are determined by
infrared effects. In four dimensions, a spacetime is considered in which a black hole forms from the collapse of a spherically symmetric null shell. A method is developed to compute the semi-classical stress-energy tensor for a massless minimally-coupled scalar field in the $in$ vacuum state outside the event horizon of the black hole. The computation of the stress-energy tensor involves finding the mode solutions of the scalar field and renormalizing the stress-energy tensor. The former has been done by expanding the modes of the scalar field in terms of a complete set of modes for an eternal Schwarzschild black hole. The latter is done by subtracting the stress-energy tensor for the scalar field in the Unruh state. The computation of the mode solutions of the scalar field in the $in$ vacuum state is discussed. The results are shown for three cases. They are the continuity of the modes for the scalar field across the null shell surface, their high-frequency behaviors for a fixed spacetime point, and their late-time behaviors for a fixed frequency and a fixed spatial point. The preliminary results for the contribution to the stress-energy tensor from the part of the scalar field modes that are ingoing inside the null shell are presented and an approximation for the contribution to the stress-energy tensor from the part of the modes that are outgoing inside the null shell is discussed. 
\end{abstract}

\mainmatter

\setcounter{secnumdepth}{2}

\chapter{Introduction} \label{chap:intro}
In the absence of a complete theory of quantum gravity that enables physicists to unite general relativity and quantum mechanics, the semi-classical theory of quantum gravity is one of the approximate theories that can provide important insights into quantum effects in gravity. In this theory, the matter and radiation fields are described by Quantum Field Theory, while the geometry remains classical. This theory is represented by the semi-classical Einstein equation
\bea
G_{\mu \nu}=8\pi G \langle T_{\mu \nu}\rangle,\label{Intro-T}
\eea
where $ \langle T_{\mu \nu}\rangle$ is the renormalized expectation value of the stress-energy tensor operator (RSET). Once an expression for this quantity is derived for some class of background geometries, the backreaction of quantum fields on the geometry can be found using ~\eqref{Intro-T}. Compared to semi-classical quantum gravity, there is a simpler approximation called Quantum Field
Theory in Curved Space (QFTCS) in which the RSET and other quantum effects are computed on a fixed
background spacetime and the backreaction on the spacetime geometry is ignored.

In 1974, Hawking used QFTCS to show that unlike classical black holes, which are not thought to radiate, "quantum black holes" \footnote{This quote is from ~\cite{Birrel-Davis}.} create and radiate particles \cite{Hawking-Nature}. He showed that the initial vacuum state for the quantum field, which is called the $in$ vacuum state, is not the same as the vacuum state at late times after the black hole has formed. The latter vacuum state is usually called the $out$ vacuum state. He computed the expectation value of the particle number operator for the $out$ vacuum state and found it was not zero. The produced particles are called Hawking Radiation. His calculation showed that at late times, the particles come out in a thermal distribution \cite{hawking:1975}. Therefore, they do not carry information about how the black hole formed. The question of what happens to the information about how the black hole formed is called the information issue. To better understand Hawking radiation and gain insight into the information issue, one might need a complete theory of quantum gravity. Without such a theory, the computation of the RSET might shed some light on these problems. The  RSET has been calculated for eternal black holes but no full numerical calculation has been done for a black hole that forms from gravitational collapse in four dimensions.

For static spherically-symmetric black holes, the RSET  for various quantum fields in different quantum states has been computed. For such spacetimes, three main vacuum states can be defined: the Boulware state \cite{boulware}, the Hartle-Harking state \cite{hartle-hawking}, and the Unruh state \cite{unruh:1976}. The Boulware state can be qualitatively interpreted as a natural vacuum state for an inertial observer at infinity. The Hartle-Hawking state is a thermal state at the black hole's temperature \footnote{The black hole temperature is equal to the temperature of Hawking radiation. This temperature is equal to $\frac{\kappa}{2\pi}$, where $\kappa$ is the surface gravity of the black hole which depends on the mass, electric charge, and angular momentum of the black hole. For a spherically symmetric, non-rotating, uncharged black hole, the surface gravity is $\ \frac {1}{4\pi M}$.}.  The Unruh state is designed to mimic the Hawking radiation at late times.

In the case of a static spherically-symmetric spacetime, the RSET for a conformally invariant scalar field has been computed for the Hartle-Hawking state in Schwarzschild spacetime \cite{fawcett,howard-candelas, howard}. The computation of the RSET for the electromagnetic field in the Hartle-Hawking state in Schwarzschild spacetime has been done in ~\cite{jensen-ottewill} and the RSET for an electromagnetic field in the Unruh state in Schwarzschild spacetime was computed in ~\cite{jmo}. Also, the RSET for a massless conformally coupled scalar field in the Boulware state was computed in Schwarzschild spacetime in \cite{jensen-et-al}. In \cite{ahs1, ahs2}, the RSET was computed for both massless and massive scalar fields with arbitrary coupling to the scalar curvature in both Schwarzschild and Reinsser-Nordstrs\"{o}m spacetimes in the Hartle-Hawking state. The RSET was computed for a massless arbitrarily-coupled scalar field in an extreme Reissner-Nordstrom spacetime
\cite{ahl}. The RSET was calculated for a spin-$\frac{1}{2}$ field in an extreme Reissner-Nordstrom spacetime in \cite{choag}. 
In Schwarzschild spacetime, the RSET was computed for a massless arbitrarily-coupled scalar field in the Boulware vacuum state \cite{abf}. 
In \cite{breen-ottewill}, the calculation of the RSET has been done for a massless conformally-coupled scalar field in the Hartle-Hawking state in a Reissner-Nordstrom-de Sitter spacetime for which the temperature associated with the black hole horizon is the same as that associated with the cosmological horizon. In ~\cite{levi-ori, levi}, for a massless minimally-coupled scalar field in the Unruh state, the RSET was computed numerically in Schwarzschild and Reissner-Nordstrom spacetimes. For the same type of quantum field in the Unruh state in Kerr spacetime, the RSET was computed in \cite{levi-et-all-kerr-st}, and in both the Unruh and Hartle-Hawking states, the RSET was computed near the inner horizon of a spherical charged black hole in \cite{Zilberman-Levi-Ori}. More recently, in the same spacetimes, the RSET for a scalar field in the Hartle-Hawking, Boulware, and Unruh states was numerically computed in \cite{RNRSET}

In the eternal black holes mentioned above, because of the lack of time-dependency of the geometry, Hawking radiation does not occur, although the Unruh state can be used to approximate it. One simple model of a black hole that forms from collapse is the null shell model. In this model, a spherically symmetric shell of photons originating from infinity implodes and forms a black hole. Inside the null shell, the spacetime is flat. Therefore, the natural vacuum state inside the shell is the Minkowski vacuum. Outside the null shell, by Birkhoff's theorem, the spacetime is Schwarzschild. The entire $in$ vacuum state for the null-shell spacetime is specified in Chapter 4.

 While the collapsing null shell model is one of the simplest models that can describe the formation of a black hole, a full numerical computation of the RSET in 4D has not been accomplished yet. In 2D, an analytic expression for the RSET for a massless minimally-coupled scalar field has been computed in \cite{hiscock} and \cite{Fabbri:2005mw}.
 
We have developed a method that can be used to compute the RSET for a massless minimally-coupled scalar field in a 4D collapsing null-shell spacetime. In this method, instead of computing the solutions to the mode equation in the region outside the null shell and outside the event horizon, we find the equivalent solution in an eternal Schwarzschild black hole spacetime. In the new problem, we expand the modes of the scalar field for the $in$ vacuum state in terms of a complete set of solutions that can be obtained using separation of variables. To renormalize the stress-energy tensor associated with the scalar field, we remove the divergence in the stress-energy tensor by subtracting the stress-energy tensor for the field in the Unruh state. Therefore, since the states we are working with possess the Hadamard property \cite{Birrel-Davis}, the difference between the unrenormalized stress-energy tensor in the $in$ vacuum state and the Unruh state is equal to the difference between their renormalized counterparts. The RSET for a massless scalar field in the Unruh state in Schwarzschild spacetime is numerically computed in \cite{levi-ori} and \cite{levi}. One can then add back this quantity to the difference between the unrenormalized stress-energy tensors. One of the advantages of this method of subtraction is that one can find out how quickly the RSET in the $in$ vacuum state approaches its Unruh counterpart. 

The aforementioned method can be used to numerically compute the modes of a massless minimally-coupled scalar field in the $in$ vacuum state. In this thesis, the focus is on the contribution of the spherically symmetric modes. For this contribution, the mode solution can be decomposed into a part that was originally ingoing and a part that was originally outgoing \cite{null-shell-method, null-shell-proceedings}. On the other hand, the Unruh state consists of modes that are positive frequency with respect to the Kruskal time coordinate on the past horizon and modes that are positive frequency with respect to the usual time coordinate on past null infinity. We have studied the late-time behavior of these modes and their approach to their Unruh counterparts. We show that the ingoing part of the $in$ modes approaches the subset of the modes in the Unruh state that are positive frequency with respect to the usual time coordinate on past null infinity. 

To gain a deeper insight into the late-time behaviors of the Kruskal modes, we have studied them in various 2D eternal black hole spacetimes \cite{unruh-paper}. The Kruskal modes can be expanded in terms of Boulware modes. We have shown that for the cases considered, the late-time behaviors of the Kruskal modes are governed by the presence or absence of infrared divergences in the Boulware modes. Particularly, we studied the solutions to a mode equation in the presence of a potential that has been modeled by a Dirac delta function. It was shown that because of the absence of infrared divergences in the Boulware modes due to scattering in the mode equation, the Kruskal modes decay at late times.

In Chapter 2, we summarize some of the key concepts of Quantum Field Theory in Curved Spacetime. In Chapter 3, we present the results of our study regarding the late-time behaviors of the Kruskal modes and their corresponding symmetric two-point correlation function in two dimensions. Chapter 4 contains a discussion of the method of computing the stress-energy tensor outside a 4D black hole that has formed from the collapse of a null shell.
In Chapter 5, we use the method to compute the modes for a scalar field in the $in$ vacuum and the Unruh states. In Chapter 6, we apply the method of computing the stress-energy tensor discussed in Chapter 4 to compute part of the spherically symmetric contribution to the RSET. 
 
\chapter{Quantum Field Theory in Curved Spacetime}
This chapter summarizes the preliminaries and key concepts in Quantum Field Theory in Curved Space that will be used in future chapters. We mostly use the notations in ~\cite{Birrel-Davis} for this chapter.

\section{Quantization in Curved Space}
The Lagrangian density for a real massive scalar field $\phi(x)$ of mass $m$ is 
\bea
\mathscr{L}(x)=\frac{1}{2}[-g(x)]^{\frac{1}{2}}\big\{-g^{\mu \nu}(x)\phi(x)_{,\mu}\phi(x)_{,\nu}-[m^2+\xi R(x)]\phi^2(x)\big\},\label{ch2-Lagrangian}
\eea
\bea
[- \scalebox{1.5}{$\Box$}_x+m^2+\xi R(x)]\phi(x)=0\label{ch2-mode-eq}
\eea
with $\scalebox{1.5}{$\Box$}_x=g^{\mu \nu}\nabla_{\mu}\nabla_{\nu}\phi=(-g)^{\frac{1}{2}}\partial_{\mu}[(-g)^{\frac{1}{2}}g^{\mu \nu}\partial_{\nu}\phi]$.

For this thesis, we are mainly interested in a massless minimally-coupled scalar field. Therefore $m=\xi=0$ and ~\eqref{ch2-mode-eq}  reduces to 
\bea
\scalebox{1.5}{$\Box$}_x
\phi(x)=0\label{ch2-mode-eq-massless-minimally-coupled}.
\eea
One of the useful concepts here is the scalar product that is defined for any two solutions of the Klein-Gordon equation, ~\eqref{ch2-mode-eq}, and is given by
\bea
(\phi_1,\phi_2)=-i\int_{\Sigma}\phi_1(x)\overset{\leftrightarrow}{\partial}_{\mu} \phi_2^{*}(x)[g_{\Sigma}(x)]^{\frac{1}{2}} d\Sigma^{\mu}\label{ch2-scalar-product}
\eea
where $\phi_1(x)\overset{\leftrightarrow}{\partial}_{\mu} \phi_2^{*}(x)=\phi_1(x)\partial_{\mu} \phi_2^*(x)-\phi_2^*(x)\partial_{\mu}\phi_1(x)$ and
$d\Sigma^{\mu}=d\Sigma\; n^{\mu}$ with $d\Sigma$ the volume element in a particular Cauchy surface $\Sigma$ and $n^\mu$ is a timelike unit vector normal to $\Sigma$. Note that $g_{\Sigma}$ is the determinant of the induced metric on the surface $\Sigma$. It is proved in \cite{LSS} that the scalar product is independent of the hypersurface $\Sigma$.

One can expand the field in terms of a complete set of solutions $\{f,f^*\}$ to the mode equation as follows
\bea
\phi=\underset{i}{\sum}[a_{i}f_{i}+ a^{\dagger}_{i}f^*_{i}].\label{ch2-expansion-first}
\eea
where $a_i|0\rangle=0$ for all $i$ with $a_i$ an annihilation operator. 

We require that $\{f,f^*\}$ satisfy the orthonormality condition with respect to the scalar product ~\eqref{ch2-scalar-product}, i.e.,
\bea
(f_i,f_j)=\delta_{ij}, \quad (f_i^*,f_j^*)=-\delta_{ij}, \quad (f_i,f_j^*)=0.
\eea
The null-shell spacetime is not static. Inside the collapsing null shell, the spacetime is flat, and outside of it, the spacetime is
Schwarzschild spacetime. Both are static and spherically symmetric. However, as we will show in Chapter 4, outside the null shell and the horizon, one can expand the $in $ modes in the null-shell spacetime in terms of a complete set of modes in an eternal Schwarzschild spacetime. In Schwarzschild spacetime, one can use separation of
variables to obtain solutions to the mode equation of the following form
\bea
f_{\w \ell m}=N\frac{Y_{\ell m}(\theta,\phi)}{r}\chi_{\w \ell}(r)e^{-i\w t_s}\label{ch2-mode-general}
\eea
where $N$ is a normalization constant and $t_s$ is the Schwarzschild time coordinate. The radial function in ~\eqref{ch2-mode-general} satisfies the following equation
\bea
\frac{d^2 \chi_{\w \ell}}{d{r_*}^2}=-\Bigg[\w^2-\Big(1-\frac{2M}{r}\Big)\Big(\frac{2M}{r^3}+\frac{\ell(\ell+1)}{r^2}\Big)\Bigg]\chi_{\w \ell},
\eea
where $r_*=r+2M \log \big(\frac{r-2M}{2M}\big)$.

The mode solutions $f_{\w \ell m}$ are normalized under the conditions
\bea
(f_{\w \ell m},f_{\w' \ell' m'})=\delta(\w-\w')\delta_{\ell \ell'}\delta_{mm'}.
\eea

Unlike Minkowski spacetime, in curved spacetime, there is no favored choice of the set $\{f,f^*\}$, or in other words, no unique choice of the vacuum state. In Schwarzschild spacetime, three different vacuum states
are usually considered. They are the Boulware state, the Hartle-Hawking state, and the
Unruh state. Qualitative descriptions of these states are given in the Introduction.  Throughout this thesis, we particularly work with the Boulware and Unruh states.  More details about these states can be found in Chapters 3 and 4. Among these three states, the Boulware state is the only one for which the mode function can be written in the form of ~\eqref{ch2-mode-general}.
\section{Stress-energy Tensor in Curved Spacetime}
The stress-energy tensor for a scalar field in curved spacetime can be found by variation of the action with respect to the metric tensor,
\bea
T_{\mu \nu}=-\frac{2}{\sqrt{-g(x)}}\frac{\delta S}{\delta g^{\mu \nu}}
\eea
For a massless minimally-coupled scalar field, the Lagrangian density in ~\eqref{ch2-Lagrangian} gives 
\bea
T_{\mu \nu}&=&\phi_{;\mu}\phi_{;\nu}-\frac{1}{2}g_{\mu \nu}g^{\rho \sigma}\phi_{;\rho}\phi_{;\sigma}.
\eea
The expectation value of $T_{\mu \nu}$ is defined by
\bea
\langle T_{\mu \nu}\rangle=\frac{1}{4}\underset{x\to x'}{\lim} \big\{(g^{\rho'}_{\mu}G^{(1)}_{;\rho'\nu}(x,x')+g^{\rho'}_{\nu}G^{(1)}_{;\mu \rho'}(x,x'))-g_{\mu \nu}g^{\rho \sigma'}G^{(1)}_{;\rho \sigma'}(x,x')\big\}\label{ch2-T}
\eea
where $g^{\rho'}_{\mu}$ is a bivector of parallel transport that is defined in Section 4.7, and  $G^{(1)}(x,x')$ is the Hadamard two-point function that is given by
\bea
G^{(1)}(x,x')\equiv \langle 0 | \phi(x)\phi(x')+\phi(x')\phi(x)|0\rangle.
\eea

The expression in ~\eqref{ch2-T} is divergent in the limit $x\to x'$. There are several methods of renormalization for ~\eqref{ch2-T}. For more discussion, see \cite{Birrel-Davis}. The divergence in ~\eqref{ch2-T} is due to the short distance behavior of the expectation value of the product of the fields and their derivatives appearing in the stress-energy tensor. This behavior is called "Ultraviolet Behavior". However, this behavior does not depend on the state of the field. Therefore, one can renormalize the stress-energy tensor in the $in$ vacuum state by subtracting the unrenormalized stress-energy tensor in a different vacuum state. Then, one adds to this the RSET for the second vacuum state to find the RSET for the first vacuum state.

In this thesis, we subtract the expectation value of the unrenormalized stress-energy tensor in the Unruh state from the expectation value of the stress-energy tensor in the $in$ vacuum state. Then, we find the RSET in the $in$ vacuum state by adding the RSET for the Unruh state to the difference between the unrenormalized stress-energy tensors in two vacua. The RSET for the Unruh state is computed in \cite{levi-ori, levi}. 
\chapter{Infrared Effects and the Unruh State}
\section{Preface}
This chapter provides the content of the article "Infrared Effects and the Unruh State" which was published in \textit{Classical and Quantum Gravity} \cite{unruh-paper}. This paper, which was the result of a collaboration between Paul Anderson, myself, and Zachary Scofield, discusses the properties of the modes of some quantum scalar fields and the corresponding symmetric two-point correlation function in the Unruh state in some two-dimensional (2D) black hole spacetimes. Particularly, we studied solutions to
\begin{itemize}
    \item a mode equation with no effective potential,
    \item a mode equation with a Dirac delta function effective potential, and
    \item a mode equation with an effective potential associated with a massive scalar field in SdS spacetime.
\end{itemize}

To understand the strange late-time behavior of the symmetric two-point function in the Unruh state, Paul Anderson used the solutions to the corresponding mode equations in the 2D black hole spacetime in the absence of a potential to show that the two formulations of the Unruh state given in ~\cite{unruh:1976} result in different expressions for the symmetric two-point correlation function associated with these modes. I showed that in the presence of scattering by a Dirac delta potential in the mode equation, the Kruskal modes vanish at future timelike infinity. However, the corresponding correlation function has a non-zero value at late times. Paul Anderson and I worked closely together to show that the two different formulations of the Unruh state give equivalent results for the two-point function when scattering effects due to a delta function potential remove the infrared divergences in the Boulware modes. The numerical work done by Zachary Scofield also showed that in the presence of a more realistic potential that occurs for a massive scalar field in
Schwarzschild-de Sitter spacetime, the Kruskal modes approach zero at late times. 

Paul Anderson posed the question as to whether there is a connection between the late-time behavior of the Kruskal modes and the infrared divergences associated with the Boulware modes. We were able to answer this question by looking at the time dependency of the Kruskal modes in the three aforementioned cases. We found that the existence or absence of infrared divergences in the Boulware modes strongly affects the late-time behavior of the Kruskal modes. We found that the existence of a non-zero potential such as a Dirac delta in an arbitrary 2D eternal black hole or a potential corresponding to a massive scalar field in a SdS spacetime with a 2D metric removes the infrared divergences in the Boulware modes. Consequently, at late times, the Kruskal modes approach zero for a fixed space coordinate.

In the published version of the following paper, there are three misprints. In (2.15), the variable of integration $\w$ should be replaced by $\w'$. In (3.16), the second integral is over $\w$, and in (3.9), $I1$, $I2$, and $I3$ should be replaced $I_1$, $I_2$, and $I_3$.

The content of this chapter can be found at \href{https://iopscience.iop.org/article/10.1088/1361-6382/acd0fd/pdf}{\textit{Classical and Quantum Gravity}}. 
\newpage
\chapter{Method to compute the stress-energy tensor for a quantized scalar field
when a black hole forms from the collapse of a null shell}
\section{Preface}
This chapter presents the content of the article "\textit{Method to compute the stress-energy tensor for a quantized scalar field when a black hole forms from the collapse of a null shell}", which was a collaboration between Paul Anderson, Alessandro Fabbri, Raymond Clark, and myself. It was published in \textit{Physical Review D} \cite{null-shell-method}. In this paper, we present a method that can be used to compute the stress-energy tensor associated with a massless minimally-coupled quantum scalar field in the $in$ vacuum state in a spacetime in which a 4D black hole forms from the collapse of a spherically-symmetric null shell"

The original idea for this method was given by Paul Anderson. This method is based on the idea that, in the region outside the null shell and outside the horizon of the black hole, one can expand the $in$ modes for a scalar field in terms of a complete set of solutions to the mode equation in the exact Schwarzschild geometry. The stress-energy tensor is renormalized by subtracting the unrenormalized stress-energy tensor for the scalar field in the Unruh state. To find the renormalized stress-energy tensor for the $in$ modes, one can add this difference to the renormalized stress-energy tensor in the Unruh state that was computed outside the future horizon of a 4D Schwarzschild black hole.
Paul Anderson mathematically formulated this method in a 2D null-shell spacetime and found an expression for the difference between the two-point functions associated with a scalar field in the $in$ vacuum state and the Unruh state. To check the validity of this method, I continued the calculations started by Paul Anderson. I found an expression for the stress-energy tensor and numerically evaluated its component in a 2D null-shell spacetime. I showed the numerical results for the stress-energy tensor agreed with the analytical results for the 2D stress-energy tensor in ~\cite{Davies-Fulling-Unruh, hiscock, Fabbri:2005mw}. Paul Anderson and I worked together to mathematically formulate the method in a 4D null-shell spacetime and we found a general form for the matching coefficients that appear in the expressions for the $in$ modes. 

To check the validity of our formulation, we used the general expressions for the matching coefficients and reconstructed the $in$ modes in the 2D case. In addition, I numerically reconstructed the $in$ modes outside the null shell and on the future horizon and checked the continuity of the $in$ mode functions on the null-shell surface. In the 4D null-shell spacetime, the presence of an effective potential in the mode equation causes scattering effects that make the matching more difficult. Therefore, to gain more insight into the properties of the $in$ modes in 4D, Paul Anderson suggested that I formulate the method for a toy model in which the effective potential in the mode equation is in the form of a Dirac delta function. Together with Paul Anderson, I computed the matching coefficients in this case and found partially analytical expressions for them. We were able to partially reconstruct the $in$ modes in this case and to check their continuity on the null shell.

The argument made by Paul Anderson in Section 4.6.3 to close the contour for the integral in (4.67) in the lower half plane is incorrect. The argument relies on the form the gamma function in the matching coefficients takes using Stirling's approximation. What is incorrect is that the dominant factor at complex infinity is canceled by another term to leading order. Without this argument, there is no obvious reason why one should close the contour in the lower half plane for all values of the coordinate $u$. If one justifies choosing such a contour, the derivations that follow this argument are valid. Work is in progress to resolve this problem.

In the version of the paper that follows, some misprints and other minor errors in the published version, \cite{null-shell-method}, have been corrected.

\newpage

\begin{center}
\Large \textbf{Method to compute the stress-energy tensor for a quantized scalar field
when a black hole forms from the collapse of a null shell}
\end{center}
\begin{center}
Paul R. Anderson${}^{1}$, Shohreh Gholizadeh Siahmazgi${}^{1}$, \\
Raymond D. Clark${}^{1}$, and Alessandro Fabbri${}^{2,3}$\\
\textit{${}^1$Department of Physics, Wake Forest University}\\
 \textit{Winston-Salem, North Carolina 27109, USA}\\
 \textit{${}^2$Departamento de F\'isica Te\'orica and IFIC, Universidad de Valencia-CSIC, C. Dr. Moliner 50, 46100 Burjassot, Spain and} \\
 \textit{${}^3$Universit\'e Paris-Saclay, CNRS/IN2P3, IJC Lab, 91405 Orsay Cedex, France}
\end{center}
\newpage
\begin{center}
  \textbf{Abstract}\\
  \end{center} 
  A method is given to compute the stress-energy tensor for a massless minimally coupled scalar field in a spacetime where a black hole forms from the collapse of a spherically symmetric null shell in four dimensions.  Part of the method involves matching the modes for the {\textit {in}} vacuum state
to a complete set of modes in Schwarzschild spacetime.  The other part involves subtracting from the unrenormalized expression for the stress-energy
tensor when the field is in the {\textit{in}} vacuum state, the corresponding expression when the field is in the Unruh state, and adding to this the renormalized
stress-energy tensor for the field in the Unruh state.  The method is shown to work in the two-dimensional case where the results are known.\\

\section{Introduction}

The stress-energy tensor of a quantized field is an extremely useful tool for studying quantum effects in curved space because it takes both particle
production and vacuum polarization into account.  It can be computed in a background spacetime to obtain the energy density, pressure, etc.,
for a quantum field in that spacetime.  It can also be used in the context of semiclassical gravity to compute the backreaction of the quantum field on the
spacetime geometry.

For black holes in four-dimensional, 4D, spacetimes, the full stress-energy tensor must be computed numerically.  This is a difficult task that has to date only been done without other approximations for the cases of static spherically symmetric black holes~\cite{fawcett,howard-candelas,howard,jensen-ottewill,jmo,jensen-et-al,ahs1,ahs2,ahl,choag,abf,breen-ottewill,levi-ori,levi,Zilberman-Levi-Ori} and the stationary Kerr metric~\cite{duffy-ottewill,levi-et-al-kerr}.
However, because of the difficulty involved, to our knowledge, no one has numerically computed the full stress-energy tensor for a quantized field in a 4D spacetime in which a black hole forms from collapse.  This is important because there can be a significant difference between the stress-energy tensor for a quantum field in a 2D versus a 4D spacetime such as that found for a massless minimally coupled scalar field in an extreme Reissner-Nordstrom spacetime~\cite{trivedi,ahl}.

In this paper we present a  method to compute the renormalized stress-energy tensor, $\la {in}| T_{ab} | {in} \ra$, for a massless minimally coupled scalar field in the case that a black hole forms from the collapse of a spherically symmetric null shell.  This model has been previously used to derive the Hawking effect~\cite{Vilkovisky,Fabbri:2005mw}, investigate how the stress-energy tensor is affected by the production of a pair of particles due to the Hawking effect~\cite{m-p},
 study some details of how the spectrum and number of produced particles changes in time during and after the collapse~\cite{mirror-bh,late-time}, and in 2D to compute the stress-energy tensor for a massless minimally coupled scalar field~\cite{hiscock,Fabbri:2005mw}.  While this is not a realistic model for collapse because the shell begins with an infinite size, this is probably the simplest model to work with that involves collapse in 4D to form a black hole.  Thus it is a reasonable first choice for the full numerical computation of the stress-energy tensor of a quantized field in a 4D spacetime in which a black hole forms from collapse.  Further, since the Hawking effect is independent of how the black hole forms~\cite{Hawking-Nature}, and since it is expected that the stress-energy tensor at late times will also be independent of the formation process, studying
how the stress-energy tensor evolves in time and approaches its late time behavior can provide insight into what is likely to happen in a more realistic model.

The method we have developed works in the region outside both the null shell and the event horizon.
In the region outside the shell, Birkhoff's theorem ensures that the metric is that for Schwarzschild spacetime~\eqref{metric-sch}.
In the region inside the shell, the space is flat.  Thus in both regions, the mode equation for the quantum field is separable and inside the shell, its solutions are
known analytically.  This allows for a numerical computation of the stress-energy tensor for the field in which only ordinary differential equations need to be solved
numerically.

For the collapsing null shell model, the initial vacuum state of the quantum field is well defined and
the main complication that occurs is due to the propagation of the modes across the null shell surface.  The crux of our method involves the expansions of the $in$
modes in terms of a complete set of solutions to the mode equation in the region outside the shell.

The stress-energy tensor for the quantum field is obtained by expanding the quantum field in terms of a complete set of modes.
This expansion is substituted into the formula for the stress-energy tensor of the corresponding classical field and the expectation value is computed.
If the field is in the $in$ vacuum state then the result is an expression which involves sums and integrals over the mode functions for the $in$ state and their derivatives.  After the renormalization counterterms are subtracted off, the resulting stress-energy tensor is finite and can be computed. This
is straightforward inside the null shell since the mode functions are known analytically and for the $in$ state, the result is that the stress-energy tensor is equal to zero.

Outside the null shell and the event horizon, the $in$ modes do not assume a simple form in 4D.  One approach to computing them would be
to use the analytically known values for the modes inside the shell and on past null infinity to provide initial data for a numerical integration of the mode equation in the region exterior to the shell.  However, outside the shell, the $in$ modes will not factorize into a product of a function that depends only on time and a function that depends only on the radial coordinate $r$.  Thus the part of the mode equation that depends on both $r$ and $t$ must be solved numerically.

We have developed an alternative method which involves expanding each of the $in$ modes in terms of a complete set of modes in Schwarzschild spacetime.
The radial parts of these modes and the matching parameters must be computed numerically.  The mode matching has been tested in the 2D case where there is no effective potential in the mode equation.  It has also been partially tested for spherically symmetric modes in 4D both when the effective potential is modeled as a delta function and when the exact effective potential is used.

One advantage of the first method is that there are no matching parameters. A disadvantage is that one must solve a partial differential equation directly using numerical techniques.  Conversely the chief advantage of the method developed here is that one only needs to numerically solve the radial mode equation, which is an ordinary differential equation.  A second advantage is that the properties of the solutions to this equation are well understood.  One disadvantage of our method is that the formulas for the matching parameters involve certain integrals that must be computed numerically.  A second disadvantage is that the computation of the stress-energy tensor involves the numerical computation of triple integrals rather than single integrals over various products of the mode functions and their derivatives.  It is not obvious to us which approach is more efficient. However, since no full numerical computation of the stress-energy tensor has been previously done for a quantized scalar field in a 4D spacetime where a black hole forms from collapse, we think the most important thing is to develop one viable method to do the calculation and that is what we present here.

When the expansions for the $in$ modes are substituted into the formula for the unrenormalized stress-energy tensor one finds
a combination of sums and integrals over various combinations of the modes and their derivatives.
Renormalization of the stress-energy tensor can be accomplished by subtracting the corresponding
expression that occurs in Schwarzschild spacetime for the Unruh state~\cite{unruh:1976}, adding that expression back and subtracting the renormalization counterterms.
  The result is the sum of two finite tensors.  The first is the difference between the expressions for the unrenormalized stress-energy tensors in the $in$ state
  and the Unruh state.  The second is the renormalized stress-energy tensor for the Unruh state.  The latter has been numerically computed for the massless minimally coupled scalar field in~\cite{levi-ori,levi}.  Thus one can simply add that result to the difference between the two stress-energy tensors to obtain the full renormalized stress-energy tensor for the scalar field in the $in$ state in the collapsing null shell spacetime.
This type of renormalization scheme has been used to compute the stress-energy tensors in Schwarzschild spacetime in the Unruh state for the conformally coupled massless scalar field~\cite{elster,jmo} and for the massless spin $1$ field~\cite{jmo}.  It has also been used to compute a late time approximation to $\la T_{t r} \ra$ for the case of a massive minimally coupled scalar field in a spacetime consisting of a massive thin shell that is initially static and then collapses to form a black hole~\cite{akhmedov-godazgar-popov}.

We have tested our method by numerically computing the difference between the stress-energy tensor for the $in$ state in the collapsing null shell spacetime and the stress-energy tensor for the Unruh state in 2D.  The results are compared with an analytic expression for the difference obtained from
previous analytic calculations of the stress-energy tensor for the Unruh state~\cite{Davies-Fulling-Unruh} and the $in$ vacuum state for the collapsing null shell spacetime~\cite{hiscock, Fabbri:2005mw}.
Our results are in agreement with those calculations.

In Section {ch4-scalar-field-null-shell}, we introduce the collapsing null shell model and then discuss the modes for a massless minimally coupled scalar field in the null shell spacetime.  A description of the method of computing the stress-energy tensor is given in Section {sec:method-Tab}. Various mode functions in Schwarzschild spacetime that are used in the computation of the stress-energy
tensor are discussed in Section {ch4-sec-mode-functions}.  In Section ~{sec:matching}, general expressions for the matching coefficients in the 4D case are derived followed by examples where the matching method is tested.
Formulas needed for the computation of the stress-energy tensor in the 4D case are derived in the first part of Section {ch4-sec-Tab-method}.  In the second part, the difference between the stress-energy tensor in the $in$ vacuum state and the Unruh state is numerically computed for the 2D case and compared with the difference obtained from previous analytic calculations.
Section {ch4-summary} contains a summary of our results.  The appendixes contain some details of a proof and some derivations that are used in the 2D examples in Secs. ~{sec:matching} and {ch4-sec-Tab-method}.
Throughout the paper, we use the sign conventions of~\cite{Misner} and units are chosen such that $\hslash=c=G=1$.

\section{Massless minimally coupled scalar field in a spacetime with a collapsing null shell}\label{ch4-scalar-field-null-shell}

\subsection{Collapsing null shell model}

We consider a model in which a spherically symmetric black hole forms from the collapse of a null shell.  Our analysis of the spacetime follows that in~\cite{Fabbri:2005mw}.  The metric inside the shell is the flat space metric
\be
ds^2 = -dt^2 + dr^2 + r^2 d \Omega^2    \;, \label{metric-flat}
\ee
and, by Birkhoff's theorem, the metric outside the shell is the Schwarzschild metric
\be
ds^2 = -\left(1-\frac{2M}{r} \right) dt_s^2
+ \left( 1-\frac{2M}{r} \right)^{-1} dr^2 + r^2 d \Omega^2  \;. \label{metric-sch}
\ee
The two metrics need to be matched along the trajectory of the null shell.  An obvious way to do this is to let the angular coordinates be continuous across the shell
along with the radial coordinate $r$ that is related to the area of a two-sphere.  Then the time coordinate is not continuous across the shell trajectory which is why we distinguish in
the above metrics between the time coordinate $t$ inside the shell and the time coordinate $t_s$ outside the shell.

The actual matching is easier in terms of radial null coordinates which can be defined inside the shell as
\bes \bea
u &=& t - r \;, \label{u-flat}\\
 v &=& t + r  \;,\, \label{v-flat} \eea
\label{u-v-flat}
\ees
and outside the shell as
\bes \bea
u_s &=& t_s - r_* \;, \label{u-sch-def}\\
  v &= & t_s + r_* \;, \label{v-sch-def}
 \eea   \label{u-v-sch}
\ees
where
\bea
  r_* &= & r + 2M \log \left(\frac{r-2M}{2M} \right)  \;,
\label{rstar-def}
\eea
is the usual tortoise coordinate in Schwarzschild spacetime.
It is easiest to let $v$ be continuous across the shell trajectory which is denoted as $v = v_0$.
The outgoing radial null coordinate is then discontinuous across the shell trajectory which is why it is denoted as $u$ inside the shell and $u_s$ outside.
The relationship between $u$ and $u_s$ is~\cite{m-p,Fabbri:2005mw}
\be
u_s = u - 4M \log \left( \frac{v_H-u}{4 M} \right)  \;,
\label{us-u}
\ee
with
\be
v_H \equiv v_0 - 4M  \;.
\label{vH-def}
\ee
Note that the value of the flat space coordinate $u$ on the event horizon is $v_H$ as can be seen from Fig.~{fig:Null-Shell-Penrose}.  Inverting, one finds that~\cite{mirror-bh}
\be
u = v_H - 4 M \, \text{W}\left[\exp\left(\frac{v_H - u_s}{4 M}\right) \right]  \;,
\label{u-us}
\ee
with $\text{W}$ the Lambert W function.
A Penrose diagram for the resulting spacetime is sketched in Fig.~{fig:Null-Shell-Penrose}.
\begin{figure}[h]
\centering
\includegraphics [trim=2.7cm 20cm 0cm 0cm,clip=true,totalheight=0.5\textheight]{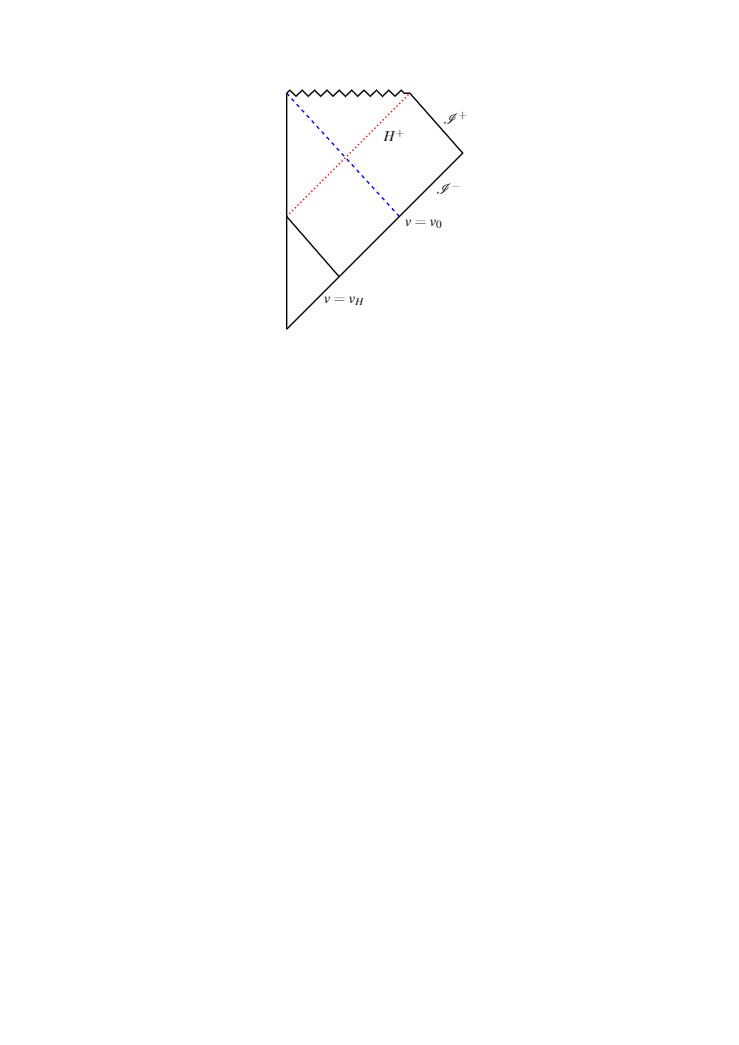}
\caption{Penrose diagram for a spacetime in which a null shell collapses to form a spherically symmetric black hole.  The vertical line on the left corresponds to the
surface $r = 0$ which is also the surface where $u = v$.  The trajectory of the shell (dashed blue curve) is $v = v_0$.  The horizon, $H^{+}$, is the dotted red curve.  Inside the shell trajectory it corresponds to the surface $u = v_H$ and outside the shell trajectory it corresponds to $u_s = \infty$. }
\label{fig:Null-Shell-Penrose}
\end{figure}

\subsection{Massless minimally coupled scalar field}

The type of quantum field we consider is a massless minimally coupled scalar field which in a general spacetime satisfies the wave equation
\be \Box \phi = 0  \;. \label{Box-phi} \ee
In the null shell spacetime the field can be expanded in terms of a complete set of modes such that
\be \phi = \sum_{\ell = 0}^\infty  \sum_{m = -\ell}^\ell \int_0^\infty d \w \, [a_{\w \ell m} f_{\w \ell m} + a^\dagger_{\w \ell m} f^{*}_{\w \ell m}] \;, \label{Phi-expansion} \ee
with $a_{\w \ell m}$ an annihilation operator.  The modes are solutions to~\eqref{Box-phi} which have the form
\be f_{\w \ell m} = N \frac{Y_{\ell, m}(\theta, \phi)}{r } \psi_{\w \ell}(\tau,r) \;, \label{f-def-1} \ee
with $N$ a normalization constant and $\tau = t$ inside the shell trajectory and $\tau = t_s$ outside.
Inside the shell trajectory the equation for $\psi_{\w \ell}$ is
\be -\frac{\partial^2 \psi_{\w\ell}}{\partial t^2}   + \frac{\partial^2 \psi_{\w \ell}}{\partial r^2} - \frac{\ell(\ell+1)}{r^2} \psi_{\w \ell} =0 \;, \label{psieq-flat} \ee
while outside the shell the equation is
\be -\frac{\partial^2 \psi_{\w\ell}}{\partial t_s^2}   +
 \frac{\partial^2 \psi_{\w \ell}}{\partial r_{*}^2} -  \left(1 - \frac{2M}{r} \right) \left( \frac{2 M}{r^3}+  \frac{\ell(\ell+1)}{r^2} \right) \psi_{\w \ell} = 0 \;. \label{psieq-Sch} \ee

The $in$ vacuum state is defined by requiring that on $\mathscr{I}^{-}$
\be \psi^{in}_{\w \ell} = e^{-i \w v}  \;. \label{in-state-def} \ee
The modes must also be regular on the surface $r = 0$ inside the shell trajectory which implies that $\psi^{in}_{\w \ell} = 0$ there.

The normalization constant $N$ is fixed using the scalar product which is defined by the relation
\bea
  (f_1, f_2) = -i\int_{\Sigma} d \Sigma \, n^\mu [f_1(x)\overset{\leftrightarrow}{\partial_\mu}f_2^{*}(x)]
 \;. \label{scalar-products}
\eea
Here $n^{\mu}$ is a future-directed unit vector orthogonal to the spacelike (or null) hypersurface $\Sigma$ and $d\Sigma$ is the volume element in $\Sigma$. The hypersurface $\Sigma$ is taken to be a Cauchy surface.  To normalize the $in$ modes it is easiest to use past null infinity, $\mathscr{I}^-$, as the Cauchy surface.
If the orthonormal condition
\be  (f_{\w \ell m}, f_{\w' \ell' m'}) = \delta_{\ell,\ell'} \delta_{m,m'} \delta(\w-\w') \;, \label{f-norm} \ee
is imposed then it is straightforward to show that
\be N = \frac{1}{\sqrt{4 \pi \w}} \;. \label{N} \ee

For the modes in the $in$ state in the region inside the null shell trajectory
\be \psi^{in}_{\w \ell} = e^{-i \w t} \chi^{ in}_{\w \ell}(r) \;. \label{psi-in-in} \ee
Substituting this into~\eqref{psieq-flat} gives
\be     \frac{d^2 \chi^{in}_{\w \ell}}{d r^2} = - \left[\w^2 - \frac{\ell(\ell+1)}{r^2} \right] \chi^{in}_{\w \ell} \;. \label{mode-eq-inside}
\ee
The solution for which $\psi^{in}_{\w \ell}$ vanishes at $r = 0$ is
\be \psi^{in}_{\w \ell} =  C_{\ell} \, e^{-i \w t}\, \w r  j_\ell(\w r) \;, \label{psi-in-a} \ee
where $C_{\ell}$ is a normalization constant and $j_\ell $ is a spherical Bessel function.  The condition~\eqref{in-state-def} on $\mathscr{I}^{-}$ fixes the value of $C_{ \ell}$.  For example, for $\ell = 0$, it is easy to show that $C_{0} = -2 i $ and
\be \psi^{in}_{\w 0} = e^{-i \w v} - e^{-i \w u} \;. \label{psi-in-0} \ee

In the region outside of the null shell trajectory $v = v_0$, the $in$ modes still have the boundary condition~\eqref{in-state-def}.  However, their other boundary condition is that $\psi^{in}_{\w \ell}$ and its first derivatives must be continuous across $v = v_0$.  The fact that the time coordinates are different on either side of this surface makes it impossible to have a solution of the form $\psi^{in}_{\w \ell} = e^{-i \w t_s} \chi^{in}_{\w \ell}(r)$ outside the null shell trajectory.  However, it is possible to write $\psi^{in}_{\w \ell}$ in terms of a complete set of mode functions of the form
$\psi_{\w \ell} = e^{-i \w t_s} \chi_{\w \ell}(r)$ outside the null shell trajectory as is shown in Sec.~{sec:matching}.


\section{Method to compute the stress-energy tensor}
\label{sec:method-Tab}

The stress-energy tensor for the quantized massless minimally coupled scalar field, $\la T_{ab} \ra$, is to be computed for the $in$ vacuum state in the region outside the null shell and outside the event horizon. The stress-energy tensor for the classical field is
\be T_{ab} = \partial_a \Phi  \partial_b \Phi - \frac{1}{2} g_{ab} g^{cd} \partial_c \Phi  \partial_d \Phi  \;. \label{Tab-class} \ee
To compute $\la {in}| T_{ab} |{in} \ra$, one can substitute~\eqref{Phi-expansion} into~\eqref{Tab-class}, use the complete set of modes for the $in$ vacuum state $f^{in}_{\w \ell m}$, and compute the expectation value.  There are two things which make this difficult. One is computing the modes $f^{in}_{\w \ell m}$ in the region outside the shell and the other is renormalizing the stress-energy tensor.  Our method to compute the stress-energy tensor provides one way to overcome these difficulties.

First, we renormalize by subtracting from the unrenormalized expression for the stress-energy tensor for the $in$ vacuum state, the unrenormalized stress-energy tensor for the Unruh state.  Since the renormalization counterterms are local and thus do not depend on the state of the quantum field, this quantity will be finite.  Then we add back the
 unrenormalized stress-energy tensor for the Unruh state and then subtract from it the renormalization counter terms.
 Schematically one can write
 \bea \la {in}| T_{ab}| {in} \ra_{ren} &=& \Delta \la T_{ab} \ra +   \la U| T_{ab} | U \ra_{ren} \;,  \nonumber \\
               \Delta \la T_{ab} \ra  &=&  \la { in}| T_{ab} |{in} \ra_{unren} - \la U| T_{ab} | U \ra_{unren} \;. \label{Tab-Unruh-sub} \eea
The quantity $\la U| T_{ab} | U \ra_{ren}$ has been numerically computed for a massless minimally coupled scalar field in Schwarzschild spacetime~\cite{levi-ori, levi}.
Thus what remains is to compute the difference between the unrenormalized expressions.  To do that it is necessary to discuss the computation of the mode functions for the quantum field that are relevant for the $in$ and Unruh states.
It is worth pointing out that the computation of $\la U| T_{ab} | U \ra_{ren}$ done in~\cite{levi-ori,levi} was done for pure Schwarzschild spacetime outside the event horizon.  However, the computation we wish to do for
$\la {in}| T_{ab}| {in} \ra_{ren}$ is for the null shell spacetime outside both the shell and the horizon.  The reason that there is no problem is that the renormalization counterterms are local and so are the same in this part of the null shell spacetime as they are in pure Schwarzschild spacetime.

Analytic expressions for the mode functions in the $in$ vacuum state, $f^{in}_{\w \ell m}$
inside the shell are given in~\eqref{psi-in-a}.  However, it is not easy to continue these to the region outside the shell because the time coordinate $t$ and the right-moving radial null coordinate $u$ are not continuous across the shell.  
However, the known solutions inside the null shell along with their behavior on $\mathscr{I}^{-}$ can be used to fix the initial data on a Cauchy surface in the null shell spacetime.  The Cauchy surface we consider here, consists of the part of $\mathscr{I}^{-}$ with $v_0 \le v < \infty$ along with the trajectory of the null shell.  This initial data could be used for a numerical calculation of the partial differential equation satisfied by $f^{in}_{\w \ell m}$ outside the shell.  Alternatively, one can expand  $f^{in}_{\w \ell m}$ in terms of a complete set of modes in the region outside the shell and use the data on the Cauchy surface to determine the matching coefficients.

Here we take a variation of the latter approach by noting that the spacetime geometry outside the shell is the Schwarzschild geometry.  Because of this, it is possible to do the matching in the corresponding part of Schwarzschild spacetime.  The advantage of this is that the matching can be to a complete set of modes in the region outside the horizon in Schwarzschild spacetime.  These modes are well understood and straightforward to work with numerically.  The disadvantage is that the relevant part of the Cauchy surface in the null shell spacetime discussed above does not form a Cauchy surface in the Schwarzschild spacetime.  This can be remedied by adding a segment along the future horizon with $-\infty < v \le v_0$.  The result is a Cauchy surface for the part of Schwarzschild spacetime that is outside of the past and future horizons.  It is illustrated in Fig.~{fig:Cauchy}.  It is worth noting that the part of the Cauchy surface on the future horizon is not causally connected with the region outside the future horizon and outside the surface $v = v_0$.  The corresponding region in the null shell spacetime is the region where we want to compute the stress-energy tensor.  Thus any initial data
can be used for the mode function $f^{in}_{\w \ell m}$ on that surface so long as $f^{in}_{\w \ell m}$ is continuous at the point where the future horizon intersects the part of the Cauchy surface with $v = v_0$.

\begin{figure}[h]
\centering
\includegraphics   [trim=0cm 0cm 0cm 0cm,clip=true,totalheight=0.25\textheight]{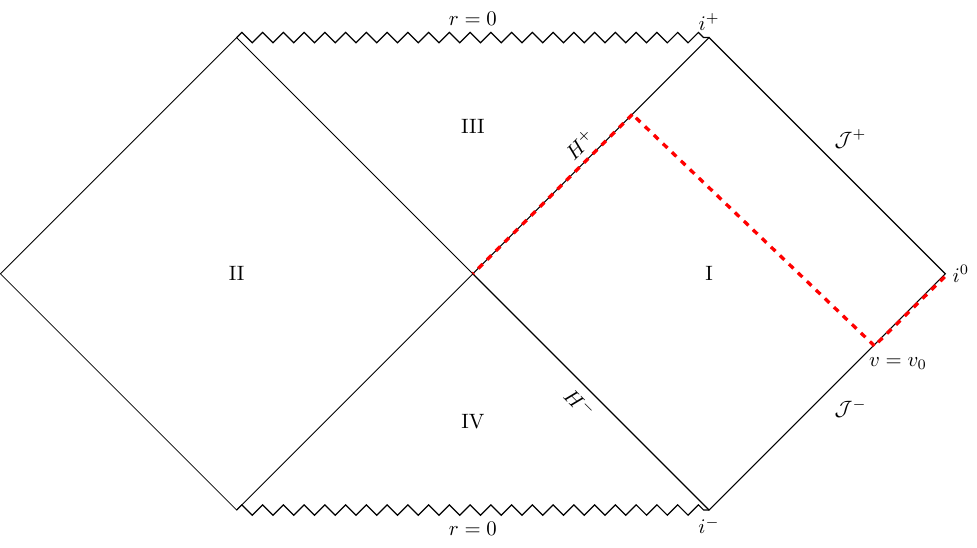}
\caption{Penrose diagram for Schwarzschild spacetime showing the Cauchy surface used for matching the $in$ modes in the null shell spacetime
to a complete set of modes in Schwarzschild spacetime in the region outside the past and future horizons.  The Cauchy surface is denoted by the dashed red curve. }
\label{fig:Cauchy}
\end{figure}

\section{Complete sets of mode functions in Schwarzschild spacetime}\label{ch4-sec-mode-functions}

In this paper, we work with four complete sets of mode functions for the part of Schwarzschild spacetime that is outside both the past and future horizons.
The frequencies of all of the modes that we consider are taken to be non-negative.

\subsection{Modes used for the $in$ state}

To expand the modes for the $in$ state in terms of a complete set of modes in Schwarzschild spacetime we find it most convenient to choose
 the complete set of modes that consists of the union of modes that are positive frequency on the future horizon $H^{+}$  and zero on future null infinity, $\mathscr{I}^{+}$, (labeled by $f^{H^{+}}_{\w \ell m}$) and modes that are positive frequency on $\mathscr{I}^{+}$ and zero on $H^{+}$ (labeled by $f^{\mathscr{I}^{+}}_{\w \ell m}$).
Both sets of modes are of the general form
\be \psi_{\w \ell} = e^{-i \w t_s} \chi_{\w \ell}(r) \;, \label{chi-def-2} \ee
with $0 \le \w < \infty$.
Substituting into~\eqref{psieq-Sch} gives the radial mode equation for Schwarzschild spacetime
\be \frac{d^2 \chi_{\w \ell}}{d r_{*}^2} = -  \left[\w^2 - \left(1 - \frac{2M}{r} \right) \left( \frac{2 M}{r^3}+  \frac{\ell(\ell+1)}{r^2} \right) \right] \chi_{\w \ell} \;. \label{chi-eq-Sch} \ee
These modes are normalized on the Cauchy surface consisting of $H^{+}$ and $\mathscr{I}^{+}$ with the result~\eqref{N}.

It is useful to consider a different complete set of mode functions of the form~\eqref{chi-def-2} which are defined by
two linearly independent solutions to the radial mode equation~\eqref{chi-eq-Sch} with the properties
\bes \bea \chi^\infty_R &\to& e^{i \w r_*} \;,  \qquad r_* \to \infty \;,\label{chi-infinity-R-def} \\
\chi^\infty_L &\to& e^{ -i \w r_*} \;, \qquad r_* \to \infty \label{chi-infinity-L-def}\;. \eea \label{chiRL-def} \ees
Near the event horizon they have the behaviors~\cite{rigorous}
\bes \bea \chi^\infty_R &\to& E_R(\w) e^{i \w r_*} + F_R(\w) e^{-i \w r_*} \;, \qquad r_* \to -\infty \;, \label{chiR-hor} \\
      \chi^\infty_L &\to& E_L(\w) e^{i \w r_*} + F_L(\w) e^{-i \w r_*} \;, \qquad r_* \to - \infty\;, \label{chiL-hor} \eea \label{chiRL-hor} \ees
where $E_R$, $E_L$, $F_R$, and $F_L$ are scattering parameters that can be determined numerically.\footnote{The subscripts $r$ and $l$ in~\cite{rigorous} have been changed here to $R$ and $L$ respectively.} They satisfy the relation $E_R F_L-E_L F_R = 1$.

For the modes $f^{H^{+}}_{\w \ell m}$, on the future horizon $\psi_{\w \, \ell}^{H^{+}} = e^{-i \w v}$ while $\psi_{\w \, \ell}^{H^{+}} =0$ on $\mathscr{I}^{+}$. The radial mode function which has these properties is
\be \chi^{H^{+}}_{\w \ell} = \frac{1}{F_L} \chi^\infty_L \;. \label{chi-H-plus-def} \ee
This is easily verified by evaluating the resulting mode function  $\psi^{H^{+}}_{\w \ell}$ on $H^{+}$ and $\mathscr{I}^{+}$.
To see how this works consider the behavior near $\mathscr{I}^{+}$:
\be \psi^{H^{+}}_{\w \ell} \to \frac{1}{F_L} e^{-i \w v - \epsilon v}  \to 0 \;, \qquad v \to \infty \;. \ee
Here we have used a positive integrating factor $\epsilon$ to explicitly show that this mode function vanishes on $\mathscr{I}^{+}$ where $v = \infty$.

For the modes $f^{\mathscr{I}^{+}}_{\w \ell m}$, on $H^{+}$, $\psi^{\mathscr{I}^{+}}_{\w \ell} = 0$ while on  $\mathscr{I}^{+}$, $\psi^{\mathscr{I}^{+}}_{\w \ell}  \to e^{-i \w u_s}$.  The radial mode function which results in these properties is
\be  \chi^{\mathscr{I}^{+}}_{\w \ell} = \chi^\infty_R - \frac{F_R}{F_L} \chi^\infty_L  \;.  \label{chi-scri-plus-def} \ee

\subsection{Complete sets of modes used to define the Unruh state}

Before discussing the modes that can be used to define the Unruh state, it is useful to consider a complete set of mode functions that are positive frequency on either the past horizon $H^{-}$ and vanish on $\mathscr{I}^{-}$ (denoted by $f^{H^{-}}_{\w \ell m}$   ) or which vanish on $H^{-}$ and are positive frequency on $\mathscr{I}^{-}$ (denoted by   $f^{\mathscr{I}^{-}}_{\w \ell m})$.
For the modes $f^{H^{-}}_{\w \ell m}$, on $H^{-}$, $\psi_{\w \, \ell}^{H^{-}} = e^{-i \w u_s}$ while $\psi_{\w \, \ell}^{H^{-}} =0$ on $\mathscr{I}^{-}$. The radial mode function which has these properties is~\cite{rigorous}
\be  \chi^{H^{-}}_{\w \ell} = \frac{ \chi^\infty_R}{E_R} \;. \label{chi-H-minus-def} \ee
For the modes $f^{\mathscr{I}^{-}}_{\w \ell m}$, on $H^{-}$, $\psi^{\mathscr{I}^{-}}_{\w \ell} = 0$ while on  $\mathscr{I}^{-}$, $\psi^{\mathscr{I}^{-}}_{\w \ell}  \to e^{-i \w v}$.  The radial mode function which has these properties is~\cite{rigorous}
\be \chi_{\w \ell}^{\mathscr{I}^{-}} =  \chi^\infty_L - \frac{E_L}{E_R} \chi^\infty_R  \;. \label{chi-scri-minus-def} \ee
These modes are normalized on the Cauchy surface consisting of $H^{-}$ and $\mathscr{I}^{-}$ with the result~\eqref{N}.

The Unruh state in Schwarzschild spacetime consists of a complete set of modes consisting of the modes $f^{\mathscr{I}^{-}}_{\w \ell m}$ and the modes 
(denoted by $f^{K}_{\w \ell m}$) that on $H^{-}$ have the form
\be \psi^K_{\w_K \ell} = e^{-i \w_K U}  \;, \label{psi-K} \ee
with
\be U = - \frac{e^{-\kappa u}}{\kappa} \;.  \label{U-Kruskal}  \ee
Here $\kappa = (4 M)^{-1}$ is the surface gravity of the black hole and $0 \le \w_K < \infty$.
These modes vanish on $\mathscr{I}^{-}$ and can be normalized on the Cauchy surface that is the union between $H^{-}$ and $\mathscr{I}^{-}$ with
the result that\footnote{Here we use the entire surface $H^{-}$ which extends from $U = -\infty$ to $U = + \infty$.}
\be f^{K}_{\w \ell m} = \frac{Y_{\ell m}}{\sqrt{4 \pi \w_K }\, r} \psi^K_{\w \ell}  \;. \label{fK-form} \ee
They can be expanded in terms of the $f^{H^{-}}_{\w \ell m}$ modes.  The result
is given in~\eqref{fK} and \eqref{alphaK-betaK}.

Since the method used to compute the stress-energy tensor involves subtracting the unrenormalized stress-energy tensor for the Unruh state it is useful to write the modes associated with this state, $f^{H^{-}}_{\w \ell m}$ and $f^{\mathscr{I}^{-}}_{\w \ell m}$ in terms of $f^{H^{+}}_{\w \ell m}$  and $f^{\mathscr{I}^{+}}_{\w \ell m}$.  The result is
\bes \bea f^{H^{-}}_{\w \ell m} &=& \frac{1}{E_R} ( F_R \,  f^{H^{+}}_{\w \ell m} +  f^{\mathscr{I}^{+}}_{\w \ell m} )  \;, \label{f-H-minus-2} \\
f_{\w \ell m}^{\mathscr{I}^{-}} &=& \frac{1}{E_R} (f^{H^{+}}_{\w \ell m} - E_L \, f^{\mathscr{I}^{+}}_{\w \ell m})  \;. \label{f-scri-minus-2}  \eea \label{f-scri-H-minus} \ees

\section{Matching Coefficients}
\label{sec:matching}

\subsection{General formulas}
\label{sec:matching-general}

In this section general formulas are derived for the matching coefficients used in an expansion of the modes of a massless minimally coupled scalar field for the $in$ vacuum state in the collapsing null shell spacetime in terms of a complete set of modes in Schwarzschild spacetime in the region outside the past and future horizons.  These can be used in the computation of the stress-energy tensor, $\left\langle in \middle| T_{ab} \middle| in \right\rangle$, for the scalar field in the part of the collapsing null shell spacetime that is outside of the shell and outside of the event horizon.

The expansion of the $in$ mode functions has the form
\bea
f^{in}_{\w \ell m} = \sum_{\ell'=0}^{\infty}\sum_{m'=-\ell'}^{\ell'} & \int_{0}^{\infty} d \w'
\Big[ A^{\mathscr{I^+}}_{\w \ell m\w' \ell' m'} f^{\mathscr{I}^+}_{\w'\ell' m'} + B^{\mathscr{I^+}}_{\w l m\w' \ell' m'} (f^{\mathscr{I}^+}_{\w' \ell' m'})^{*}
 \nonumber \\ & + A^{H^+}_{\w \ell m \w' \ell' m'} f^{H^+}_{\w' \ell' m'} + B^{H^+}_{\w \ell m\w'\ell' m'} (f^{H^+}_{\w' \ell' m'})^{*}\Big]\;. \label{General-in-modes}
\eea
The matching coefficients are found using the scalar product in~\eqref{scalar-products} and the orthonormality of the modes $f^{(\mathscr{I}^{+}, \, H^{+})}$ with respect to this scalar product.  The result is
\bes \bea
A^{(\mathscr{I^+}, H^+)}_{\w \ell m\w' \ell' m'} &=&( f^{in}_{\w \ell m} , f^{(\mathscr{I}^+, H^+)}_{\w' \ell' m'}) \;, \\
B^{(\mathscr{I^+}, H^+)}_{\w \ell m\w' \ell' m'} &=& -( f^{in}_{\w \ell m} , (f^{(\mathscr{I}^+, H^+)}_{\w' \ell' m'})^{*}) \;.
\eea \ees
For the Cauchy surface we consider, ~\eqref{scalar-products} reduces to integrals of the form
\bea
    \int du\int d\Omega r^2\overset{\leftrightarrow}{\partial_u}\;, \quad  \quad \int dv\int d\Omega r^2\overset{\leftrightarrow}{\partial_v}\;.
\eea
On the hypersurfaces where these integrals are computed, the following properties for spherical harmonics can be used:
\bes \bea
    \int d\Omega Y_{lm}(\theta, \phi)Y_{l'm'}^{*}(\theta, \phi)&=&\delta_{l,l'} \delta_{m,m'},  \\
    \int d\Omega Y_{lm}(\theta, \phi)Y_{l'm'}(\theta, \phi)&=&(-1)^{m}\delta_{l,l'} \delta_{m,-m'} \;.
\eea \ees
As a result we can write
\bes \bea A^{(\mathscr{I^+}, H^+)}_{\w l m\w' l' m'} & =& \delta_{l,l'} \delta_{m,m'} A^{(\mathscr{I^+}, H^+)}_{\w \w' \ell}  \;, \label{A-short-def} \\
          B^{(\mathscr{I^+}, H^+)}_{\w l m\w' l' m'} & =& (-1)^m \delta_{l,l'} \delta_{m,-m'} B^{(\mathscr{I^+}, H^+)}_{\w \w' \ell}  \;, \label{B-short-def}
\eea \label{A-B-short-def} \ees
and
\be
f^{in}_{\w \ell m} =  \frac{Y_{\ell m}}{r \sqrt{4 \pi}} \int_{0}^{\infty} \frac{d \w'}{\sqrt{\w'}}
\Big[ A^{\mathscr{I^+}}_{\w \w' \ell} \psi^{\mathscr{I}^+}_{\w'\ell} + B^{\mathscr{I^+}}_{\w \w' \ell} (\psi^{\mathscr{I}^+}_{\w' \ell})^{*}
 + A^{H^+}_{\w \w' \ell} \psi^{H^+}_{\w' \ell} + B^{H^+}_{\w \w'\ell} (\psi^{H^+}_{\w' \ell})^{*}\Big]\;. \label{General-in-modes-2}
\ee
From this expression, one can see that if, at small $\w'$, the matching coefficients go like $\frac{1}{\sqrt{\w'}}$ then there is an infrared divergence
in the integral and it is not obvious how to deal with it.  For this reason, we use integrations by parts in some of the computations of the matching coefficients below to avoid this difficulty.  For Schwarzschild spacetime in 4D, our results when substituted into~\eqref{General-in-modes-2} do not give infrared divergences.

The contribution to the matching coefficients from the three segments of the Cauchy surface in Fig.~{fig:Cauchy} are
\bes \bea
A^{(\mathscr{I^+}, H^+)}_{\w \w' l} & =& \left(A^{(\mathscr{I^+}, H^+)}_{\w \w' l} \right)_{H^{+}} + \left(A^{(\mathscr{I^+}, H^{+})}_{\w \w' l} \right)_{v_0}
   +  \left(A^{(\mathscr{I^+}, H^+)}_{\w \w' l} \right)_{\mathscr{I}^{-}} \;, \label{A-parts} \\
\left(A^{(\mathscr{I^+}, H^+)}_{\w \w' l} \right)_{H^{+}} &=&
-\frac{i}{4\pi\sqrt{\omega\omega'}}\int_{-\infty}^{v_{0}}dv\;\psi^{in}_{\omega l}(u = v_H,v) \overset{\leftrightarrow}{\partial_v}[\psi_{\omega' l}^{(\mathscr{I^+}, H^+)}(u_s = \infty,v)]^{*} \;,  \label{A-Hplus} \\
\left(A^{(\mathscr{I^+}, H^{+})}_{\w \w' l} \right)_{v_0} &=& -\frac{i}{4\pi\sqrt{\omega\omega'}} \int_{-\infty}^{v_{H}}du\;\psi^{in}_{\omega l}(u,v_0) \overset{\leftrightarrow}{\partial_u}[\psi_{\omega' l}^{(\mathscr{I^+}, H^+)}(u_s(u),v_0)]^{*} \;, \label{A-v0} \\
 \left(A^{(\mathscr{I^+}, H^+)}_{\w \w' l} \right)_{\mathscr{I}^{-}} &=& -\frac{i}{4\pi\sqrt{\omega\omega'}}\int_{v_{0}}^{\infty}dv\;\psi^{in}_{\omega l}(u =-\infty,v) \overset{\leftrightarrow}{\partial_v}[\psi_{\omega' l}^{(\mathscr{I^+}, H^+)}(u_s=-\infty,v) ]^{*}\;, \label{A-Iminus} \eea \ees
and
\bes \bea
B^{(\mathscr{I^+}, H^+)}_{\w \w' l} & =& \left(B^{(\mathscr{I^+}, H^+)}_{\w \w' l} \right)_{H^{+}} + \left(B^{(\mathscr{I^+}, H^{+})}_{\w \w' l} \right)_{v_0}
   +  \left(B^{(\mathscr{I^+}, H^+)}_{\w \w' l} \right)_{\mathscr{I}^{-}} \;, \label{B-parts} \\
\left(B^{(\mathscr{I^+}, H^+)}_{\w \w' l} \right)_{H^{+}} &=&
\frac{i}{4\pi\sqrt{\omega\omega'}}\int_{-\infty}^{v_{0}}dv\;\psi^{in}_{\omega l}(u = v_H,v) \overset{\leftrightarrow}{\partial_v} \psi_{\omega' l}^{(\mathscr{I^+}, H^+)}(u_s = \infty,v) \;, \label{B-Hplus} \\
\left(B^{(\mathscr{I^+}, H^{+})}_{\w \w' l} \right)_{v_0} &=& \frac{i}{4\pi\sqrt{\omega\omega'}} \int_{-\infty}^{v_{H}}du\;\psi^{in}_{\omega l}(u,v_0) \overset{\leftrightarrow}{\partial_u}\psi_{\omega' l}^{(\mathscr{I^+}, H^+)}(u_s(u),v_0) \label{B-v0} \;, \\
 \left(B^{(\mathscr{I^+}, H^+)}_{\w \w' l} \right)_{\mathscr{I}^{-}} &=& \frac{i}{4\pi\sqrt{\omega\omega'}}\int_{v_{0}}^{\infty}dv\;\psi^{in}_{\omega l}(u=-\infty,v) \overset{\leftrightarrow}{\partial_v}\psi_{\omega' l}^{(\mathscr{I^+}, H^+)}(u_s=-\infty,v) \;. \label{B-Iminus} \eea \ees

It is important to note that the integrals in~\eqref{A-Hplus} and~\eqref{B-Hplus} are computed with the integrands evaluated on $H^{+}$.  Since
$\psi_{\omega' l}^{\mathscr{I^+}} = 0$ on $H^{+}$
\be \left(A^{\mathscr{I^+}}_{\w \w' l} \right)_{H^{+}} = \left(B^{\mathscr{I^+}}_{\w \w' l} \right)_{H^{+}} = 0 \;. \ee
On $H^{+}$,  $\psi^{H^{+}}_{\w \ell} = e^{-i \w v}$.  The part of the Cauchy surface in Schwarzschild spacetime
which is on $H^{+}$ has no counterpart in the collapsing null shell spacetime and, as discussed in Sec.~{sec:method-Tab}, is causally disconnected from the region outside the collapsing null shell
and outside the event horizon.  Thus the only restriction on the mode functions $\psi^{in}_{\w \ell}$
for this surface is continuity at $v = v_0$.
The simplest mode function to use on this surface is then
\be \psi^{in}_{\w \ell}(u = v_H, v) = \psi^{in}_{\w \ell}(u = v_H, v_0) \;. \ee
With this choice, it turns out to be useful to write the contribution to the matching coefficients from $H^+$ in the form
\bes \bea \left(A^{ H^+}_{\w \w' l} \right)_{H^{+}} &=&  \frac{i}{4 \pi \sqrt{\w \w'}} \psi^{in}_{\w \ell}(v_H,v_0) e^{i \w' v_0}
  - \frac{i}{2 \pi} \sqrt{\frac{\w'}{\w}} \frac{e^{i \w' v_0}}{\w' - i \epsilon} \psi^{in}_{\w \ell}(v_H,v_0)
   \;, \\
  \left(B^{ H^+}_{\w \w' l} \right)_{H^{+}} &=&  -\frac{i}{4 \pi \sqrt{\w \w'}} \psi^{in}_{\w \ell}(v_H,v_0) e^{-i \w' v_0}
 + \frac{i}{2 \pi} \sqrt{\frac{\w'}{\w}} \frac{e^{-i \w' v_0}}{\w' + i \epsilon} \psi^{in}_{\w \ell}(v_H,v_0)  \;, \eea \ees
where for each integral an integration by parts has been done and an integrating factor $0 < \epsilon \ll 1$ has been included to make the integrals converge.

The integrals in~\eqref{A-v0} and~\eqref{B-v0} are computed with the integrands evaluated on the surface $v = v_0$.  In this case $\psi^{in}_{\w \ell}$ is
given by~\eqref{psi-in-a} while analytic expressions for $\psi^{\mathscr{I}^{+}, H^{+}}_{\w \ell}$ are only known for the limits $u_s \to \pm \infty$.  For all intermediate
values of $u_s$ these modes must be computed numerically.

The integrals in~\eqref{A-Iminus} and~\eqref{B-Iminus} are computed with the integrands evaluated on the surface $\mathscr{I}^{-}$.  In this case $\psi^{in}_{\w \ell}$ is given by~\eqref{in-state-def}.  From~\eqref{chi-H-plus-def} and~\eqref{chi-scri-plus-def} one can deduce that on $\mathscr{I}^{-}$
\bea \psi^{ H^{+}}_{\w \ell} &=& \frac{1}{F_L} e^{-i \w v}  \;, \nonumber \\
     \psi^{ \mathscr{I}^{+}}_{\w \ell} &=& - \frac{F_R}{F_L} e^{-i \w v}  \;. \label{psi-Hplus-Iplus-Iminus} \eea

To avoid infrared divergences in~\eqref{General-in-modes-2} it is useful to subtract and then add back the quantity $e^{-i \w v_0}$ from $\psi^{in}_{\w \ell}$
in~\eqref{A-v0} and~\eqref{B-v0}.  Then after integrations by parts the contributions from the surface $v = v_0$ can be written as
\bes \bea  \left(A^{ H^{+}}_{\w \w' l} \right)_{v_0} &=& - \frac{i}{4 \pi \sqrt{\w \w'} }  \psi^{in}_{\w \ell}(v_H,v_0) e^{i \w' v_0} +  \frac{i}{4 \pi \sqrt{\w \w'}\, F_L^{*}(\w',\ell)}  e^{-i (\w-\w') v_0}  \nonumber \\
   & & + \frac{i}{2 \pi \sqrt{\w \w'} } \int_{-\infty}^{v_H} du \, \left[\partial_u \psi^{in}_{\w \ell}(u,v_0) \right] \psi^{H^{+} *}_{\w' \ell}(u_s(u),v_0) \;, \\
   \left(B^{ H^{+}}_{\w \w' l} \right)_{v_0} &=&  \frac{i}{4 \pi \sqrt{\w \w'} }  \psi^{in}_{\w \ell}(v_H,v_0) e^{-i \w' v_0} -  \frac{i}{4 \pi \sqrt{\w \w'}\, F_L(\w',\ell)}  e^{-i (\w+\w') v_0}  \nonumber \\
   & & - \frac{i}{2 \pi \sqrt{\w \w'} } \int_{-\infty}^{v_H} du \, \left[\partial_u \psi^{in}_{\w \ell}(u,v_0) \right] \psi^{H^{+}}_{\w' \ell} (u_s(u),v_0)\;, \\
  \left(A^{ \mathscr{I}^{+}}_{\w \w' l} \right)_{v_0} &=& - \frac{i}{4 \pi \sqrt{\w \w'} } \frac{F^{*}_R(\w',\ell)}{F^{*}_L(\w', \ell)} e^{-i (\w - \w') v_0} \nonumber \\
 && - \frac{i}{2 \pi \sqrt{\w \w'}} \int_{-\infty}^{v_H} du \,  [\psi^{in}_{\w \ell}(u, v_0) - e^{-i \w v_0}] \partial_u \psi^{\mathscr{I}^+ *}_{\w' \ell}(u_s(u),v_0) \;, \\
  \left(B^{ \mathscr{I}^{+}}_{\w \w' l} \right)_{v_0} &=&  \frac{i}{4 \pi \sqrt{\w \w'} } \frac{F_R(\w',\ell)}{F_L(\w', \ell)} e^{-i (\w + \w') v_0} \nonumber \\
 && + \frac{i}{2 \pi \sqrt{\w \w'}} \int_{-\infty}^{v_H} du \, \left[ \psi^{in}_{\w \ell}(u, v_0) - e^{-i \w v_0} \right] \partial_u \psi^{\mathscr{I}^+}_{\w' \ell}(u_s(u),v_0) \;.
      \eea \ees

The modes on $\mathscr{I}^{-}$ take on the simple forms~\eqref{in-state-def} and~\eqref{psi-Hplus-Iplus-Iminus}.  This makes it possible to evaluate the integrals for the contributions to the matching parameters from that surface. After integrating by parts, we find
\bes \bea  \left(A^{ H^{+}}_{\w \w' l} \right)_{\mathscr{I}^{-}} &=& - \frac{i}{4 \pi \sqrt{\w \w'} \, F^{*}_L(\w',\ell)} e^{-i (\w-\w') v_0}
 + \frac{i}{2 \pi} \sqrt{\frac{\w'}{\w}} \frac{1}{F_L^{*}(\w',\ell)}\frac{e^{i(\w'-\w)v_0}}{\w'-\w+i \epsilon}  \;, \\
\left(B^{ H^{+}}_{\w \w' l} \right)_{\mathscr{I}^{-}} &=& \frac{i}{4 \pi \sqrt{\w \w'} \, F_L(\w',\ell)} e^{-i (\w+\w') v_0}
 - \frac{i}{2 \pi} \sqrt{\frac{\w'}{\w}} \frac{1}{F_L(\w',\ell)}\frac{e^{-i(\w+\w')v_0}}{\w'+\w-i \epsilon}  \;, \\
\left(A^{ \mathscr{I}^{+}}_{\w \w' l} \right)_{\mathscr{I}^{-}} &=& -\frac{i}{4 \pi \sqrt{\w \w'}}\frac{F_R^{*}(\w',\ell)}{ F_L^{*}(\w',\ell)} e^{-i (\w-\w') v_0}
 - \frac{i}{2 \pi} \sqrt{\frac{\w'}{\w}} \frac{F_R^{*}(\w',\ell)}{F_L^{*}(\w',\ell)}\frac{e^{-i(\w-\w')v_0}}{\w'-\w+i \epsilon}  \;, \\
\left(B^{ \mathscr{I}^{+}}_{\w \w' l} \right)_{\mathscr{I}^{-}} &=& -\frac{i}{4 \pi \sqrt{\w \w'}}\frac{F_R(\w',\ell)}{ F_L(\w',\ell)} e^{-i (\w+\w') v_0}
 + \frac{i}{2 \pi} \sqrt{\frac{\w'}{\w}} \frac{F_R(\w',\ell)}{F_L(\w',\ell)}\frac{e^{-i(\w+\w')v_0}}{\w'+\w-i \epsilon}  \;.
\eea \ees

Combining these results together, the general formulas for the matching coefficients are
\bes \bea A^{ H^{+}}_{\w \w' l} &=& - \frac{i}{2 \pi} \sqrt{\frac{\w'}{\w}} \frac{e^{i \w' v_0}}{\w' - i \epsilon} \psi^{in}_{\w \ell}(v_H,v_0)
   + \frac{i}{2 \pi} \sqrt{\frac{\w'}{\w}} \frac{1}{F_L^{*}(\w',\ell)}\frac{e^{i(\w'-\w)v_0}}{\w'-\w+i \epsilon}  \nonumber \\
   & & + \frac{i}{2 \pi \sqrt{\w \w'} } \int_{-\infty}^{v_H} du \, \left[\partial_u \psi^{in}_{\w \ell}(u,v_0) \right] \psi^{H^{+} *}_{\w' \ell}(u_s(u),v_0) \;, \label{A-H-gen-mat}\\
      B^{ H^{+}}_{\w \w' l} &=& \frac{i}{2 \pi} \sqrt{\frac{\w'}{\w}} \frac{e^{-i \w' v_0}}{\w' + i \epsilon} \psi^{in}_{\w \ell}(v_H,v_0)
      - \frac{i}{2 \pi} \sqrt{\frac{\w'}{\w}} \frac{1}{F_L(\w',\ell)}\frac{e^{-i(\w+\w')v_0}}{\w'+\w-i \epsilon} \nonumber \\
      & & - \frac{i}{2 \pi \sqrt{\w \w'} } \int_{-\infty}^{v_H} du \, \left[\partial_u \psi^{in}_{\w \ell}(u,v_0) \right] \psi^{H^{+}}_{\w' \ell} (u_s(u),v_0)\;, \label{B-H-gen_mat}\\
     A^{ \mathscr{I}^{+}}_{\w \w' l} &=&  - \frac{i}{2 \pi} \sqrt{\frac{\w'}{\w}} \frac{F_R^{*}(\w',\ell)}{F_L^{*}(\w',\ell)}\frac{e^{-i(\w-\w')v_0}}{\w'-\w+i \epsilon} \nonumber \\  & &
       -\frac{i}{2 \pi \sqrt{\w \w'}} \int_{-\infty}^{v_H} du \, \left[  \psi^{in}_{\w \ell}(u, v_0) - e^{-i \w v_0} \right] \partial_u \psi^{\mathscr{I}^+  *}_{\w' \ell}(u_s(u),v_0) \;,  \label{A-I-gen-mat} \\
     B^{ \mathscr{I}^{+}}_{\w \w' l} &=&  \frac{i}{2 \pi} \sqrt{\frac{\w'}{\w}} \frac{F_R(\w',\ell)}{F_L(\w',\ell)}\frac{e^{-i(\w+\w')v_0}}{\w'+\w-i \epsilon} \nonumber \\
  & &     + \frac{i}{2 \pi \sqrt{\w \w'}} \int_{-\infty}^{v_H} du \, \left[ \psi^{in}_{\w \ell}(u, v_0) - e^{-i \w v_0} \right] \partial_u \psi^{\mathscr{I}^+}_{\w' \ell}(u_s(u),v_0) \;.
     \label{B-I-gen-mat} \eea \label{gen-mat}\ees

\subsection{Expansions of modes for the Unruh state}

As discussed in Sec.~{sec:method-Tab}, our method for renormalizing the stress-energy tensor involves subtracting the unrenormalized stress-energy tensor for the Unruh modes.  Recall these modes include the set of modes $f^K_{\w \ell m}$ that are positive frequency on the past horizon with respect to the Kruskal time coordinate along with the set of modes $f^{\mathscr{I}^{-}}_{\w \ell m}$ that on $\mathscr{I}^{-}$ have $\psi_{\w \ell} = e^{-i \w v}$.
Before subtracting the contribution from the $f^{\mathscr{I}^{-}}_{\w \ell m}$ modes we first write them in terms of $f^{\mathscr{I}^{+}}_{\w \ell m}$ and $f^{H^{+}}_{\w \ell m}$ using~\eqref{f-scri-minus-2}.

For the contributions of the $f^K_{\w \ell m}$ modes we first write down the Bogolubov transformation
\be f^K_{\w \ell m} =  \int_0^\infty d \w' \, \left[  \alpha^K_{\w \w' \ell} f^{H^{-}}_{\w' \ell m} + \beta^K_{\w \w' \ell}   f^{H^{-}*}_{\w' \ell m} \right]  \;. \label{fK} \ee
 The coefficients can be obtained using the scalar product~\eqref{scalar-products}
   with a Cauchy surface consisting of the union of past null infinity and the past horizon in Schwarzschild spacetime.  Integrating over the angular coordinates one finds that
   the Bogolubov coefficients can be written in the form~\eqref{A-B-short-def} with $\alpha$ replacing $A$ and $\beta$ replacing $B$.
   Integrating the remaining integrals over $u_s$ by parts one finds that\footnote{This calculation was originally done in~\cite{bec} but note that there is a mistake in the results.  The expressions in that paper are missing a factor of $(4M)^{\pm  i 4M\w'}$. }
\bes \bea \alpha^K_{\w_K \w' \ell} &=&  \frac{1}{2 \pi} \sqrt{\frac{\w'}{\w_K}} (4 M)^{1+i 4 M \w'}  \int_{-\infty}^0 d U_K e^{-i \w_K U_K} (-U_K)^{-1 - i 4 M \w'} \nonumber \\
&=& \frac{1}{2 \pi} \sqrt{\frac{\w'}{\w_K}} \, (4M)^{1 +i 4 M \w'} \frac{\Gamma(\delta - i 4 M \w)}{ (-i \w_K + \epsilon)^{-i 4 M \w'}} \;, \label{alphaK} \\
\beta^K_{\w_K \w' \ell} &=&  \frac{1}{2 \pi} \sqrt{\frac{\w'}{\w_K}} (4 M)^{1-i 4 M \w'}  \int_{-\infty}^0 d U_K e^{-i \w_K U_K} (-U_K)^{-1 + i 4 M \w'} \nonumber \\
&=& \frac{1}{2 \pi} \sqrt{\frac{\w'}{\w_K}} \, (4M)^{1 -i 4 M \w'} \frac{\Gamma(\delta + i 4 M \w)}{ (-i \w_K + \epsilon)^{i 4 M \w'}} \;. \label{betaK}
\eea \label{alphaK-betaK} \ees
Here $\delta$ and $\epsilon$ are integrating factors with $0 < \delta \ll 1$ and $0 < \epsilon \ll 1$.  Note that the Bogolubov coefficients are independent of the value of $\ell$.  This is because the effective potential vanishes on $H^{-}$ which is the surface where the integrals are being computed.
Then we use~\eqref{f-scri-H-minus} to express the modes $f^{(H^{-},\,\mathscr{I}^{-})}_{\w \ell m}$ in terms of the modes $f^{(\mathscr{I}^{+},\;H^{+})}_{\w \ell m}$.

\subsection{2D example}

In this section we will illustrate the matching for the case of a 2D spacetime which has a perfectly reflecting mirror at $r = 0$.  The metric
inside the shell is the flat space metric
\be
ds^2 = -dt^2 + dr^2    \;, \label{metric-flat-2D}
\ee
and the metric outside the shell is the Schwarzschild metric
\be
ds^2 = -\left(1-\frac{2M}{r} \right) dt_s^2
+ \left( 1-\frac{2M}{r} \right)^{-1} dr^2   \;. \label{metric-sch-2D}
\ee
The Penrose diagram is the same as in the 4D case as is the definition of the radial null coordinates $u$, $u_s$, and $v$ and the relation between $u$ and $u_s$.

The general form of the mode functions is
\be f_\w = \frac{\psi_\w}{\sqrt{4 \pi \w}}  \;.  \label{f-form-2D} \ee
There is no scattering for the massless minimally coupled scalar field modes in 2D so
\be E_R = F_L = 1\;, \qquad E_L = F_R = 0 \;. \label{scattering-coeff-2D} \ee

Inside the shell the $in$ modes are
\be \psi^{in}_\w =   e^{-i \w v} - e^{-i \w u} \;.  \label{psi-in-2D} \ee
In the region outside the shell the spacetime is the 2D version of Schwarzschild spacetime and the modes are
\bes \bea   \psi^{\mathscr{I}^{+}}_\w &=& \psi^{H^{-}}_\w = e^{- i \w u_s}  \;, \label{right-moving-2D} \\
           \psi^{H^{+}}_\w &=& \psi^{\mathscr{I}^{-}}_\w  = e^{-i \w v} \;. \label{left-moving-2D} \eea \label{sch-modes-2D} \ees

The expansion for the $in$ modes is similar to the 4D case except there are no parameters $\ell$ and $m$ related to the spherical harmonics.  Thus
\bea
f^{in}_{\w} & =  & \int_{0}^{\infty} d \w'
\Big[ A^{H^+}_{\w \w' } f^{H^+}_{\w' } + B^{H^+}_{\w \w'} (f^{H^+}_{\w' })^{*}
   + A^{\mathscr{I^+}}_{\w \w'} f^{\mathscr{I}^+}_{\w'} + B^{\mathscr{I^+}}_{\w \w' } (f^{\mathscr{I}^+}_{\w' })^{*}\Big]\;.\label{2D-in-mode-expansion}
\eea
The matching coefficients are given by substituting~\eqref{scattering-coeff-2D},~\eqref{psi-in-2D}, and~\eqref{sch-modes-2D} into~\eqref{gen-mat}.  It is then easy to show that
\bea \left[ B^{H^{+}}_{\w,  \w'} f^{H^+\, *}_{\w' } \right]_{\w' \to - \w'} &=&  A^{H^{+}}_{\w,  \w'} f^{H^+}_{\w' } \;, \nonumber \\
     \left[ B^{\mathscr{I}^{+}}_{\w,  \w'} f^{\mathscr{I}^{+} \, *}_{\w' } \right]_{\w' \to - \w'} &=&  A^{ \mathscr{I}^{+}}_{\w,  \w'} f^{\mathscr{I}^{+}}_{\w' } \;,
   \label{B-w-minus-w} \eea
where the quantities on the right-hand side are to be evaluated at $\w' < 0$.  As a result
\bea
f^{in}_{\w} & =  & \int_{-\infty}^{\infty} d \w'
\Big[   A^{H^+}_{\w \w' } f^{H^+}_{\w' } + A^{\mathscr{I^+}}_{\w \w'} f^{\mathscr{I}^+}_{\w'} \Big]\;. \label{2D-in-mode-expansion-2}
\eea
Because $\psi^{H^{+}}_{\w'}$ does not depend on $u$, the integral in~\eqref{A-H-gen-mat} is trivial to evaluate and one finds that
\be A^{H^+}_{\w \w'} = - \frac{i}{2 \pi} \sqrt{\frac{\w'}{\w}} \frac{e^{-i (\w - \w') v_0}}{\w' - i \epsilon}
   + \frac{i}{2 \pi} \sqrt{\frac{\w'}{\w}} \frac{e^{i(\w'-\w)v_0}}{\w'-\w+i \epsilon}
  \;. \label{A-H-2D} \ee

To see what the contribution to $f^{in}_\w$ is from the $f^{H^{+}}_\w$ modes, first substitute ~\eqref{A-H-2D} into the first two terms of~\eqref{2D-in-mode-expansion} along with~\eqref{f-form-2D}
and~\eqref{psi-in-2D} with the result
\bea \left(f^{in}_\w \right)_{H^{+}}  &=&  \frac{i e^{-i \w v_0}}{2 \pi \sqrt{4 \pi \w}} \int_{-\infty}^\infty d \w' \; \left[ e^{i \w' (v_0 - v)} \left(-\frac{1}{\w' - i \epsilon}  +   \frac{1}{\w' - \w + i \epsilon} \right) \right] \nonumber \\
 & = &  \frac{ e^{-i \w v_0}}{\sqrt{4 \pi \w}}  \theta(v_0-v) + \frac{ e^{-i \w v}}{\sqrt{4 \pi \w}}  \theta(v-v_0) \;. \label{f-in-H+-2D}\eea

We next consider the contribution of the $f^{\mathscr{I}^{+}}$ modes.  The matching coefficient in~\eqref{A-I-gen-mat} is
\bea A^{\mathscr{I}^{+}}_{\w \w'} &=& -\frac{1}{2 \pi }\sqrt{\frac{\w'}{\w}} \int_{-\infty}^{v_H} du \, e^{-i \w u}  e^{i \w' u_s(u)} \frac{d u_s}{du} \nonumber \\
&=&
-\frac{1}{2 \pi} \sqrt{\frac{\w'}{\w}}  \int_{-\infty}^{v_H} du \, e^{-i (\w - \w') u}  \left(\frac{v_H-u}{4 M} \right)^{-i 4 M \w'} \left[ 1 + \frac{4M}{v_h-u} \right]  \;. \eea
Changing variables to $x = v_H -u$ and performing an integration by parts gives
\bea A^{\mathscr{I}^{+}}_{\w \w'} &=& \frac{i}{2 \pi} \sqrt{\w \w'} e^{-i(\w-\w') v_H} \frac{(4 M)^{1+i 4 M \w'}}{i (\w'-\w)+\epsilon} \int_0^\infty dx \; e^{i(\w-\w')x-\epsilon x}  \;
  x^{-i 4 M \w' -1 + \delta} \nonumber \\
  &=& \frac{i}{2 \pi} \sqrt{\w \w'} e^{-i(\w-\w') v_H} (4 M)^{1+i 4 M \w'} \frac{\Gamma(\delta - i 4 M \w')}{[i (\w'-\w)+ \epsilon]^{1-i4M\w'} }  \;. \label{Ascrp2D} \eea
Note that two integrating factors have been used with $0 < \epsilon \ll 1$ and $0 < \delta \ll 1$.

To find the contribution to $f^{in}_\w$ from the $f^{\mathscr{I}^{+}}_\w$ modes, first substitute~\eqref{Ascrp2D}  into~\eqref{2D-in-mode-expansion-2} with the result
\bea \left(f^{in}_\w\right)_{\mathscr{I}^{+}} &=& \frac{i 4 M \sqrt{\w}}{2 \pi \sqrt{4 \pi}} e^{-i \w v_H}
\int_{-\infty}^\infty d \w'   e^{i \w' (v_H-u_s)} (4 M)^{i 4 M \w'} \frac{\Gamma(\delta - i 4 M \w')}{[i (\w'-\w) + \epsilon]^{1-i 4 M \w'}} \;. \label{I1-a} \eea
Note that the denominator has an essential singularity in the upper half $\w'$ plane while the Gamma function has simple poles in the lower half plane at
 \be  \w' = - \frac{i \delta}{4 M} \;, \ee
  and
  \be \w' = - \frac{i n}{4 M} \;, \qquad n = 1, 2, \ldots \ee
In the complex plane at large $|\w'|$ Sterling's approximation gives
\be  \Gamma( -i 4 M \w') \approx \sqrt{2 \pi} e^{i 4 M \w'} e^{(-i 4 M \w' - 1/2) \log(-i 4 M \w')}   \;. \ee
Using the usual change of variables $\w' = R e^{i \theta}$, with $R > 0$, it is straightforward to show that the dominant contribution to the integrand of~\eqref{I1-a} in the large $R$ limit comes from the factor $e^{4 M R \sin \theta \, \log R}$, and therefore one must close in the lower half plane.  This means there is no contribution from the essential singularity but there is a contribution from each pole of the Gamma function.  At these poles it is straightforward to show that
\be \Gamma(\delta -i 4 M \w) \to \frac{(-1)^n}{n! (n - i 4 M \w)}  \;, \qquad n = 0, 1, 2, \ldots  \ee
Then
\be  \left(f^{in}_\w\right)_{\mathscr{I}^{+}}  = \frac{4Mi\sqrt{\w}}{\sqrt{4 \pi}} e^{- i \w v_H} \sum_{n=0}^\infty \frac{(-1)^n}{n!} (n-i 4 M \w)^{n-1}
    \left[\exp\left(\frac{(v_H-u_s)}{4 M} \right)\right]^n  \;. \label{finu} \ee
Because the general solutions to the 2D mode equation in Schwarzschild spacetime are of the form $\psi = g(u_s) + h(v)$ with $g$ and $h$ arbitrary functions, the exact solution for the $in$ modes is
\be \left(f^{in}_\w\right)_{\mathscr{I}^{+}} = - \frac{e^{-i \w u(u_s)}}{\sqrt{4\pi\w}}  = -\frac {e^{-i \w v_H}}{\sqrt{4\pi\w}} \, \exp \left\{ i 4 M \w\;  \text{W}\left[\exp\left(\frac{(v_H-u_s)}{4M}\right)\right] \right\} \;, \label{f(u)Lambert}\ee
where~\eqref{u-us} has been used and $\text{W}(z)$ is the Lambert W function.
  To make a comparison between ~\eqref{finu} and ~\eqref{f(u)Lambert}, one needs to write the latter in terms of a series. This has been done in~\cite{Corless}.
  An alternative derivation is given in Section ~{appendix-A}.  The result is
\be e^{- c \text{W}(z)}=\sum_{n=0}^\infty \frac {c (n+c)^{n-1}}{(n)!} (-z)^n\;. \label{final-Lambert}\ee
Taking $c=-4iM\w$ and $z=\exp\left(\frac{v_H-u_s}{4M}\right)$ in ~\eqref{final-Lambert}, one can see that~\eqref{f(u)Lambert} and ~\eqref{finu} are equivalent.

\subsection{Delta function potential}
\label{sec:delta-function}

In this section, we apply our matching method to the case where the potential term in~\eqref{chi-eq-Sch} is replaced by
\be V = \lambda \delta(r_*)  \;, \label{V-delta} \ee
with $\lambda$ a positive real constant.  This can serve as a model for the original potential which has a single peak and vanishes at the horizon and infinity.
The resulting mode equation can be solved analytically and the solutions are simple enough that the matching coefficients can be computed analytically. Some of these matching coefficients will be used to partially reconstruct the mode functions $f^{in}_{\w \ell}$ in the case that $\ell = 0$.

For $\ell = 0$ in 4D the $in$ modes inside the null shell take on the particularly simple form~\eqref{psi-in-0}.
In the region outside the shell the mode functions in the complete set with $\ell = 0$  have the general form
\be f^{(H^{+}, \mathscr{I}^{+})}_{\w' 0 0} = \frac{Y_{00}}{r \sqrt{4 \pi \w'}} \psi^{(H^{+}, \mathscr{I}^{+})}_{\w'0} \;, \qquad \psi^{(H^{+}, \mathscr{I}^{+})}_{\w'0} = e^{-i \w' t_s} \chi^{(H^{+}, \mathscr{I}^{+})}_{\w'0} \;.  \label{f-H-I-delta} \ee
The radial parts of the modes satisfy the following equation
\be
    \frac{d^{2}\chi}{dr_{*}^2}+(\omega^2-\lambda\delta(r_{*}))\chi=0 \;.
\label{DiracDeltaModeEquation} \ee
In the region where $r_*>0$, two linearly independent solutions are
\bes \bea
\chi^\infty_R &=& e^{i\omega r_{*}}\;, \label{RightMoving1}\\
\chi^\infty_L &=& e^{-i\omega r_{*}}\;. \label{LeftMoving1}
\eea \label{modes-delta-rs-gt-0} \ees
 For $r_*<0$, $\chi_R$ and $\chi_L$ can be expressed in the following way
\bes \bea
  \chi^\infty_{R}&=& E_R e^{i\omega r_{*}}+ F_R e^{-i\omega r_{*}}\;,  \label{RightMoving2}\\
  \chi^\infty_{L}&=& E_L e^{i\omega r_{*}}+ F_L e^{-i\omega r_{*}}\;.  \label{LeftMoving2}
\eea \label{modes-delta-rs-lt-0}\ees
Imposing the continuity of the mode function and discontinuity of its first derivative in the usual way at the spacelike curve $r_{*}=0$, the following analytic expressions are found for the scattering coefficients
\bes \bea
E_R &=& 1+\frac{i\lambda}{2\w}\; , \label{Dirac-ER}\\
F_R &=& -\frac{i\lambda}{2\w}\; ,\label{Dirac-FR}\\
E_L &=& F_R ^{*}=\frac{i\lambda}{2\w}\; ,\label{Dirac-EL}\\
F_L &=& E_R ^{*}=1-\frac{i\lambda}{2\w}\; .\label{Dirac-FL}
\eea \label{Scattering-Coefficients} \ees

Then the mode functions that we are using for the matching can be obtained from~\eqref{chi-H-plus-def} and~\eqref{chi-scri-plus-def} with the result
\bes \bea \psi^{H^{+}}_{\w'0}  &=& \theta(-r_*) \left[ e^{-i \w' v} + \frac{\frac{i \lambda}{2}}{\left(\w' - \frac{i \lambda}{2} \right)} e^{-i \w' u_s} \right]
  + \theta(r_*) \frac{\w'}{\left(\w' - \frac{i \lambda}{2} \right)} e^{-i \w' v} \;, \label{psi-H-delta} \\
  \psi^{\mathscr{I}^{+}}_{\w'0} &=& \theta(-r_*) \frac{\w'}{\left(\w' - \frac{i \lambda}{2} \right)} e^{-i \w' u_s}  + \theta(r_*) \left[  \frac{\frac{i \lambda}{2}}{\left(\w' - \frac{i \lambda}{2} \right)} e^{-i \w' v} + e^{-i \w' u_s} \right] \;. \label{psi-I-delta} \eea \label{psi-delta} \ees

To verify that the matching coefficients can be used to reconstruct the original mode functions for the case $\ell = 0$ it is useful to break them up into
contributions that come from the term proportional to  $e^{-i \w v}$ in~\eqref{psi-in-0} and the term proportional to $-e^{-i \w u}$.  In what follows we compute
the matching coefficients for both terms but then focus only on those that come from the term proportional to $e^{-i \w v}$.
Substituting~\eqref{psi-delta}, and~\eqref{Scattering-Coefficients} into~\eqref{gen-mat} one finds the matching coefficients
\bes \bea
A^{H^{+}}_{\w \w' 0} &=& (A^{H^{+}}_{\w \w' 0})_v  + (A^{H^{+}}_{\w \w' 0})_u  \;, \nonumber \\
(A^{H^{+}}_{\w \w' 0})_v &=& - \frac{i}{2 \pi} \sqrt{\frac{\w'}{\w}} \frac{e^{i \w' v_0}}{\w' - i \epsilon} e^{-i \w v_0}
   + \frac{i}{2 \pi} \sqrt{\frac{\w'}{\w}} \frac{\w'}{\left(\w'+\frac{i \lambda}{2} \right)}\frac{e^{i(\w'-\w)v_0}}{(\w'-\w+i \epsilon)} \;, \label{AHv-delta} \\
 (A^{H^{+}}_{\w \w' 0})_u &=&   \frac{i}{2 \pi} \sqrt{\frac{\w'}{\w}} \frac{e^{i \w' v_0}}{\w' - i \epsilon} e^{-i \w v_H}
    - \frac{1}{2 \pi} \sqrt{\frac{\w}{ \w'} } \int_{-\infty}^{v_H} du \,  e^{-i \w u}  \left[ \theta(r_{*}) \frac{\w'}{\w' + \frac{i \lambda}{2}} e^{i \w' v_0} \right.  \nonumber \\
 & &  \left. \qquad    + \, \theta(-r_{*}) \left( e^{i \w' v_0}
     - \frac{\frac{i \lambda}{2}}{\w' + \frac{i \lambda}{2}} e^{i \w' u_s(u)} \right) \right] \;, \label{AHu-delta}
\eea \ees
\bes \bea
A^{\mathscr{I}^{+}}_{\w \w' 0} &=& (A^{\mathscr{I}^{+}}_{\w \w' 0})_v  + (A^{\mathscr{I}^{+}}_{\w \w' 0})_u  \;, \nonumber \\
(A^{\mathscr{I}^{+}}_{\w \w' 0})_v &=&  - \frac{i}{2 \pi} \sqrt{\frac{\w'}{\w}} \frac{\frac{i \lambda}{2}}{(\w' + \frac{i \lambda}{2})}\frac{e^{-i(\w-\w')v_0}}{(\w'-\w+i \epsilon)}\;,  \label{AIv-delta}   \\
  (A^{\mathscr{I}^{+}}_{\w \w' 0})_u &=&  -\frac{1}{2 \pi} \sqrt{\frac{\w'}{ \w}} \int_{-\infty}^{v_H} du  \frac{du_s(u)}{du} \, e^{-i \w u} e^{i \w' u_s(u)} \left[ \theta(r_{*})  +
  \theta(-r_{*}) \frac{\w'}{\w'+\frac{i \lambda}{2}}  \right]
   \;.
  \label{AIu-delta} \eea \ees
Note that the relations~\eqref{B-w-minus-w}  are satisfied by these matching coefficients so the relation~\eqref{2D-in-mode-expansion-2} also holds. Thus
 \bea
 (f^{in}_{\w 0 0})_v &=&  \int_{-\infty}^\infty d \w'   \left[ (A^{H^{+}}_{\w \w' 0})_v  f^{H^+}_{\w'00} +
    (A^{\mathscr{I}^{+}}_{\w \w' 0})_v  f^{\mathscr{I}^+}_{\w'00} \right] \;.
    \label{in-modes-v-part}
    \eea
Substituting~\eqref{AHv-delta},~\eqref{AIv-delta}, and~\eqref{psi-delta} into~\eqref{in-modes-v-part} gives after some algebra
\bes \bea  (f^{in}_{\w00})_v &=& \frac{Y_{00}}{r \sqrt{4 \pi \w}} \left[\theta(-r_*) I_1 + \theta(r_*) I_2 \right] \;, \nonumber \\
   I_1 &=& - \frac{i}{2 \pi } e^{-i \w v_0} \int_{-\infty}^\infty d \w' \left[ \frac{\frac{i \lambda}{2} e^{i \w'(v_0-u_s)}}{(\w'-i\epsilon) \left(\w' - \frac{i \lambda}{2} \right)}
     + \frac{e^{i \w'(v_0-v)}}{(\w'-i \epsilon)} - \frac{\w' e^{i \w' (v_0-v)}}{\left(\w' + \frac{i \lambda}{2} \right) \left( \w' - \w + i \epsilon \right)} \right] \nonumber \\
     &=&  \theta(v_0-u_s) e^{-i \w v_0} \left[e^{-\frac{\lambda}{2}(v_0-u_s)} - 1 \right] + \theta(v_0-v) e^{-i \w v_0} \nonumber \\
   & &     + \frac{ \theta(v-v_0)}{\left( \w + \frac{i \lambda}{2} \right)} \left[ \frac{i \lambda}{2} e^{-i \w v_0} e^{- \frac{\lambda}{2}(v-v_0)} + \w e^{- i \w v} \right]\;, \\
   I_2 &=& - \frac{i}{2 \pi } e^{-i \w v_0} \int_{-\infty}^\infty d \w' \left[ \frac{\w' e^{i \w'(v_0-v)}}{(\w'-i \epsilon)\left(\w' - \frac{i \lambda}{2} \right)}
   -  \frac{ e^{i \w'(v_0-v)}}{\w'-\w + i \epsilon} + \frac{\frac{i \lambda}{2} e^{i \w'(v_0-u_s)}}{(\w'- \w + i\epsilon)\left(\w' + \frac{i \lambda}{2} \right)} \right] \nonumber \\
   &=&   \theta(v_0-v) e^{-i \w v_0} e^{-\frac{\lambda}{2}(v_0-v)} + \theta(v-v_0) e^{-i \w v}  \nonumber \\
    & & - \theta(u_s-v_0) \frac{ \frac{i \lambda}{2}}{\left( \w + \frac{i \lambda}{2} \right)} \left[
   - e^{-i \w v_0} e^{-\frac{\lambda}{2}(u_s-v_0)} + e^{-i \w u_s} \right]\;.
 \eea \label{f-in-v}\ees
It is easy to verify that~\eqref{f-in-v} gives the correct values for $(f^{in}_{\w00})_v$ on the future horizon for $v \le v_0$, on the null shell surface $v = v_0$, and  on past null infinity for $v \ge v_0$.

\subsection{Partial analytic results for the matching coefficients in 4D for $\ell = 0$}

Because of the simple form of the $in$ modes for $\ell = 0$ inside the null shell~\eqref{psi-in-0}, it is possible to compute the matching coefficients
for the $e^{-i \w v}$ part analytically.  To do so we begin by
substituting~\eqref{psi-in-0} into~\eqref{gen-mat} with the result
\bes \bea A^{ H^{+}}_{\w \w' 0} &=& - \frac{i}{2 \pi} \sqrt{\frac{\w'}{\w}} \frac{e^{i \w' v_0}}{\w' - i \epsilon} (e^{-i \w v_0} - e^{-i \w v_H})
   + \frac{i}{2 \pi} \sqrt{\frac{\w'}{\w}} \frac{1}{F_L^{*}(\w',0)}\frac{e^{i(\w'-\w)v_0}}{\w'-\w+i \epsilon}  \nonumber \\
   & & - \frac{1}{2 \pi} \sqrt{\frac{\w }{\w'}}  \int_{-\infty}^{v_H} du \,  e^{-i \w u}  \psi^{H^{+} *}_{\w' 0}(u_s(u),v_0) \;, \label{A-H-ell-0-mat}\\
      B^{ H^{+}}_{\w \w' 0} &=& \frac{i}{2 \pi} \sqrt{\frac{\w'}{\w}} \frac{e^{-i \w' v_0}}{\w' + i \epsilon} (e^{-i \w v_0} - e^{-i \w v_H})
      - \frac{i}{2 \pi} \sqrt{\frac{\w'}{\w}} \frac{1}{F_L(\w',0)}\frac{e^{-i(\w+\w')v_0}}{\w'+\w-i \epsilon} \nonumber \\
      & & + \frac{1}{2 \pi} \sqrt{\frac{\w}{ \w'}}  \int_{-\infty}^{v_H} du \,e^{-i \w u} \psi^{H^{+}}_{\w' 0} (u_s(u),v_0)\;, \label{B-H-ell-0-mat}\\
     A^{ \mathscr{I}^{+}}_{\w \w' 0} &=&  - \frac{i}{2 \pi} \sqrt{\frac{\w'}{\w}} \frac{F_R^{*}(\w',0)}{F_L^{*}(\w',0)}\frac{e^{-i(\w-\w')v_0}}{\w'-\w+i \epsilon}
      \nonumber \\
       && \;\; +\frac{i}{2 \pi} \frac{1}{\sqrt{\w \w'}} \int_{-\infty}^{v_H} du \, e^{-i \w u} \partial_u \psi^{\mathscr{I}^+  *}_{\w' 0}(u_s(u),v_0) \;,  \label{A-I-ell-0-mat} \\
     B^{ \mathscr{I}^{+}}_{\w \w' 0} &=&  \frac{i}{2 \pi} \sqrt{\frac{\w'}{\w}} \frac{F_R(\w',0)}{F_L(\w',0)}\frac{e^{-i(\w+\w')v_0}}{\w'+\w-i \epsilon}
     \nonumber \\
  && \;\;     - \frac{i}{2 \pi} \frac{1}{\sqrt{\w \w'}} \int_{-\infty}^{v_H} du \,e^{-i \w u} \partial_u \psi^{\mathscr{I}^+}_{\w' 0}(u_s(u),v_0) \;.
     \label{B-I-ell-0-mat} \eea \label{ell-0-mat}\ees

Note that the integrals have to be computed numerically because the mode functions in Schwarzschild spacetime must be computed numerically.
However, because of the simple form that $\psi^{ in}_{\w 0}$ takes, it is possible to separate the matching coefficients into separate matching
coefficients for the part that goes like $e^{- i \w v}$ inside the null shell and the part that goes like $e^{-i \w u}$ there.  The matching coefficients
for $e^{-i \w v}$ do not depend on the integrals.  In what follows we focus on these matching coefficients.  Examination of~\eqref{ell-0-mat} gives for these coefficients
\bes \bea \left(A^{ H^{+}}_{\w \w' 0}\right)_v &=& - \frac{i}{2 \pi} \sqrt{\frac{\w'}{\w}} \frac{e^{i (\w'-\w) v_0}}{\w' - i \epsilon}
   + \frac{i}{2 \pi} \sqrt{\frac{\w'}{\w}} \frac{1}{F_L^{*}(\w',0)}\frac{e^{i(\w'-\w)v_0}}{\w'-\w+i \epsilon},  \nonumber \\
   & & \label{A-H-ell-0-v-mat}\\
      \left(B^{ H^{+}}_{\w \w' l}\right)_v &=& \frac{i}{2 \pi} \sqrt{\frac{\w'}{\w}} \frac{e^{-i (\w'+\w) v_0}}{\w' + i \epsilon}
   - \frac{i}{2 \pi} \sqrt{\frac{\w'}{\w}} \frac{1}{F_L(\w',0)}\frac{e^{-i(\w+\w')v_0}}{\w'+\w-i \epsilon}, \nonumber \\
      & & \; \label{B-H-ell-0-v-mat}\\
     \left(A^{ \mathscr{I}^{+}}_{\w \w' l}\right)_v &=&  - \frac{i}{2 \pi} \sqrt{\frac{\w'}{\w}} \frac{F_R^{*}(\w',0)}{F_L^{*}(\w',0)}\frac{e^{-i(\w-\w')v_0}}{\w'-\w+i \epsilon}
      \;,  \label{A-I-ell-0-v-mat} \\
     \left(B^{ \mathscr{I}^{+}}_{\w \w' l}\right)_v &=&  \frac{i}{2 \pi} \sqrt{\frac{\w'}{\w}} \frac{F_R(\w',0)}{F_L(\w',0)}\frac{e^{-i(\w+\w')v_0}}{\w'+\w-i \epsilon}
     \;.
     \label{B-I-ell-0-v-mat} \eea \label{ell-0-v-mat}\ees

These matching coefficients can be used to reconstruct the part of the mode function which goes like $e^{-i \w v}$ inside the shell by substituting the expressions into~\eqref{General-in-modes}.  To check them we shall compute the resulting integral on $H^{+}$.  Recall that we are working in
the exact Schwarzschild spacetime rather than the null shell spacetime when we do the matching.  The same applies to the reconstruction.  Thus the results for the reconstruction for which $v \ge v_0$ also apply to the null shell spacetime,  but the results for $v < v_0$ do not apply to the null shell spacetime.

Recall that the modes $f^{\mathscr{I}^{+}}$ vanish on $H^{+}$.
\bes \bea \left(f^{in}_{\w 0 0} \right)_v &=& \frac{Y_{0 0}}{r \sqrt{4 \pi \w}} (I_1 + I_2)\;,  \nonumber \\
   I_1 &=&  - \frac{i}{2 \pi} e^{-i \w v_0} \int_0^\infty d \w' \,\left[ \frac{e^{i(v_0-v)\w'}}{\w'-i \epsilon} -  \frac{e^{-i(v_0-v)\w'}}{\w'+i \epsilon} \right]\;, \label{ch4-I1} \\
   I_2 &=& \frac{i}{2 \pi} e^{-i \w v_0} \int_0^\infty d \w' \, \left[\frac{1}{F_L^{*}(\w',0)} \frac{e^{-i \w'(v-v_0)}}{\w'-\w + i \epsilon}
   - \frac{1}{F_L(\w',0)} \frac{e^{i \w'(v-v_0)}}{\w'+\w - i \epsilon} \right] \;. \label{ch4-I2}
 \eea \label{fin-ell-0-v}\ees
If in the second term of $I_1$ a change of variables is made so that $\w' \to - \w'$ then one finds that
\be I_1 =   - \frac{i}{2 \pi} e^{-i \w v_0} \int_{-\infty}^\infty d \w' \frac{e^{i(v_0-v)\w'}}{\w'-i \epsilon} = e^{-i \w v_0}\, \theta(v_0-v) \;,\ee
with $\theta$ the step function.  It is thus clear that the initial data on $H^{+}$ for $-\infty < v < v_0$ does not affect the mode functions on the
part of the future horizon for which $v_0 < v < \infty$.

It can be shown from the properties of the scattering coefficients given in~\cite{rigorous}, that $F_L(\w')=F^{*}_L(-\w')$. Using this identity and changing the variable of integration in the second integral in the same way as was done for $I_1$, one obtains
\be
(f^{in}_{\w 0 0})_v =\frac{Y_{00} e^{-i\w v_0}}{r\sqrt{4\pi\w}}\theta(v_{0}-v)+\frac{i Y_{00}}{r 2\pi \sqrt{4\pi\w}}\int_{-\infty}^{\infty} d\w' \frac{e^{-i\w'v}}{
F^*_{L}(\w',0)}\frac{e^{iv_{0}(\w'-\w)}}{\w'-\w+i\epsilon} \;.\ee
To compute this integral using complex integration techniques one must know the singularity structure of $\frac{1}{F_L^{*}}$ which is difficult since this scattering coefficient
must be computed numerically.  However, one can at least test whether it has one or more singularities in the complex plane by assuming it does not and computing the integral.
We will call the result $f^{test}$ because there is no guarantee that this method will give the correct answer.
The result of such an integration is
\[ f^{test} =\frac{e^{-i\w v_0}}{\sqrt{4\pi\w}}\theta(v_{0}-v)+\frac{e^{-i\w v}}{ F_{L}^{*}(\w,0)\sqrt{4\pi\w}}\theta(v-v_{0})\;.
\]
Here complex integration has been performed using a contour in the lower half of the complex plane.
It is obvious that at $v=v_{0}$, the continuity condition for $(f^{in}_{\w 0 0})_v$ is not satisfied so $f^{test} \ne (f^{in}_{\w 0 0})_v$
which implies that $\frac{1}{F_{L}^{*}(\w')}$ has one or more singularities in the complex plane.

Alternatively, one can work with $I_2$ in the form~\eqref{ch4-I2}, use the relation $(\w'\mp\w \pm i\epsilon)^{-1} = \mp i \pi \delta(\w'\mp\w) + (\w' \mp \w)^{-1}$, and compute the principal value parts of the integral numerically for $v > v_0$.    This has been done and the result is shown in Fig.~{fig:I2}.  It is clear from the plots in this figure that on the future horizon $(f^{in}_{\w 0 0})_v$ is continuous at $v = v_0$.
\begin{figure}[h]
\centering
\includegraphics[totalheight=0.25\textheight]{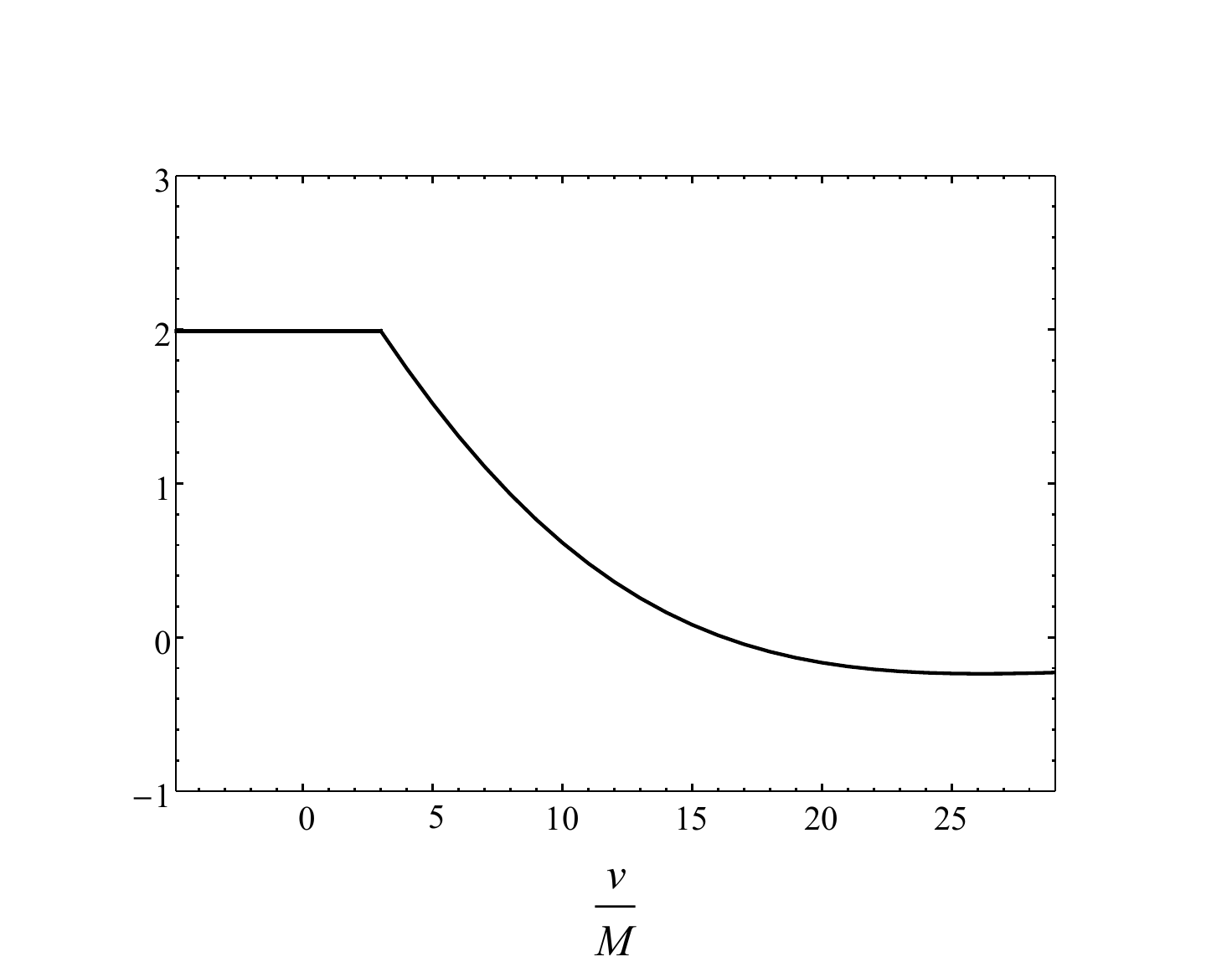}
\includegraphics[totalheight=0.25\textheight]{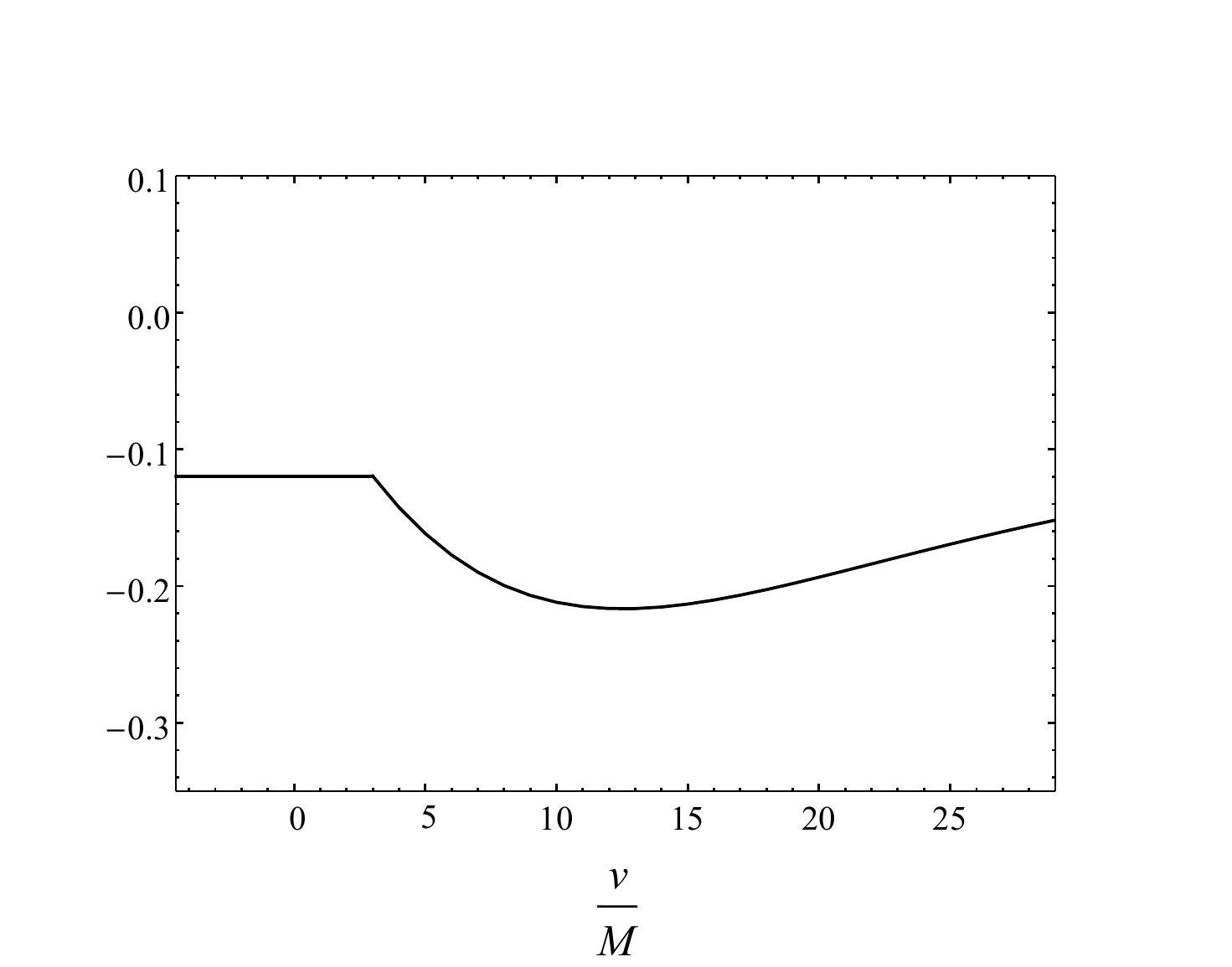}
\caption{The real (left) and the imaginary (right) parts of  $\sqrt{\frac{4 \pi}{M}} (f^{in}_{\w 00})_v $ on the future horizon have been plotted. In both plots, $M\omega = 0.02$ and $\frac{v_0}{M}=3$. The plots clearly show that $(f^{in}_{\w 00})_v$ is continuous at $v = v_0$. }
\label{fig:I2}
\end{figure}

\section{Stress-Energy Tensor}\label{ch4-sec-Tab-method}

\subsection{Method in 4D}
\label{sec-4D-method}

For the massless minimally coupled scalar field the classical stress-energy tensor in a general curved spacetime is given in~\eqref{Tab-class} and a renormalization expression for $\la T_{ab} \ra$ is given in~\eqref{Tab-Unruh-sub}. 
To compute $\la T_{ab} \ra$ using~\eqref{Tab-Unruh-sub} it is useful to begin with the points split and to write the stress-energy tensor in terms of derivatives of the
Hadamard Green's function
\be  G^{(1)}(x,x') = \la \{ \phi(x), \phi(x') \}\ra  \;. \label{G1-def} \ee
We adopt the notation
\be \Delta G^{(1)}(x,x') = \la {in}| \{ \phi(x), \phi(x') \} | {in}\ra - \la {U}| \{ \phi(x), \phi(x') \} | {U}\ra \;, \label{Delta-G-def} \ee
with $|{in} \ra$ representing the $in$ vacuum state and $|U \ra$ the Unruh state.  The corresponding difference in the stress-energy tensors is then
\be  \Delta \la T_{ab} \ra = \frac{1}{4} \lim_{x' \to x} \left[ \left( g_a^{c'} \Delta G^{(1)}_{;c'; b} + g_b^{c'}  \Delta G^{(1)}_{;a; c'} \right)  - \, g_{a b}\, g^{c d'} \Delta G^{(1)}_{;c ;d'} \right]  \;. \label{Delta-Tab-1} \ee
Here the quantity $g_a^{b'}$ parallel transports a vector from $x'$ to $x$ and is called the bivector of parallel transport~\cite{christensen-76}.
To leading order when the point separation is small
\be g_a^{b'} = g_a^b = \delta_a^b  \;. \label{gabp-leading}\ee
The subleading orders all vanish in the limit $x' \to x$.
Since there are no ultraviolet divergences in the quantity $\Delta \la T_{ab} \ra$ one can use~\eqref{gabp-leading} in~\eqref{Delta-Tab-1} with the result
\be  \Delta \la T_{ab} \ra =   \frac{1}{4}  \left[\lim_{x' \to x} \left( \Delta G^{(1)}_{;a' ;b} +   \Delta G^{(1)}_{;a ;b'} \right)  - g_{a b}\, g^{c d}  \lim_{x' \to x} \Delta G^{(1)}_{;c ;d'} \right]  \;, \label{Delta-Tab-2} \ee
where a slight abuse of notation has been used for the implied sum over $d$ and $d'$ in the last term.  It is important to note that this expression is valid in both two and four dimensions.

Expanding the field in terms of modes as in~\eqref{Phi-expansion} one finds for the $in$ modes
that
\be   \la 0 \, {in} |  \{ \phi(x), \phi(x') \} |0 \, {in} \ra = \sum_{\ell = 0}^\infty \sum_{m = - \ell}^\ell \int_0^\infty d \w \; [f^{in}_{\w \ell m}(x) (f^{in}_{\w \ell m}(x'))^{*} + f^{in}_{\w \ell m}(x') (f^{{in}}_{\w \ell m}(x))^{*}]  \;.  \label{G1-in} \ee
The Unruh state in Schwarzschild spacetime consists of modes that are positive frequency with respect to the usual time coordinate on $\mathscr{I}^{-}$ along with modes that are positive frequency with respect to the Kruskal time coordinate on $H^{-}$ so that
\bea   \la U |  \{ \phi(x), \phi(x') \} |U \ra &=& \sum_{\ell = 0}^\infty \sum_{m = - \ell}^\ell \left\{ \int_0^\infty d \omega_K \, [f^{K}_{\w_K \ell m}(x) (f^{K}_{\w_K \ell m}(x'))^{*} + f^{K}_{\w_K \ell m}(x') (f^{{K}}_{\w_K \ell m}(x))^{*}] \right.  \nonumber \\
  && \; \left.  + \,  \int_0^\infty d \w \; [f^{\mathscr{I}^{-}}_{\w \ell m}(x) (f^{\mathscr{I}^{-}}_{\w \ell m}(x'))^{*} + f^{\mathscr{I}^{-}}_{\w \ell m}(x') (f^{{\mathscr{I}^{-}}}_{\w \ell m}(x))^{*}]  \right\}  \;.  \label{G1-U}  \eea

The next step is to find expansions for these two-point functions in terms of the complete set of modes $f^{(\mathscr{I}^{+}, H^{+})}$ that we are using.
For $ \la 0 \, {in} |  \{ \phi(x), \phi(x') \} |0 \, {in} \ra$ one can substitute~\eqref{General-in-modes} into~\eqref{G1-in}.  This results in integrals of the form
\bea  &&\sum_{\ell = 0}^\infty \sum_{m = - \ell}^\ell \int_0^\infty d \w \; \int_0^\infty d \w_1  \; \int_0^\infty d\w_2 \left\{ \left[  A^{(H^{+}, \mathscr{I}^{+})} f^{(H^{+}, \mathscr{I}^{+})} +  B^{(H^{+}, \mathscr{I}^{+})} (f^{(H^{+}, \mathscr{I}^{+})})^{*} \right] \right. \nonumber \\
&& \qquad \qquad \left.  \times \left[  (A^{(H^{+}, \mathscr{I}^{+})})^{*} (f^{(H^{+}, \mathscr{I}^{+})})^{*} +  (B^{(H^{+}, \mathscr{I}^{+})})^{*} f^{(H^{+}, \mathscr{I}^{+})} \right] \right\}
 \;, \label{triple-integral-form} \eea
where the subscripts on the matching coefficients and mode functions have been suppressed.
For $\la U |  \{ \phi(x), \phi(x') \} |U \ra$ one can first substitute~\eqref{fK} and~\eqref{alphaK-betaK} into~\eqref{G1-U} to obtain an expression in terms of
$f_{\w \ell m}^{(H^{-},\mathscr{I}^{-})}$.  Then~\eqref{f-scri-H-minus} can be used to obtain an expression for $\la U |  \{ \phi(x), \phi(x') \} |U \ra$ that depends only on $f_{\w \ell m}^{(\mathscr{I}^{+}, H^{+})}$.

\subsection{2D Example}
\label{sec:Tab-2D}

In this section, the method discussed above is tested by using it to compute the stress-energy tensor for the scalar field in the corresponding 2D spacetime where the answer is known.  The computation will be done in the region $v > v_0$ outside the null shell and outside the horizon.
From~\eqref{f-in-H+-2D} it is clear that for $v > v_0$ the contribution from the $f^{H^{+}}_{\w'}$ modes to $f^{in}_\w$ is
\be \left(f^{in}_\w \right)_{H^{+}} = \frac{e^{-i \w v}}{\sqrt{4 \pi \w}}  = f^{H^{+}}_{\w}  \;. \label{f-in-H+-2} \ee
Thus
\bea f^{in}_\w = f^{H^+}_\w +  \int_0^\infty d\w \; [A^{\mathscr{I}^{+}}_{\w \w'} f^{\mathscr{I}^{+}}_{\w'}  + B^{\mathscr{I}^{+}}_{\w \w'} (f^{\mathscr{I}^{+}}_{\w'})^{*}]\;, \label{fin-A-B}
\eea
with $A^{\mathscr{I}^{+}}_{\w \w'}$ given in~\eqref{Ascrp2D}. Using the relation $\Gamma(x) = \frac{\Gamma(1+x)}{x}$ one obtains the form used for the numerical computations
\bea
A^{\mathscr{I}^{+}}_{\w \w'} &=& -\frac{1}{2 \pi} \sqrt{\frac{\w}{ \w'}} (4M)^{i 4M \w'} e^{-i(\w-\w') v_H} \frac{\Gamma(1-i4M\w')}{[-i(\w-\w')+\epsilon]^{1-i4M\w'}}\;.  \label{AI2}
\eea
Then, using the relations~\eqref{B-w-minus-w} one finds
\be B^{\mathscr{I}^{+}}_{\w \w'} =  \frac{1}{2 \pi} \sqrt{\frac{\w}{ \w'}} (4M)^{-i 4M \w'} e^{-i(\w+\w') v_H} \frac{\Gamma(1+i4M\w')}{[-i(\w+\w')+\epsilon]^{1+i4M\w'}}  \;. \label{BI2}  \ee
In what follows the superscript $\mathscr{I}^+$ on the matching coefficients A and B will be suppressed.

Next, with the aim of finding the components of the stress-energy tensor using~\eqref{Delta-Tab-2}, we construct the Hadamard form of Green's function which in 2D is
\bea
 G^{(1)}(x,x') = \int_0^\infty d \w \; [f^{in}_\w(x) f^{in \;*}_{\w}(x') + f^{in \;*}_\w(x) f^{in}_{\w}(x') ]\;.
  \label{Hadamard}
 \eea
Substituting ~\eqref{fin-A-B} into ~\eqref{Hadamard} gives
 \bea
 G^{(1)}(x,x') &=& \int_0^\infty d \w \; \Big\{ \Big[f^{H^{+}}_\w(x) +  \int_0^\infty d\w_1 \; [A_{\w \w_1} f^{\mathscr{I}^{+}}_{\w_1}(x)  + B_{\w \w_1} f^{\mathscr{I}^{+}*}_{\w_1}(x)]\Big] \nonumber \\
 & & \left. \times  \Big[f^{H^{+} *}_\w(x') +  \int_0^\infty d\w_2 \;[ [A^{*}_{\w \w_2} f^{\mathscr{I}^{+}\;*}_{\w_2}(x')  + B^{*}_{\w \w_2} f^{\mathscr{I}^{+}}_{\w_2}(x')]\Big] \right.  \nonumber \\
 & & \left. + \Big[f^{H^{+}}_\w(x') +  \int_0^\infty d\w_1 \; [A_{\w \w_1} f^{\mathscr{I}^{+}}_{\w_1}(x')  + B_{\w \w_1} f^{\mathscr{I}^{+}*}_{\w_1}(x')]\Big] \right.  \nonumber \\
 & & \times \Big[f^{H^{+} *}_\w(x) +  \int_0^\infty d\w_2 \; [A^{*}_{\w \w_2} f^{\mathscr{I}^{+} *}_{\w_2}(x)  + B^{*}_{\w \w_2} f^{\mathscr{I}^{+}}_{\w_2}(x)]\Big] \Big\}\;. \label{G1-1} \eea
  Expanding the integrand of the integral over $\w$ results in three types of expressions: an integral consisting of products of the modes $f^{H^+}_\w$, which we  call $G_A$, another integral which includes cross products between the modes $f^{H^+}_\w$ and  $f^{\mathscr{I}^+}_\w$, which we call $G_B$, and finally an integral consisting of products of the modes  $f^{\mathscr{I}^+}_\w$, which we called $G_C$\;.

 To renormalize we follow a procedure equivalent to that outlined in Sec.~{sec-4D-method}.
 We begin by subtracting off the integrals with the integrand evaluated in the large $\w$ limit.  When we add them back, we get contributions that are identical to those obtained for the Unruh state.  We are not quite subtracting off the Unruh modes, because the large $\w$ solutions have a dependence on $v_H$.  However, when the subtracted terms are added back and the integral over $\w$ is computed, then factors of $\delta(\w_1 - \w_2)$ and $\delta(\w_1 + \w_2)$ are obtained.  Note that terms proportional to $\delta(\w_1 + \w_2)$ vanish.  For the ones that do not vanish, once one integrates over say $\w_2$,  the dependence on $v_H$ vanishes.

 In Sec.~{appendix-B} it is shown that when this method is applied to $ G^{(1)}(x,x')$, the $\Delta G_A$ term vanishes.  It is also shown that, while the $\Delta G_B$ term does not vanish, it does not contribute to the stress-energy tensor.  As a result, the only term that contributes to $\Delta \la T_{ab} \ra$ is $\Delta G_C(x,x')$ which has the form
\bea
\Delta G_C(x,x')&=& \int_0^{\infty} d\w_1 \int_0^{\infty} d\w_2 \Big\{ [f_{\w_1}^{\mathscr{I^+}}(x)f_{\w_2}^{\mathscr{I^+}\; *}(x')+f_{\w_1}^{\mathscr{I^+}}(x')f_{\w_2}^{\mathscr{I^+}\;*}(x)]\Delta I_1 \nonumber \\ & &
+\; [f_{\w_1}^{\mathscr{I^+}}(x)f_{\w_2}^{\mathscr{I^+}}(x')+f_{\w_1}^{\mathscr{I^+}}(x')f_{\w_2}^{\mathscr{I^+}}(x)]\Delta I_2 \nonumber \\ & &
+\;[f_{\w_1}^{\mathscr{I^+}\;*}(x)f_{\w_2}^{\mathscr{I^+}\;*}(x')+f_{\w_1}^{\mathscr{I^+}\;*}(x')f_{\w_2}^{\mathscr{I^+}\;*}(x)]\Delta I_3 \nonumber \\ & &
+\;[f_{\w_1}^{\mathscr{I^+}\;*}(x)f_{\w_2}^{\mathscr{I^+}}(x')+f_{\w_1}^{\mathscr{I^+}\; *}(x')f_{\w_2}^{\mathscr{I^+}}(x)]\Delta I_4 \Big\} \;, \label{Del-GC}
\eea
with
\bes \bea
\Delta I_1 &=& \int _0^{\infty} d\w \left\{A_{\w \w _1}A_{\w \w _2}^{*}-\mathcal{O}(A_{\w \w _1}A_{\w \w _2}^{*})\right\} \label{Del-I1}\;,\\
\Delta I_2 &=& \int _0^{\infty} d\w \left\{ A_{\w \w _1}B_{\w \w _2}^{*}-\mathcal{O}(A_{\w \w _1}B_{\w \w _2}^{*})\right\} \label{Del-I2}\;,\\
\Delta I_3 &=& \int _0^{\infty} d\w \left\{B_{\w \w _1}A_{\w \w _2}^{*}-\mathcal{O}(B_{\w \w _1}A_{\w \w _2}^{*})\right\}\label{Del-I3}\;,\\
\Delta I_4 &=& \int _0^{\infty} d\w \left\{B_{\w \w _1}B_{\w \w _2}^{*}-\mathcal{O}(B_{\w \w _1}B_{\w \w _2}^{*})\right\}\; \label{Del-I4}.
\eea  \label{In}  \ees
Here $\mathcal{O}$ indicates the asymptotic behavior of the matching coefficients for $\w \gg \w_{1,2}$.

The integrals in~\eqref{In} can be computed analytically. Substituting the explicit expression for A from~\eqref{AI2} into~\eqref{Del-I1} gives
\bes \bea  \Delta I_{1} &=&  \frac{1}{4 \pi^2  \sqrt{ \w_1 \w_2}} (4 M)^{i 4 M (\w_1-\w_2)} e^{i v_H(\w_1 - \w_2)} \Gamma(1-i 4M\w_1) \Gamma(1+i4M\w_2)
  \Delta K_1 \;, \label{Del-I1-1} \\
  \Delta K_{1} &=&   \lim_{\Lambda \to \infty} (-i)^{i 4 M \w_1} (i)^{-i 4 M \w_2}\bigg\{ \left[ \int_{0}^\Lambda d\w \frac{\w}{ ( \w-\w_1 + i \epsilon_1)^{1-i4M\w_1} ( \w-\w_2 - i \epsilon_2)^{1+i4M\w_2}}  \right. \nonumber \\   && \left. \qquad
- \int_{1}^\Lambda d\w \,\w^{-1+i4M(\w_1-\w_2)} \right]- \int_0^1  d\w \, \w^{-1+i4M(\w_1-\w_2)} \bigg\} \;. \label{DelK1-1}
\eea \ees
First, we compute the indefinite integrals and evaluate them at the limits. Since $\epsilon_1$ and $\epsilon_2$  go to $0^+$ at the end of the calculation, it is acceptable to add terms containing them to the exponents. The first integral is
 \bea \Delta K_{1a} &=& (-i)^{i 4 M \w_1} (i)^{-i 4 M \w_2} \int_{0}^\Lambda d\w \frac{\w}{ ( \w-\w_1 + i \epsilon_1)^{1-i4M(\w_1- i \epsilon_1)} ( \w-\w_2 - i \epsilon_2)^{1+i4M(\w_2+i \epsilon_2)}}\nonumber \\
                        &=&  (-i)^{i 4 M \w_1} (i)^{-i 4 M \w_2} \left[-i \frac{(\w-\w_1+i \epsilon_1)^{i4M(\w_1- i \epsilon_1)} (\w-\w_2-i \epsilon_2)^{-i 4M(\w_2+ i \epsilon_2)}}{4 M (\w_1-\w_2 - i \epsilon_1 - i \epsilon_2)} \right]_0^\Lambda  \nonumber \\
                        &=& -i (-i)^{i 4 M \w_1} (i)^{-i 4 M \w_2} \left[ \frac{(\Lambda-\w_1)^{i4M\w_1} (\Lambda-\w_2)^{-i 4M\w_2}}{4 M (\w_1-\w_2) - i \epsilon_1 - i \epsilon_2} \right.
                         \nonumber \\
                         & & \left. \; - \frac{(-\w_1)^{i4M\w_1} (-\w_2)^{-i 4M\w_2}}{4 M (\w_1-\w_2 - i \epsilon_1 - i \epsilon_2)} \right] \;.  \eea
Note that after evaluating the integral at the limits, $\epsilon_1$ and $\epsilon_2$ are set to zero in the exponents because they have no effect there.
Also, each term is a combination of a principal value and a term proportional to $\delta(\w_1-\w_2)$, thus
\bes \bea \Delta K_{1a} &=&   e^{2 \pi M (\w_1+\w_2)} \left[-i \frac{(\Lambda-\w_1)^{i4M\w_1} (\Lambda-\w_2)^{-i 4M\w_2}}{4 M (\w_1-\w_2)} + \frac{\pi}{4M} \delta(\w_1-\w_2) \right]
   \nonumber \\
 & &  +  e^{-2 \pi M (\w_1+\w_2)} \left[ i \frac{\w_1^{i4M\w_1} \w_2^{-i 4M\w_2}}{4 M (\w_1-\w_2)}  - \frac{\pi}{4M} \delta(\w_1-\w_2) \right] \;.  \label{K1a}
    \eea
Here we adopt the notation that the principal value of a term such as $\frac{1}{a \pm i \epsilon}$ is written as $\frac{1}{a}$.
   The second and third integrals in~\eqref{DelK1-1} are
  \bea \Delta K_{1b} &=& -(-i)^{i 4 M \w_1} (i)^{-i 4 M \w_2} \int_{1}^\Lambda d\w \w^{-1+i4M(\w_1-\w_2)} \nonumber \\
     &=&      i \frac{e^{2 \pi M(\w_1+\w_2)}}{4 M(\w_1-\w_2)} \left[ \Lambda^{i 4M(\w_1-\w_2)} - 1\right]  \;, \label{Del-K1b} \\
           \Delta K_{1c} &=&  -(-i)^{i 4 M \w_1} (i)^{-i 4 M \w_2} \int_0^1  d\w \, \w^{-1+i4M(\w_1-\w_2)} \nonumber \\
              &=&   - e^{2 \pi M(\w_1+\w_2)} \int_{-\infty}^0 dz e^{[i 4M (\w_1-\w_2)+ \epsilon] z}  =  -  \frac{e^{2 \pi M(\w_1+\w_2)}}{i 4M (\w_1-\w_2)+ \epsilon}  =  \frac{i e^{2 \pi M(\w_1+\w_2)}}{4M (\w_1-\w_2) -i \epsilon} \nonumber \\  & & \qquad =   \frac{ie^{2 \pi M(\w_1+\w_2)}}{4M (\w_1-\w_2)} -e^{2 \pi M(\w_1+\w_2)} \frac{\pi}{4M} \delta(\w_1-\w_2) \;, \label{Del-K1c}
\eea \ees
where in the integral for $\Delta K_{1c}$ the change of variable $z = \log \w$ has been made and an integrating factor $\epsilon$ has been inserted.
Combining these results, one finds
\bes \bea \Delta K_1 &=&    e^{-2 \pi M (\w_1+\w_2)} \left[ i \frac{\w_1^{i4M\w_1} \w_2^{-i 4M\w_2}}{4 M (\w_1-\w_2)}  - \frac{\pi}{4M} \delta(\w_1-\w_2) \right] \nonumber \\
   & =&  i  e^{-2 \pi M (\w_1+\w_2)} \frac{\w_1^{i4M\w_1} \w_2^{-i 4M\w_2}}{4 M (\w_1-\w_2 - i \epsilon)} \;. \label{Del-K1-1} \eea
Substituting ~\eqref{Del-K1-1} into~\eqref{Del-I1-1} gives
\bea \Delta I_1 &=& \frac{i}{4 \pi^2  \sqrt{ \w_1 \w_2}} (4 M)^{i 4 M (\w_1-\w_2)} e^{i v_H(\w_1 - \w_2)}   \Gamma(1-i 4M\w_1) \Gamma(1+i4M\w_2) \nonumber \\
  & & \qquad \times \, e^{-2 \pi M( \w_1+\w_2)} \frac{\w_1^{i4M\w_1} \w_2^{-i 4M\w_2}}{4 M (\w_1-\w_2 - i \epsilon)} \;. \label{Del-I1-2}
       \eea \ees
Note that this is a finite contribution to $\Delta G_c$ because of the factor of $e^{-2 \pi M( \w_1+\w_2)}$.

Next, consider  $\Delta I_4$ which is the other term with nonvanishing delta functions.
\bes \bea  \Delta I_{4} &=&  \frac{1}{4 \pi^2  \sqrt{ \w_1 \w_2}} (4 M)^{-i 4 M (\w_1-\w_2)} e^{-i v_H(\w_1 - \w_2)} \Gamma(1+i 4M\w_1) \Gamma(1-i4M\w_2)
  \Delta K_4 \;, \label{Del-I4-1} \\
  \Delta K_{4} &=&   \lim_{\Lambda \to \infty} (-i)^{-i 4 M \w_1} (i)^{i 4 M \w_2}\bigg\{ \left[ \int_{0}^\Lambda d\w \frac{\w}{ ( \w+\w_1 + i \epsilon_1)^{1+i4M\w_1} ( \w+\w_2 - i \epsilon_2)^{1-i4M\w_2}}  \right. \nonumber \\   && \left. \qquad
- \int_{1}^\Lambda d\w \,\w^{-1-i4M(\w_1-\w_2)} \right]- \int_0^1  d\w \, \w^{-1-i4M(\w_1-\w_2)} \bigg\} \;. \label{DelK4-1} \;
\eea \ees
The integrals in $ \Delta K_{4}$ can be computed analytically with the result
\bea  \Delta K_{4a} &=& e^{-2\pi M(\w_1+\w_2)} \int_{0}^{\Lambda} d\w  \frac{\w}{( \w+\w_1+i \epsilon_1)^{1+i4M(\w_1+i \epsilon_1)} (\w+\w_2-i \epsilon_2)^{1-i4M(\w_2-i \epsilon_2)}} \nonumber \\
              &=& e^{-2\pi M(\w_1+\w_2)} \frac{i}{4 M} \left[ \frac{(\Lambda+\w_1)^{-i4M\w_1} (\Lambda+\w_2)^{i4M\w_2}}{\w_1-\w_2+ i (\epsilon_1+\epsilon_2)}
               -  \frac{\w_1^{-i4M\w_1} \w_2^{i4M\w_2}}{\w_1-\w_2+ i (\epsilon_1+\epsilon_2)}   \right], \nonumber \\
              \Delta K_{4b}& =&  e^{-2\pi M(\w_1+\w_2)} \left[ -\frac{\Lambda^{-i 4 M  (\w_1-\w_2)}}{-i 4M(\w_1-\w_2)} + \frac{1}{-i 4M(\w_1-\w_2)} \right] \,, \nonumber \\
              \Delta K_{4c} &= &  - e^{-2\pi M(\w_1+\w_2)} \int_{-\infty}^0 dz \, e^{[-i 4M(\w_1-\w_2)+\epsilon]z} = -\frac{ e^{-2\pi M(\w_1+\w_2)}}{-i 4M(\w_1-\w_2)+\epsilon}
             \nonumber \\
              &=& -i \frac{e^{-2\pi M(\w_1+\w_2)}}{4 M (\w_1-\w_2) + i \epsilon}  = -i  \frac{e^{-2\pi M(\w_1+\w_2)}}{4 M (\w_1-\w_2)} - e^{-2\pi M(\w_1+\w_2)} \frac{\pi}{4M}   \delta(\w_1-\w_2)  \;.
                \eea
Both terms in $\Delta K_{4a}$ can be written in terms of their principal values added to a term proportional to $\delta(\w_1-\w_2)$. Combining these terms, the following expression for $\Delta K_4$ is obtained
\bes \bea  \Delta K_4 &=&   e^{-2\pi M(\w_1+\w_2)}\left[-i \frac{\w_1^{-i4M\w_1} \w_2^{i4M\w_2}}{4M(\w_1-\w_2)} - \frac{ \pi }{4M}   \delta(\w_1-\w_2)\right]
   \nonumber \\
     &=& - i e^{-2\pi M(\w_1+\w_2)} \; \frac{\w_1^{-i4M\w_1} \w_2^{i4M\w_2}}{4M(\w_1-\w_2 + i \epsilon)}  \;.
     \eea
Finally
\bea \Delta I_4 &=& -\frac{i}{4 \pi^2  \sqrt{ \w_1 \w_2}} (4 M)^{-i 4 M (\w_1-\w_2)} e^{-i v_H(\w_1 - \w_2)}   \Gamma(1+i 4M\w_1) \Gamma(1-i4M\w_2) \nonumber \\
    & & \qquad \times\, e^{-2\pi M(\w_1+\w_2)} \; \frac{\w_1^{-i4M\w_1} \w_2^{i4M\w_2}}{4M(\w_1-\w_2 + i \epsilon)}  \;. \label{Del-I4-2} \eea \ees
Note that if we let $\w_1 \leftrightarrow \w_2$ in the expression~\eqref{Del-I1-2} for $\Delta I_1$, then we get $\Delta I_4$ in~\eqref{Del-I4-2}.
It is also true that if this switch is made in the entire contribution to the two-point function from $\Delta I_1$ then that is equal to the contribution from $\Delta I_4$.
Finally, the total contribution from $\Delta I_4$ can be shown to be the complex conjugate of the total contribution from $\Delta I_1$.  The sum of both contributions is real.       Next consider $\Delta I_2$
\bes \bea  \Delta I_{2} &=&  -\frac{1}{4 \pi^2  \sqrt{ \w_1 \w_2}} (4 M)^{i 4 M (\w_1+\w_2)} e^{i v_H(\w_1 + \w_2)} \Gamma(1-i 4M\w_1) \Gamma(1-i4M\w_2)
  \Delta K_2 \;, \label{Del-I2-1} \\
  \Delta K_{2} &=&   \lim_{\Lambda \to \infty} (-i)^{i 4 M \w_1} (i)^{i 4 M \w_2}\bigg\{ \left[ \int_{0}^\Lambda d\w \frac{\w}{ ( \w-\w_1 + i \epsilon_1)^{1-i4M\w_1} ( \w+\w_2 - i \epsilon_2)^{1-i4M\w_2}}  \right. \nonumber \\   && \left. \qquad
- \int_{1}^\Lambda d\w \,\w^{-1+i4M(\w_1+\w_2)} \right]- \int_0^1  d\w \, \w^{-1+i4M(\w_1+\w_2)} \bigg\} \;, \label{DelK2-1}
\eea \ees
where the integrals in $ \Delta K_{2}$ can be computed analytically
\bes \bea  \Delta K_{2a} &=&  (-i)^{i 4 M \w_1} (i)^{i 4 M \w_2} \int_{0}^\Lambda d\w \frac{\w}{( \w-\w_1 + i \epsilon_1)^{1-i4M(\w_1-i \epsilon_1)} (\w+\w_2-i \epsilon_2)^{1-i4M(\w_2-i \epsilon_2)}}  \nonumber \\
   &=& -\frac{i}{4M} (-i)^{i 4 M \w_1} (i)^{i 4 M \w_2} \left\{ \frac{(\Lambda-\w_1)^{i4M \w_1} (\Lambda+\w_2)^{i4M\w_2}}{[\w_1+\w_2 - i(\epsilon_1 + \epsilon_2)]} \right. \nonumber \\  & & \left. -  \frac{(-\w_1)^{i4M\w_1} \w_2^{i4M\w_2}}{[\w_1+\w_2 - i(\epsilon_1 + \epsilon_2)]} \right\} \nonumber \\
   &=& -\frac{i}{4M} e^{2 \pi M(\w_1 -\w_2)} \frac{(\Lambda-\w_1)^{i4M \w_1} (\Lambda+\w_2)^{i4M\w_2}}{[\w_1+\w_2 - i(\epsilon_1 + \epsilon_2)]}  \nonumber \\
     & &   + \frac{i}{4M} e^{-2 \pi M(\w_1+\w_2)} \frac{\w_1^{i4M\w_1} \w_2^{i4M\w_2}}{[\w_1+\w_2 - i(\epsilon_1 + \epsilon_2)]}\;,  \label{Del-K2a}  \\
  \Delta K_{2b} &=&  \frac{i}{4 M} e^{2 \pi M(\w_1 -\w_2)} \frac{\Lambda^{i 4M (\w_1+\w_2)}}{\w_1+\w_2}  - \frac{i}{4 M} e^{2 \pi M(\w_1 -\w_2)} \frac{1}{\w_1+\w_2} \;, \\
  \Delta K_{2c} &=&  - e^{2\pi M(\w_1-\w_2)} \int_{-\infty}^0 dz \, e^{[i 4M(\w_1+\w_2)+\epsilon]z} = -\frac{ e^{2\pi M(\w_1-\w_2)}}{i 4M(\w_1+\w_2)+\epsilon}
             \nonumber \\
              &=& i \frac{e^{2\pi M(\w_1-\w_2)}}{4 M (\w_1+\w_2) - i \epsilon}  = i  \frac{e^{2\pi M(\w_1-\w_2)}}{4 M (\w_1+\w_2)} - \frac{\pi}{4M}   \delta(\w_1+\w_2)  \;.
\eea \ees
Given that $\delta(\w_1+\w_2) = 0$ since the frequencies are all non-negative, one can set $\epsilon_1=\epsilon_2=0$. Then
\bes \be \Delta K_2 = i e^{-2 \pi M(\w_1+\w_2)} \frac{\w_1^{i4M\w_1} \w_2^{i4M\w_2}}{4M(\w_1+\w_2)} \;, \ee
and
\bea \Delta I_2 &=& -\frac{i}{4 \pi^2  \sqrt{ \w_1 \w_2}} (4 M)^{i 4 M (\w_1+\w_2)} e^{i v_H(\w_1 + \w_2)}  \Gamma(1-i 4M\w_1) \Gamma(1-i4M\w_2) \nonumber \\
   & & \times e^{-2 \pi M(\w_1+\w_2)} \frac{\w_1^{i4M\w_1} \w_2^{i4M\w_2}}{4M(\w_1+\w_2)}  \;. \label{Del-I2-2} \eea  \ees

Comparing $\Delta I_2$ in~\eqref{Del-I2} and $\Delta I_3$ in~\eqref{Del-I3}, one can immediately see that their contributions to the two-point function,~\eqref{Del-GC}, are the complex conjugate of each other if one also takes $\w_1 \leftrightarrow \w_2$ in the contribution from $\Delta I_2$
 \bea \Delta I_3 &=&  (\Delta I_2)^{*}\nonumber\\
 &=&\frac{i}{4 \pi^2  \sqrt{ \w_1 \w_2}} (4 M)^{-i 4 M (\w_1+\w_2)} e^{-i v_H(\w_1 + \w_2)}  \Gamma(1+i 4M\w_1) \Gamma(1+i4M\w_2) \nonumber \\
   & & \times e^{-2 \pi M(\w_1+\w_2)} \frac{\w_1^{-i4M\w_1} \w_2^{-i4M\w_2}}{4M(\w_1+\w_2)} \;.  \label{Del-I3-2} \eea

Substituting~\eqref{Del-I1-2},~\eqref{Del-I4-2},~\eqref{Del-I2-2}, and~\eqref{Del-I3-2} into~\eqref{Del-GC} one finds
\bea
\Delta G_C(x,x')&=&\mathfrak{R}\Bigg\{ \frac{i}{8\pi^3}\int_0^{\infty}\frac{d\w_1}{\w_1}\int_0^{\infty}\frac{d\w_2}{\w_2}e^{-2\pi M(\w_1+\w_2)} \nonumber \\ & &
\times \bigg\{ \Big[e^{-i\w_1 u_s+i\w_2 u'_s}+e^{-i\w_1 u'_s+i\w_2 u_s}\Big] \frac{(4M\w_1e^{\frac{v_H}{4M}})^{4iM\w_1}}{(4M\w_2e^{\frac{v_H}{4M}})^{4iM\w_2}}\frac{\Gamma(1-4iM\w_1)\Gamma(1+4iM\w_2)}{4M(\w_1-\w_2-i\epsilon)} \nonumber \\ & &
-\Big[e^{-i\w_1 u_s-i\w_2 u'_s}+e^{-i\w_1 u'_s-i\w_2 u_s}\Big](4M\w_1e^{\frac{v_H}{4M}})^{4iM\w_1}(4M\w_2e^{\frac{v_H}{4M}})^{4iM\w_2} \nonumber \\ & &
 \times \frac{\Gamma(1-4iM\w_1)\Gamma(1-4iM\w_2)}{4M(\w_1+\w_2)}\bigg\} \Bigg\}\;. \label{Delta-GC-fin}
 \eea
There are infrared divergences in this expression.  However, it is easy to see that the derivatives in the general formula for the stress-energy tensor~\eqref{Tab-class} bring down factors of $\w_1$ and $\w_2$ which remove these infrared divergences.  Recalling that $\Delta G_C$ is the only contribution to $\la T_{ab} \ra$ from  $\Delta G^{(1)}$, it is straightforward to show using~\eqref{Delta-Tab-2},~\eqref{rtptrp},~\eqref{rrpttp},~\eqref{rtpttp}, and~\eqref{Delta-GC-fin} that
\bea
\Delta \langle T_{tt} \rangle&=&  -(1-\frac{2M}{r}) \lim_{x' \to x} \frac{1}{4}(\Delta G_{C\; ;t';r}+\Delta G_{C \;;t;r'})\nonumber \\
&=&\mathfrak{R}\Bigg\{ \frac{i}{8\pi^3}\int_0^{\infty}d\w_1 \; \int_0^{\infty}d\w_2 \; e^{-2\pi M(\w_1+\w_2)} \nonumber \\ & &
\times \bigg\{ e^{i(\w_2-\w_1) u_s} \frac{(4M\w_1e^{\frac{v_H}{4M}})^{4iM\w_1}}{(4M\w_2e^{\frac{v_H}{4M}})^{4iM\w_2}}\frac{\Gamma(1-4iM\w_1)\Gamma(1+4iM\w_2)}{4M(\w_1-\w_2-i\epsilon)} \nonumber \\ & &
+ e^{-i(\w_2 +\w_1)u_s}(4M\w_1e^{\frac{v_H}{4M}})^{4iM\w_1}(4M\w_2e^{\frac{v_H}{4M}})^{4iM\w_2} \nonumber \\ & &
 \times \frac{\Gamma(1-4iM\w_1)\Gamma(1-4iM\w_2)}{4M(\w_1+\w_2)}\bigg\} \Bigg\}\;.
\label{2D-Ttt} \eea

 The integral over $\w_2$ of the first term inside the curly bracket can be written in the form
  \bea
 \Delta \langle T_{tt} \rangle _1 &=& \int_0^{\infty} d\w_2 \; \frac{f(\w_2)}{\w_1-\w_2-i\epsilon}=  \int_0^\infty d\w_2 \;\left[ \frac{f(\w_2)}{\w_1-\w_2}+i\pi\delta(\w_1-\w_2) \right] \nonumber \\
 &=& \lim_{\epsilon \to 0^+} \left[ \int_0^{\w_1-\epsilon} d \w_2  \frac{f(\w_2)}{\w_1-\w_2} + \int_{\w_1+\epsilon}^\infty d \w_2  \frac{f(\w_2)}{\w_1-\w_2} \right] + i \pi f(\w_1)
\;, \label{CPV}
  \eea
where the definition of the Cauchy principal value integral has been explicitly used.

Thus, extracting the explicit form of the $f(\w_2)$  from ~\eqref{2D-Ttt} and substituting it into ~\eqref{CPV} yields
\bea
\Delta \langle T_{tt} \rangle
 &=&\mathfrak{R}\Bigg\{ \frac{i}{8\pi^3}\int_0^{\infty} d\w_1\; \Big[\int_0^{\infty} d\w_2 \;  e^{-2\pi M(\w_1+\w_2)} e^{i(\w_2- \w_1) u_s} \nonumber \\ & & \times
 \frac{(4M\w_1e^{\frac{v_H}{4M}})^{4iM\w_1}}{(4M\w_2e^{\frac{v_H}{4M}})^{4iM\w_2}}\frac{\Gamma(1-4iM\w_1)\Gamma(1+4iM\w_2)}{4M(\w_1-\w_2)}\Big]\Bigg\}-
 \frac{1}{8\pi^2}\int_0^{\infty} d\w_1 \; \frac{e^{-4\pi M\w_1}}{4M} \nonumber \\ & &
\times \Gamma(1-4iM\w_1)\Gamma(1+4iM\w_1)\;.  \label{Del-Ttt}
 \eea

The stress-energy tensor for a massless minimally coupled scalar field in the 2D collapsing null shell spacetime has been previously computed analytically using a different method~\cite{hiscock, Fabbri:2005mw} and the stress-energy tensor for the Unruh state has also been computed analytically~\cite{Davies-Fulling-Unruh,Fabbri:2005mw}.  For the difference one finds
 \bea
\Delta \la T_{uu} \ra &=& -\frac{1}{24\pi}\left[\frac{8M}{(u-v_0)^3}+\frac{24M^2}{(u-v_0)^4}\right] - \frac{1}{768 \pi M^2}  \;, \nonumber  \\
\Delta \la T_{uv} \ra &=& \Delta \la T_{vv} \ra = 0 \;, \nonumber \\
\Delta \la T_{tt} \ra &=& \Delta \la T_{uu} \ra + 2 \Delta \la T_{uv} \ra + \Delta \la T_{vv} \ra \nonumber \\
 & = & -\frac{1}{24\pi}\left[\frac{8M}{(u-v_0)^3}+\frac{24M^2}{(u-v_0)^4}\right]  - \frac{1}{768 \pi M^2}  \;.
\label{Hiscock-Result} \eea

Both terms in~\eqref{Del-Ttt} have been computed numerically. In the first integral, the numerical computation has been performed by the symmetric removal of the neighborhood with radius $\epsilon$ about the singular points of the integrand, $\w_1=\w_2$. The integral of the second term in ~\eqref{CPV}  has been computed using a more straightforward numerical method.
 Our results for $\Delta \langle  T_{tt} \rangle$ in~\eqref{CPV} are shown in Fig.~{fig:Ttt}. Although it is not possible to detect this from the plot, our numerical results agree with the analytical results in~\cite{hiscock} \cite{Fabbri:2005mw} to more than ten digits.
 \begin{figure}[h]
\centering
\includegraphics [totalheight=0.45\textheight]{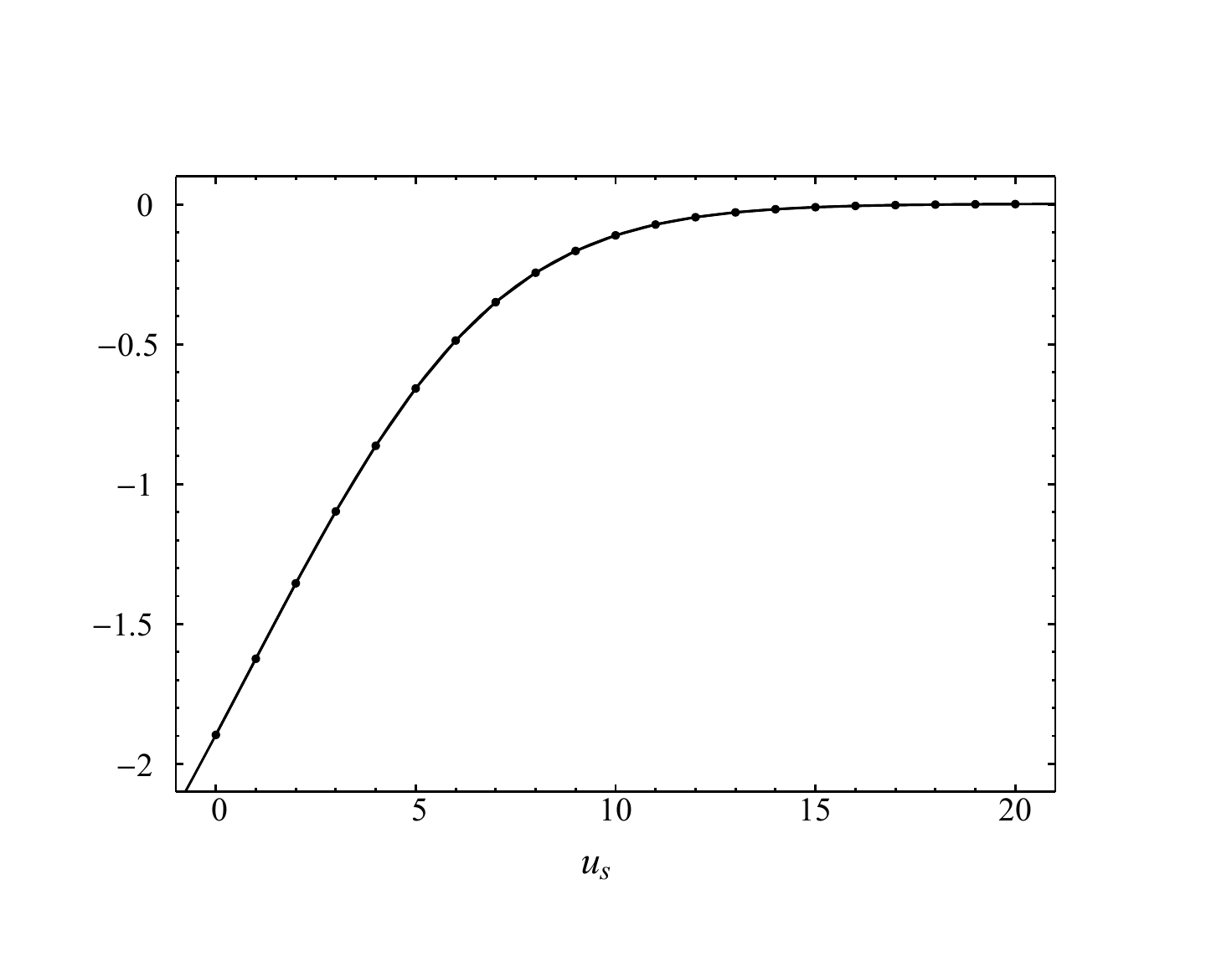}
\caption{The quantity $10^{4}M^2\Delta \langle T_{tt} \rangle$ is plotted for the massless minimally coupled scalar field in the region exterior to the null shell
and to the event horizon. The dots correspond to the results of the numerical computations.  The solid curve represents the analytic results in~\eqref{Hiscock-Result}.   }
\label{fig:Ttt}
\end{figure}

It is worth mentioning that in 2D, once  $\Delta \langle T_{tt} \rangle$ is numerically computed,  $\Delta \langle T_{rr} \rangle$ and  $\Delta \langle T_{tr} \rangle$ can be easily determined from the relations~\eqref{TrrTtt} and~\eqref{DTrt-DTtt}.

\section{Summary}\label{ch4-summary}

We have presented a method of numerically computing the stress-energy tensor for a massless minimally coupled scalar field in the case when a black hole is formed from the
collapse of a spherically symmetric null shell in four dimensions.  There are two primary parts to the method.  The first is to expand the mode functions in the natural $in$ vacuum state in terms of a complete set of mode functions in the part of Schwarzschild spacetime that is outside of the event horizon of the black hole. Expressions have been found for the matching coefficients that involve integrals of these mode functions over the trajectory of the null shell.

The second part of the method involves subtracting the unrenormalized expression for the stress-energy tensor in the Unruh state from the expression for the unrenormalized stress-energy tensor in the $in$ vacuum state.  Since the ultraviolet divergences in the stress-energy tensor are independent of the state, this difference is finite.  Then one can add to this the renormalized expression for the stress-energy tensor in the Unruh state that has already been computed~\cite{levi-ori,levi} and the result is the full renormalized stress-energy tensor for the $in$ vacuum state.

We have tested the first part of the method by analytically computing the matching coefficients in the 2D case and reconstructing the mode functions for the $in$ vacuum state.  We have also analytically computed the matching coefficients in 4D for the spherically symmetric mode functions (those with $\ell = 0$) in the $in$ vacuum state for a simple model in which the effective potential in the mode equation is proportional to a Dirac delta function.  In this case, it was possible to analytically compute the part of the mode function in the $in$ vacuum state that is proportional to $e^{-i \w v}$ inside the null shell and to verify that it gives the known result on the matching surface.  Finally, for the actual case of a collapsing null shell in 4D, we have analytically computed parts of the matching coefficients and used those parts to numerically compute part of one of the $in$ modes on the future horizon and shown that it has the correct value at the point where the future horizon intersects the null shell trajectory.

 The second part of the method has been tested by numerically computing in 2D the difference between the stress-energy tensor in the $in$ vacuum state for the collapsing null shell spacetime and the Unruh state for Schwarzschild spacetime. The result is in excellent agreement with an analytic expression for the difference obtained from prior calculations of the stress-energy tensor in these two states~\cite{Davies-Fulling-Unruh, hiscock, Fabbri:2005mw}.

These tests provide substantial evidence that the method will work and that it will be possible to numerically compute the exact renormalized stress-energy tensor for a massless minimally coupled scalar field in a 4D spacetime in which a black hole forms from the collapse of a spherically symmetric null shell.  Work on that computation is in progress.

\section{Appendix A: Lagrange Inversion Theorem applied to the Lambert W function}\label{appendix-A}

In~\cite{Corless} the relation
\be e^{-c \text{W}(x)} =  \sum_{n=0}^\infty \frac {c(n+c)^{n-1}}{n!} (-x)^n \;. \label{final-Lambert-1} \ee
is derived for any complex constant $c$.  An alternative derivation is given here.  It is based on the Lagrange inversion theorem~\cite{Jacobi}.  In~\cite{Gessel} different forms for the Lagrange inversion theorem are given, one of which we use here. To state the form that is most useful to us we use the notation in~\cite{Gessel} that
if $f(x)$ is expanded in a Laurent series then $[x^n] f(x)$ denotes the coefficient of $x^n$ in that series.
Then a statement of the theorem is as follows:
Suppose $f$ is a function of $x$ and there is a relation of the following form:
\bea f(x) = x R(f(x)) \;, \label{f-relation} \eea
where $R(t)$ is a power series in $t$.  Suppose further that $\phi(t)$ is also a Laurent series in $t$.  Then for any nonzero integer $n$, $\phi(f(x))$ can be expressed in terms of a unique power series in $x$ with coefficients
\bea
[x^n]\phi(f)\equiv  \frac{1}{n!}  \left. \frac{d^{n} \phi(f(x))}{dx^n}\right|_{x = 0} = \frac{1}{n}[t^{n-1}] \phi'(t)R(t)^n\;, \label{LIT}
\eea
where the interpretation of the far right-hand side is that one first expands the function $\frac{d\phi(t)}{dt} R(t)$ in powers of $t$, then chooses the coefficient of the term proportional to $t^{n-1}$ in that series and divides that coefficient by $n$.

To use this to obtain a power series for the function $e^{-c W(x)}$, note that the Lambert W function satisfies the relation
\be \text{W}(x) = x e^{-\text{W}(x)}  = x \sum_{n=0}^\infty \frac{(-W(x))^n}{n!}  \;. \label{Lambert-2} \ee
Thus we can choose the function $R(t)$ in~\eqref{f-relation} to be $R(t) = e^{-t}$.
We also choose $\phi(t) = e^{-c t}$.  Then
\bea [x^n] e^{-c \text{W}(x)} &=& \frac{1}{n} [t^{n-1}]\phi'(t) R^n(t)  \nonumber \\
                       &=& -\frac{c}{n} [t^{n-1}] e^{-(c+n) t} =  -\frac{c}{n!} [-(c+n)]^{n-1}  \nonumber \\
                       &=& (-1)^n \frac{c}{n!} (c +n)^{n-1}  \;. \eea
Equation~\eqref{final-Lambert-1} follows immediately from this.

\section{Appendix B: Contributions to the stress-energy tensor}\label{appendix-B}

The calculations in this appendix are done entirely for the Schwarzschild geometry.  Therefore for simplicity we use $t$ and $u$ to denote the usual time coordinate and the right moving radial null coordinate in Schwarzschild spacetime.

In Sec.~{sec:Tab-2D} it is mentioned that for the null-shell spacetime in 2D the Hadamard Green's function in~\eqref{G1-1} can be broken into three parts.  One of these, which we call $G_A^{(1)}(x,x')$, includes $f^{H^+}$ and its complex conjugate and is given by the expression
\bea
 G^{(1)}_A(x,x')= \int_0^{\infty}d\w \Big\{f^{H^+}_{\w}(x)f^{H^+ \; *}_{\w}(x')+f^{H^+}_{\w}(x')f^{H^+ \; *}_{\w}(x)\Big\}
 \;. \label{GA}
\eea
For the Unruh state the corresponding contribution to $G^{(1)}(x,x')$ is exactly the same so $\Delta G^{(1)}_A(x,x') = 0$.

A second part, $G_B^{(1)}(x,x')$, has terms involving products of $f^{H^+}$ and its complex conjugate with $f^{\mathscr{I^+}}$ and its complex conjugate such that
 \bea
 G^{(1)}_{B}(x,x') &=& \int_0^\infty d \w \;\Big\{ \int_0^\infty d\w_2 \;   \Big[A^{*}_{\w \w_2} f^{H^{+}}_\w(x) f^{\mathscr{I}^{+}\;*}_{\w_2}(x')  + B^{*}_{\w \w_2}  f^{H^{+}}_\w(x)f^{\mathscr{I}^{+}}_{\w_2}(x')
\Big] \nonumber \\ &&
+  \int_0^\infty d\w_1 \; \Big[A_{\w \w_1}  f^{H^{+} \;*}_\w(x') f^{\mathscr{I}^{+}}_{\w_1}(x)  + B_{\w \w_1}  f^{H^{+} \;*}_\w(x') f^{\mathscr{I}^{+}\;*}_{\w_1}(x)\Big]\nonumber \\ &&
 + \; \int_0^\infty d\w_2 \;\Big[A^{*}_{\w \w_2} f^{H^{+}}_\w(x')  f^{\mathscr{I}^{+}\;*}_{\w_2}(x)  + B^{*}_{\w \w_2} f^{H^{+}}_\w(x')  f^{\mathscr{I}^{+}}_{\w_2}(x)\Big] \nonumber \\ &&
+ \; \int_0^\infty d\w_1 \;\Big[ A_{\w \w_1} f^{H^{+} \;*}_\w(x) f^{\mathscr{I}^{+}}_{\w_1}(x')  + B_{\w \w_1}  f^{H^{+} \;*}_\w(x)f^{\mathscr{I}^{+}\;*}_{\w_1}(x')\Big]\Big\}
\;. \label{GB} \eea
There is no contribution to $G^{(1)}(x,x')$ which has terms of this form if the field is in the Unruh state, so there is no subtraction term and $
\Delta G_B^{(1)}(x,x') = G_B^{(1)}(x,x')$.

While $ G^{(1)}_{B}(x,x')$ contributes to the two-point function, we next show that its contribution to the stress-energy tensor is zero.
Substituting~\eqref{f-form-2D} into~\eqref{GB} and using~\eqref{right-moving-2D} and~\eqref{left-moving-2D} , one readily finds that
 \bea
\Big[ G^{(1)}_{B}(x,x') \Big]_{;t;t'}&=&\frac{1}{4\pi}\int_0^\infty d \w \; \Big\{\int_0^\infty d\w_2 \;  \sqrt{\w\w_{2}} \; (A^{*}_{\w \w_2}e^{-i\w v+i\w_{2}u'} -B^{*}_{\w \w_2} e^{-i\w v-i\w_{2} u'}) \nonumber \\ &&
+ \int_0^\infty d\w_1 \;  \sqrt{\w\w_{1}} \;(A_{\w \w_1}e^{i\w v'-i\w_{1}u} -B_{\w \w_1} e^{i\w v'+i\w_{1} u}) \nonumber \\ &&
+ \int_0^\infty d\w_2 \;  \sqrt{\w\w_{2}} \; (A^{*}_{\w \w_2}e^{-i\w v'+i\w_{2}u} -B^{*}_{\w \w_2} e^{-i\w v'-i\w_{2} u}) \nonumber \\ &&
+ \int_0^\infty d\w_1 \;  \sqrt{\w\w_{1}} \; (A_{\w \w_1}e^{i\w v-i\w_{1}u'} -B_{\w \w_1} e^{i\w v+i\w_{1} u'})\Big\}
\;, \label{G_tt'} \eea

 \bea
\Big[ G^{(1)}_{B}(x,x') \Big]_{;r;r'}&=&\frac{1}{4\pi(1-\frac{2M}{r})^2} \int_0^\infty d \w \;\Big\{ \int_0^\infty d\w_2 \;  \sqrt{\w\w_{2}} \; (-A^{*}_{\w \w_2}e^{-i\w v+i\w_{2}u'} +B^{*}_{\w \w_2} e^{-i\w v-i\w_{2} u'}) \nonumber \\ &&
+ \int_0^\infty d\w_1 \;  \sqrt{\w\w_{1}} \; (-A_{\w \w_1}e^{i\w v'-i\w_{1}u} +B_{\w \w_1} e^{i\w v'+i\w_{1} u}) \nonumber \\ &&
+\int_0^\infty d\w_2 \;  \sqrt{\w\w_{2}} \; (-A^{*}_{\w \w_2}e^{-i\w v'+i\w_{2}u} +B^{*}_{\w \w_2} e^{-i\w v'-i\w_{2} u}) \nonumber \\ &&
+\int_0^\infty d\w_1 \;  \sqrt{\w\w_{1}} \; (-A_{\w \w_1}e^{i\w v-i\w_{1}u'}+B_{\w \w_1} e^{i\w v+i\w_{1} u'})\Big\}
\;. \label{G_rr'} \eea
From~\eqref{Delta-Tab-2} one finds
\bea
\Delta \langle  T_{tt} \rangle=\frac{1}{4} \lim_{x' \to x} \Big[\Delta G_{;t;t'}+(1-\frac{2M}{r})^2 \Delta G_{;r;r'}\Big] \label{T_ttB}
\; .\eea
By substituting ~\eqref{G_tt'} and ~\eqref{G_rr'}  into ~\eqref{T_ttB},  it is easy to see that the contribution to $\langle T_{tt}\rangle$ is zero.

Next consider the contribution of $ G^{(1)}_{B}(x,x')$ to $\langle T_{rr} \rangle$.
Using~\eqref{Delta-Tab-2} it is not hard to show that
\bea
\Delta \langle  T_{rr} \rangle=\frac{1}{4} \lim_{x' \to x} \Big[\frac{\Delta G_{;t;t'}}{(1-\frac{2M}{r})^2}+ \Delta G_{;rr'}\Big] \label{T_rrB}
\;. \eea
Together with ~\eqref{T_ttB}, one obtains
\bea
\Delta \langle T_{rr} \rangle=\frac {\Delta \langle T_{tt} \rangle}{\left(1-\frac{2M}{r}\right)^2}\;. \label{TrrTtt}
\eea
Thus $ G^{(1)}_{B}(x,x') $ does not contribute to  $\langle T_{rr}\rangle$ either.

Finally, we consider the contribution of  $ G^{(1)}_{B}(x,x') $  to $\langle T_{tr}\rangle$.  From~\eqref{Delta-Tab-2} one finds
\bea
\Delta \langle T_{tr} \rangle=\frac{1}{4} \lim_{x' \to x} \Big[\Delta G_{;t';r}+ \Delta G_{;t;r'}\Big] \label{T_trB}
\;. \eea
Taking the derivative of ~\eqref{GB} with respect to $t$ and $r'$, one finds
 \bea
\Big[ G^{(1)}_{B}(x,x') \Big]_{;t;r'}&=&\frac{1}{4\pi(1-\frac{2M}{r'})} \int_0^\infty d \w \; \Big\{\int_0^\infty d\w_2 \;  \sqrt{\w\w_{2}} \; (-A^{*}_{\w \w_2}e^{-i\w v+i\w_{2}u'} +B^{*}_{\w \w_2} e^{-i\w v-i\w_{2} u'}) \nonumber \\ &&
+  \int_0^\infty d\w_1 \;  \sqrt{\w\w_{1}} \; (A_{\w \w_1}e^{i\w v'-i\w_{1}u} -B_{\w \w_1} e^{i\w v'+i\w_{1} u}) \nonumber \\ &&
+ \int_0^\infty d\w_2 \;  \sqrt{\w\w_{2}} \; (A^{*}_{\w \w_2}e^{-i\w v'+i\w_{2}u} -B^{*}_{\w \w_2} e^{-i\w v'-i\w_{2} u}) \nonumber \\ &&
+ \int_0^\infty d\w_1 \;  \sqrt{\w\w_{1}} \; (-A_{\w \w_1}e^{i\w v-i\w_{1}u'}+B_{\w \w_1} e^{i\w v+i\w_{1} u'})\Big\}
\;. \label{G_tr'} \eea
and taking the derivative of ~\eqref{GB} with respect to $t'$ and $r$ gives
 \bea
\Big[ G^{(1)}_{B}(x,x') \Big]_{;t';r}&=&\frac{1}{4\pi(1-\frac{2M}{r'})}\int_0^\infty d \w \;\Big\{ \int_0^\infty d\w_2 \;  \sqrt{\w\w_{2}} \; (A^{*}_{\w \w_2}e^{-i\w v+i\w_{2}u'} -B^{*}_{\w \w_2} e^{-i\w v-i\w_{2} u'}) \nonumber \\ &&
+ \int_0^\infty d\w_1 \;  \sqrt{\w\w_{1}} \; (-A_{\w \w_1}e^{i\w v'-i\w_{1}u} +B_{\w \w_1} e^{i\w v'+i\w_{1} u}) \nonumber \\ &&
+ \int_0^\infty d\w_2 \;  \sqrt{\w\w_{2}} \; (-A^{*}_{\w \w_2}e^{-i\w v'+i\w_{2}u} +B^{*}_{\w \w_2} e^{-i\w v'-i\w_{2} u}) \nonumber \\ &&
+\int_0^\infty d\w_1 \;  \sqrt{\w\w_{1}} \; (A_{\w \w_1}e^{i\w v-i\w_{1}u'}-B_{\w \w_1} e^{i\w v+i\w_{1} u'})\Big\}
\;. \label{G_t'r}
\eea
It is clear that $\Big[ G^{(1)}_{B}(x,x') \Big]_{;t';r} = - \Big[ G^{(1)}_{B}(x,x') \Big]_{;t;r'}$ and therefore that their contribution to  $\langle T_{tr}\rangle$ is zero.

The third part of $G^{(1)}(x,x')$ we call $G^{(1)}_C(x,x')$.  Its contribution to $\Delta \la T_{tt} \ra$ is given in Sec.~{sec:Tab-2D}.

\section{Appendix C: Relation between two components of $\Delta \la T_{ab} \ra$}\label{appendix-C}

The calculations in this appendix are done entirely for the Schwarzschild geometry.  Therefore for simplicity, we use $t$ and $u$ to denote the usual time coordinate and the right moving radial null coordinate in Schwarzschild spacetime.

In this appendix, a relation is derived between two components of $\Delta \la T_{ab} \ra$ in~\eqref{Delta-Tab-2} for the 2D collapsing null shell spacetime.
As shown in Section ~{appendix-B} only $\Delta G_C(x,x')$ in~\eqref{Del-GC} contributes to $\Delta \la T_{ab} \ra$.  The explicit form for $\Delta G_c(x,x')$ is
\bea
\Delta G_C(x,x')&=&\frac{1}{4\pi} \int_0^\infty d \w \int_0^{\infty} d\w_1 \int_0^{\infty} d\w_2 \frac{1}{\sqrt{\w_1 \w_2}}\Big\{[A_{\w \w_1}e^{-i\w_1 u}+B_{\w \w_1}e^{i\w_1 u}]\nonumber \\ &&
\times [A_{\w \w_2}^{*} e^{i\w_2 u'}+B_{\w \w_2}^{*}e^{-i\w_2 u'}]+[A_{\w \w_1}e^{-i\w_1 u'}+B_{\w \w_1} e^{i\w_1 u'}]\nonumber \\ &&
 \times [A_{\w \w_2}^{*} e^{i\w_2 u}+B_{\w \w_2}^{*} e^{-i\w_2 u}]\nonumber \\ &&
-\mbox{subtraction terms}\Big\}\;,
\eea
where the subtraction terms have exactly the same form except that the matching coefficients are replaced by the Bogolubov coefficients~\eqref{alphaK-betaK} for the Unruh state.
Then
\bea
\Big[\Delta G_C(x,x')\Big]_{;r;t'}&=&\frac{1}{4\pi}\frac{1}{(1-\frac{2M}{r})}  \int_0^\infty d \w \int_0^{\infty} d\w_1 \int_0^{\infty} d\w_2 \sqrt{\w_1 \w_2}\Big\{[iA_{\w \w_1}e^{-i\w_1 u}-iB_{\w \w_1}e^{i\w_1 u}]\nonumber \\ &&
\times [iA_{\w \w_2}^{*}e^{i\w_2 u'}-iB_{\w \w_2}^{*}e^{-i\w_2 u'}]+[-iA_{\w \w_1} e^{-i\w_1 u'}+iB_{\w \w_1} e^{i\w_1 u'}]\nonumber \\ &&
\times [-iA_{\w \w_2}^{*} e^{i\w_2 u}+iB_{\w \w_2}^{*} e^{-i\w_2 u}]\nonumber \\ &&
-\mbox{subtraction terms}\Big\}\;. \label{Delta-GCrtp}
\eea
A similar calculation for $\Big[\Delta G_C(x,x')\Big]_{;t;r'}$ gives the opposite sign for each term in square brackets and a replacement of $r$ with $r'$ in the overall factor of $(1-\frac{2M}{r})^{-1}$.
Thus
\be \lim_{x'\to x} \Big[\Delta G_C(x,x')\Big]_{;r;t'} = \lim_{x'\to x} \Big[\Delta G_C(x,x')\Big]_{;t;r'} \;. \label{rtptrp} \ee

Next consider
\bea
\Big[\Delta G_C(x,x')\Big]_{;t;t'}&=&\frac{1}{4\pi} \int_0^\infty d\w  \int_0^{\infty} d\w_1 \int_0^{\infty} d\w_2 \sqrt{\w_1 \w_2}\Big\{[-iA_{\w \w_1}e^{-i\w_1 u}+iB_{\w \w_1}e^{i\w_1 u}]\nonumber \\ &&
 \times [iA_{\w \w_2}^{*}e^{i\w_2 u'}-iB_{\w \w_2}^{*}e^{-i\w_2 u'}]+[- iA_{\w \w_1} e^{-i\w_1 u'}+ iB_{\w \w_1} e^{i\w_1 u'}]\nonumber \\ &&
 \times [iA_{\w \w_2}^{*} e^{i\w_2 u}-iB_{\w \w_2}^{*} e^{-i\w_2 u}]\nonumber \\ &&
-\mbox{subtraction terms}\Big\}\;.  \label{Delta-GCttp}
\eea
A similar computation for $\Big[\Delta G_C(x,x')\Big]_{;r;r'}$ gives the relation
\be \Big[\Delta G_C(x,x')\Big]_{;r;r'} = \frac{1}{(1-\frac{2M}{r}) (1-\frac{2M}{r'})} \Big[\Delta G_C(x,x')\Big]_{;t;t'} \;. \label{rrpttp} \ee
Also a comparison of~\eqref{Delta-GCrtp} and~\eqref{Delta-GCttp} shows  that
\bea
\Big[\Delta G_C(x,x')\Big]_{;r;t'}=-\frac{\Big[\Delta G_C(x,x')\Big]_{;t;t'}}{1-\frac{2M}{r}}\;.
\label{rtpttp}
\eea
Finally by substituting~\eqref{rrpttp} into~\eqref{T_ttB} and substituting~\eqref{rtptrp} and~\eqref{rtpttp} into~\eqref{T_trB} one can see that
\bea
  \Delta \langle  T_{rt}\rangle = - \frac{\Delta \langle T_{tt}\rangle}{1-\frac{2M}{r}}\;. \label{DTrt-DTtt}
\eea
\\[10pt]

\textbf{Acknowledgments}

P. R. A. would like to thank Eric Carlson, Gregory Cook, Charles Evans, Adam Levi, and Amos Ori for helpful conversations and Adam Levi for sharing some of his numerical data.
A.F. acknowledges partial financial support from the
Spanish Ministerio de Ciencia e Innovaci\'on grant
FIS2017-84440-C2-1-P and from the Generalitat Valenciana grant
PROMETEO/2020/079.  This work was supported in part by the National Science Foundation under Grants No.  PHY-1308325, PHY-1505875, and PHY-1912584 to Wake Forest University.
Some of the numerical work was done using the WFU DEAC cluster; we thank the WFU Provost's Office and Information
Systems Department for their generous support.
\chapter{quantized scalar field in a null-shell spacetime}
\section{Introduction}
In this chapter, we use the method of computing the modes of a quantized scalar field outlined in the previous chapter to numerically compute the spherically-symmetric modes of a scalar field in the $in$ vacuum state in the null-shell spacetime. We first write these modes in terms of some matching coefficients and a complete set of modes that can be computed using separation of variables. We next discuss the computation of the matching coefficients. Finally, we numerically compute the $in$ modes and check them in several limits where we know their exact values.

\section{$v-$dependent Modes}
We focus on deriving partially analytical expressions for the spherically-symmetric modes of a massless minimally-coupled scalar field in the collapsing null-shell spacetime. These results will be used to compute the stress-energy tensor for the massless scalar field in the null-shell spacetime. In Section 4.6.5, it was shown that the contribution to the matching coefficients from the part of the $in$ modes with $\ell=0$ that go like $e^{-i\w v}$ inside the null shell and the part of the $in$ modes that go like $e^{-i\w u}$ inside the null shell can be written separately. Similarly, one can expand the modes that go like $e^{-i\w v}$ inside the null shell  as follows
\bea
\big(f^{in}_{\w }\big)_v=\int_0^{\infty}\Big\{\big(A^{H^+}_{\w \w'}\big)_vf^{H^+}_{\w'}+\big(B^{H^+}_{\w \w'}\big)_vf^{H^{+*}}_{\w'}+\big(A^{\mathscr{I}^+}_{\w \w'}\big)_vf^{\mathscr{I}^+}_{\w'}+\big(B^{\mathscr{I}^+}_{\w \w'}\big)_vf^{\mathscr{I}^{+*}}_{\w'}\Big\}d\w',\label{ch4-fin-vpart}
\eea
where for the region outside the null shell, $v>v_0$, $\big(A^{H^+}_{\w \w'}\big)_v$ and $\big(B^{H^+}_{\w \w'}\big)_v$ are the second terms in ~\eqref{A-H-gen-mat} and ~\eqref{B-H-gen_mat} respectively;
\bes \bea
\big(A^{H^+}_{\w \w'}\big)_v&=&\frac{i}{2 \pi} \sqrt{\frac{\w'}{\w}} \frac{1}{F_L^{*}(\w',0)}\frac{e^{i(\w'-\w)v_0}}{\w'-\w+i \epsilon},\\
\big(B^{H^+}_{\w \w'}\big)_v&=& - \frac{i}{2 \pi} \sqrt{\frac{\w'}{\w}} \frac{1}{F_L(\w',0)}\frac{e^{-i(\w+\w')v_0}}{\w'+\w-i \epsilon},\label{ABH-v-part}
\eea \ees
and $\big(A^{\mathscr{I}^+}_{\w \w'}\big)_v$ and $\big(B^{\mathscr{I}^+}_{\w \w'}\big)_v$ are the first terms in ~\eqref{A-I-gen-mat} and ~\eqref{B-I-gen-mat} respectively;
\bes \bea
\big(A^{\mathscr{I}^+}_{\w \w'}\big)_v&=& - \frac{i}{2 \pi} \sqrt{\frac{\w'}{\w}} \frac{F_R^{*}(\w',0)}{F_L^{*}(\w',0)}\frac{e^{-i(\w-\w')v_0}}{\w'-\w+i \epsilon}, \\
\big(B^{\mathscr{I}^+}_{\w \w'}\big)_v&=& \frac{i}{2 \pi} \sqrt{\frac{\w'}{\w}} \frac{F_R(\w',0)}{F_L(\w',0)}\frac{e^{-i(\w+\w')v_0}}{\w'+\w-i \epsilon}.\label{ABI-v-part}
\eea \ees
Note that the first terms in \eqref{A-I-gen-mat} and ~\eqref{B-I-gen-mat} are the contributions to the matching coefficients from the data on the future horizon for $v<v_0$. Therefore, they are causally disconnected from the region both outside the null shell and outside the future horizon and do not contribute to the $in$ modes.

Since the focus is only on the spherically-symmetric modes, we have dropped the subscripts $\ell$ and $m$ in the matching coefficients, and $f^{in}_{\w}$. Substituting ~\eqref{ABH-v-part} and ~\eqref{ABI-v-part} into ~\eqref{ch4-fin-vpart} and using the identity $|F_L|^2-|F_R|^2=1$ given in Section 4.5.1, one finds
\bea
\big(f^{in}_{\w}\big)_v=&=&\frac{Y_{00}}{r\sqrt{4\pi\w}}\int_0^{\infty}\Bigg\{\frac{i}{2\pi}\frac{e^{-i(\w-\w')v_0}}{\w'-\w+i\epsilon}\chi_L^{\infty}e^{-i\w't_s}-\frac{i}{2\pi}\frac{e^{-i(\w'+\w)v_0}}{\w'+\w-i\epsilon}\chi_R^{\infty}e^{i\w't_s}\nonumber \\ &&
-\frac{i}{2\pi}\frac{F_R^*(\w')}{F_L^*(\w')}\frac{e^{-i(\w-\w')v_0}}{\w'-\w+i\epsilon}\chi_R^{\infty}e^{-i\w' t_s}\nonumber \\ &&
+\frac{i}{2\pi}\frac{1}{\sqrt{4\pi\w}}\frac{F_R(\w')}{F_L(\w')}\frac{e^{-i(\w'+\w)v_0}}{\w'+\w-i\epsilon}\chi_L^{\infty}e^{i\w' t_s}\Bigg\}d\w'.\label{ch4-fin-vpart2}
\eea
One can next use contour integration over the first and the second terms inside the curly bracket in ~\eqref{ch4-fin-vpart2} to find
\bea
\big(f^{in}_{\w}\big)_v&=&\frac{Y_{00}}{r\sqrt{4\pi \w}}\Bigg\{\chi_L^{\infty}e^{-i\w t_s}
-\frac{i}{2\pi}\int_{0}^{\infty}\frac{F_R^*(\w')}{F_L^*(\w')}\frac{e^{-i(\w-\w')v_0}}{\w'-\w+i\epsilon}\chi_R^{\infty}e^{-i\w' t_s}d\w'\nonumber \\ &&
+\frac{i}{2\pi}\int_0^{\infty}\frac{F_R(\w')}{F_L(\w')}\frac{e^{-i(\w'+\w)v_0}}{\w'+\w-i\epsilon}\chi_L^{\infty}e^{i\w' t_s}d\w'\Bigg\}.\label{ch5-fin-v-second-2}
\eea
Note that before performing the contour integration, we changed the variable of integration in the second term such that $\w'\to -\w'$ to write the contribution of the first and second terms in the form of an integral from $-\infty$ to $\infty$. 
\section{$u$-dependent Modes}
The $\big(f^{in}_{\w}\big)_u$ modes can be expanded in terms of the matching coefficients in the following way
\bea
\big(f^{in}_{\w }\big)_u=\int_0^{\infty}\Big\{\big(A^{H^+}_{\w \w'}\big)_uf^{H^+}_{\w'}+\big(B^{H^+}_{\w \w'}\big)_uf^{H^{+*}}_{\w'}+\big(A^{\mathscr{I}^+}_{\w \w'}\big)_uf^{\mathscr{I}^+}_{\w'}+\big(B^{\mathscr{I}^+}_{\w \w'}\big)_uf^{\mathscr{I}^{+*}}_{\w'}\Big\}d\w',\label{ch4-fin-upart}
\eea
where the matching coefficients  $A^{\big({H^+, \mathscr{I}^+}\big)}_{\w \w'}$ and $B^{\big({H^+, \mathscr{I}^+}\big)}_{\w \w'}$ are given in terms of the integrals in ~\eqref{ell-0-mat};\\
\bes \bea
\big(A^{H^+}_{\w \w'}\big)_u&=& - \frac{1}{2 \pi} \sqrt{\frac{\w }{\w'}}  \int_{-\infty}^{v_H} du \,  e^{-i \w u}  \psi^{H^{+} *}_{\w' 0}(u_s(u),v_0),\label{ch5-AHu}\\
\big(B^{H^+}_{\w \w'}\big)_v&=& \frac{1}{2 \pi} \sqrt{\frac{\w}{ \w'}}  \int_{-\infty}^{v_H} du \,e^{-i \w u} \psi^{H^{+}}_{\w' 0} (u_s(u),v_0),\label{ch5-BHu}\\
\big(A^{\mathscr{I}^+}_{\w \w'}\big)_u&=&\frac{i}{2 \pi} \frac{1}{\sqrt{\w \w'}} \int_{-\infty}^{v_H} du \, e^{-i \w u} \partial_u \psi^{\mathscr{I}^+  *}_{\w' 0}(u_s(u),v_0),\label{ch5-AIu}\\
\big(B^{\mathscr{I}^+}_{\w \w'}\big)_u&=&- \frac{i}{2 \pi} \frac{1}{\sqrt{\w \w'}} \int_{-\infty}^{v_H} du \,e^{-i \w u} \partial_u \psi^{\mathscr{I}^+}_{\w' 0}(u_s(u),v_0).\label{ch5-BIu}
\eea \label{ch5-ABHIu}\ees
Note that the integrals above have to be computed on the null-shell trajectory and since $\psi^{H^{+} *}_{\w' 0}$ and $\psi^{\mathscr{I}^{+} *}_{\w' 0}$ do not have an analytic form on the null-shell trajectory, the integrals have to be evaluated numerically. In the next section, we will outline the method used to calculate these matching coefficients.
\section{Numerical Computation of the Matching Coefficients}
Using ~\eqref{chi-def-2}, ~\eqref{chi-H-plus-def}, and ~\eqref{chi-scri-plus-def}, one finds that 
\bea
\psi^{H^+}_{\w}=\frac{\chi_L^{\infty}(r)}{F_L(\w)}e^{-i\w t_s}
\eea
and \bea\psi^{\mathscr{I}^+}_{\w}=\Big(\chi_R^{\infty}(r)-\frac{F_R(\w)}{F_L(\w)}\chi^{\infty}_{L}\Big)e^{-i\w t_s}.
\eea
For simplicity, we shall rewrite the integrals in ~\eqref{ch5-ABHIu} in terms of a new spatial coordinate $s$ defined by
\bea
s=\frac{r-2M}{M}.
\eea
Substituting the general form for $\psi_{\w}^{H^+}$ and $\psi_{\w}^{\mathscr{I}^+}$ into ~\eqref{ch5-ABHIu} and using the definitions ~\eqref{u-v-flat}, ~\eqref{u-v-sch}, and ~\eqref{rstar-def}, we find
\bes \bea
\big(A^{H^+}_{\w \w'}\big)_u&=& -\frac{e^{i\gamma_A}}{\pi}\sqrt{\frac{\w}{\w'}}\frac{1}{F_L^*(\w')}\int_0^{\infty}ds\; e^{i\Omega_A}\big(\chi^c_{\w'}+i\chi^s_{\w'}\big) ,\\
\big(B^{H^+}_{\w \w'}\big)_u&=&\frac{e^{i\gamma_B}}{\pi}\sqrt{\frac{\w}{\w'}}\frac{1}{F_L(\w')}\int_0^{\infty}ds e^{i\Omega_B}\big(\chi^c_{\w'}-i\chi^s_{\w'}\big),\\
\big(A^{\mathscr{I}^+}_{\w\w'}\big)_u&=&-\frac{i}{2\pi\sqrt{\w\w'}}e^{i\gamma_A}\int_0^{\infty}ds\; e^{i\Omega_A}\Big[-i\w'\big(1+\frac{2}{s}\big)\chi_{\w'}^{\mathscr{I}^{+*}}+\frac{d}{ds}\chi_{\w'}^{\mathscr{I}^{+*}}\Big],\\
\big(B^{\mathscr{I}^+}_{\w\w'}\big)_u&=&\frac{i}{2\pi\sqrt{\w\w'}}e^{i\gamma_B}\int_0^{\infty}ds\; e^{i\Omega_B}\Big[i\w'\big(1+\frac{2}{s}\big)\chi_{\w'}^{\mathscr{I}^+}+\frac{d}{ds}\chi_{\w'}^{\mathscr{I}^+}\Big],
\eea \label{mat-general-s} \ees 
where 
\bes \bea 
\gamma_A&=&-(\w-\w')v_0+4\w-2\w'+2\w'\log{2},\\
\gamma_B&=&-(\w+\w')v_0+4\w+2\w'-2\w'\log{2},
\eea \ees
and
\bes \bea 
\Omega_A&=&2\w s-\w's-2\w'\log{s},\\
\Omega_B&=&2\w s+\w's+2\w'\log{s}.
\eea \ees
\\
As mentioned earlier, the integrals in ~\eqref{mat-general-s} have to be computed numerically. However, one can still find analytical approximations for the integrand in the cases $s\gg 1$ and $\w^2\gg V_{\text{eff}}(s)$. Note that $V_{\text{eff}}(s)$ is the effective potential in ~\eqref{chi-eq-Sch}. It vanishes in the limits $s\to 0$ and $s\to \infty$. For any given value of $\w$, for small enough values of $s$,  $\w^2 \gg |V_{\text{eff}}(s)|$.  That is the near-horizon limit. In this limit, $\chi_R^{\infty}(\w,s)$ and $\chi_L^{\infty}(\w,s)$ are given by ~\eqref{chiRL-hor}. For large enough values of $s$, $\w^2\gg V_{\text{eff}}(s)$. We call this the large-distance limit.  In this limit, $\w^2$  is the dominant term inside the bracket in~\eqref{chi-eq-Sch} and $\chi_R^{\infty}(\w,s)$ and $\chi_L^{\infty}(\w,s)$ are given by ~\eqref{chiRL-def}. For intermediate values of $s$, a numerical computation must be performed to find the integrals in ~\eqref{mat-general-s}. In the following sections, we derive analytic approximations for the contributions to the matching coefficients from the near-horizon and large-distance regions.
\\
\subsection{Near-Horizon Approximation}
In this limit, the leading order terms for $\chi_R^{\infty}(\w,r(s))$ and  $\chi_L^{\infty}(\w,r(s))$ are given in  ~\eqref{chiRL-hor}. Writing   $\chi_R^{\infty}(\w,r(s))$  in terms of $s$ and substituting it into the integral for $\big(A^{H^+}_{\w \w'}\big)_u$ in ~\eqref{mat-general-s}, we find
\bea
(A^{H^+}_{\w\w'})_u&=&-\frac{e^{i\gamma_A}}{\pi}\sqrt{\frac{\w}{\w'}}\frac{E_R}{F_L^*}\int_{0}^{s_1}ds\; e^{i(2\w s-\w's-2\w'\log s)}e^{i\w' (s+2+2\log s-2\log 2)}\nonumber \\ &&
-\frac{e^{i\gamma_A}}{\pi}\sqrt{\frac{\w}{\w'}}\frac{F_R}{E_R}\int_{0}^{s_1}ds\; e^{i(2\w s-\w's-2\w'\log s)}\;e^{-i\w' (s+2+2\log s-2\log 2)}\label{AHhor}
\eea
where $s_1$ is an upper cut-off such that $\w'^2\gg V_{\text{eff}}(s_1)$. Using the fact that $F_L^{*}=E_R$, and integrating the first integral on the right-hand side, ~\eqref{AHhor} simplifies to
\bea
(A^{H^+}_{\w\w'})_u&=&-\frac{e^{i\gamma_A}}{2i\w\pi}\sqrt{\frac{\w}{\w'}}e^{i\w' (2-2\log 2)}\Big(e^{2i\w s_1}-1\Big)\nonumber \\ &&
-\frac{e^{i\gamma_A}}{\pi}\sqrt{\frac{\w}{\w'}}\frac{F_R}{E_R}e^{-i\w' (2-2\log 2)}\int_{0}^{s_1}ds\; e^{2i(\w -\w')s}s^{-4i\w'}.\label{ch5-AHsmall2}
\eea
We next change the variable of integration to $t=-2i(\w-\w')s$ to find
\bea
(A^{H^+}_{\w\w'})_u&=&-\frac{e^{i\gamma_A}}{2i\w\pi}\sqrt{\frac{\w}{\w'}}e^{i\w' (2-2\log 2)}\Big(e^{2i\w s_1}-1\Big)\nonumber \\ &&
-\frac{e^{i\gamma_A}}{\pi}\sqrt{\frac{\w}{\w'}}\frac{F_R}{E_R}e^{-i\w' (2-2\log 2)}\Big(\frac{-1}{2i(\w-\w')}\Big)^{-4i\w'+1}\int_{0}^{t_1}dt\; e^{-t}t^{-4i\w'},\label{AH-hor3}
\eea
where $t_1=-2i(\w-\w')s_1$.  The integral in ~\eqref{AH-hor3} can be written in terms of the lower incomplete gamma function as follows
\bea
(A^{H^+}_{\w\w'})_u&=&-\frac{e^{i\gamma_A}}{2i\w\pi}\sqrt{\frac{\w}{\w'}}e^{i\w' (2-2\log 2)}\Big(e^{2i\w s_1}-1\Big)\nonumber \\ &&
-\frac{e^{i\gamma_A}}{\pi}\sqrt{\frac{\w}{\w'}}\frac{F_R}{E_R}e^{-i\w' (2-2\log 2)}\Big(\frac{-1}{2i(\w-\w')}\Big)^{-4i\w'+1}\gamma\Big(-4i\w'+1,t_1\Big)\label{ch5-AHsamll}
\eea
One might think that ~\eqref{ch5-AHsamll} is not well-defined at $\w'=\w$. However, we show here that in the limit $\w'\to \w$, the second term in ~\eqref{ch5-AHsamll} is equal to the answer we get by setting $\w'=\w$ in ~\eqref{ch5-AHsmall2} and directly calculating the integral. In the limit $\w'\to \w$, the lower incomplete gamma function can be written as
\bea
\gamma(\alpha,x)=\underset{k=0}{\overset{\infty}{\sum}}\frac{(-1)^k x^{k+\alpha}}{k!(\alpha+k)}.\label{ch5-lower-gamma}
\eea
The series in ~\eqref{ch5-lower-gamma} converges for all complex $\alpha$ and $x$. Keeping only the first term in ~\eqref{ch5-lower-gamma}, and using that to write the second term in ~\eqref{ch5-AHsamll} for small values of $\w-\w'$, we find
\bea
-\frac{e^{i\gamma_A}}{\pi}\sqrt{\frac{\w}{\w'}}\frac{F_R}{E_R}e^{-i\w' (2-2\log 2)}\Big(\frac{s_1^{1-4i\w'}}{1-4i\w'}\Big)
\eea
which is equal to the second term in ~\eqref{ch5-AHsmall2} when we set $\w=\w'$.
We can similarly compute the small $s$ contribution to $(B^{H^+}_{\w\w'})_u$ with the result shown below
\bea
(B^{H^+}_{\w\w'})_u&=&\frac{e^{i\gamma_B}}{2i\w\pi}\sqrt{\frac{\w}{\w'}}e^{-i\w' (2-2\log 2)}\Big(e^{2i\w s_1}-1\Big)\nonumber \\ &&
+\frac{e^{i\gamma_B}}{\pi}\sqrt{\frac{\w}{\w'}}\frac{F_R^*}{E_R^*}e^{i\w' (2-2\log 2)}\Big(\frac{-1}{2i(\w+\w')}\Big)^{4i\w'+1}\gamma\Big(4i\w'+1,t_2\Big)
\eea
where $t_2=-2i(\w+\w')s_1$.\\
The near-horizon contributions to $(A^{\mathscr{I}^+}_{\w\w'})_u$ and $(B^{\mathscr{I}^+}_{\w\w'})_u$ can be derived in a similar way with the results
\bes \bea
(A^{\mathscr{I}^+}_{\w\w'})_u&=&
-\frac{\w'e^{-i\w'(2-\log 2)}}{\pi\sqrt{\w\w'}}\frac{e^{i\gamma_A}}{(2i(\w'-\w)+\epsilon)^{-4i\w'}}\gamma(-4i\w'+\delta,t_1)\nonumber \\ &&
-\frac{\w'e^{-i\w'(2-\log 2)}}{\pi\sqrt{\w\w'}}\frac{e^{i\gamma_A}}{(2i(\w'-\w)+\epsilon)^{-4i\w'+1}}\gamma(-4i\w'+1+\delta,t_1),\\
(B^{\mathscr{I}^+}_{\w\w'})_u&=&
+\frac{\w'e^{i\w'(2-\log 2)}}{\pi\sqrt{\w\w'}}\frac{e^{i\gamma_B}}{(-2i(\w'+\w)+\epsilon)^{4i\w'}}\gamma(4i\w'+\delta,t_2)\nonumber \\ &&
+\frac{\w'e^{i\w'(2-\log 2)}}{\pi\sqrt{\w\w'}}\frac{e^{i\gamma_B}}{(-2i(\w'+\w)+\epsilon)^{4i\w'+1}}\gamma(4i\w'+1+\delta,t_2).
\eea \ees
\subsection{Large-Distance Approximation}
For the values of $s_2\gg 1$ with $\w'^2\gg V_{\text{eff}}(s_2)$, $\chi_R^{\infty}$ and $\chi_L^{\infty}$ are given in ~\eqref{chiRL-def}. Substituting these expressions into $(A^{H^+}_{\w\w'})_u$ and $(B^{H^+}_{\w\w'})_u$ in ~\eqref{mat-general-s}, we find
\bes\bea
\big(A^{H^+}_{\w \w'}\big)_u&=& -\frac{e^{4i\w'}}{\pi}\sqrt{\frac{\w}{\w'}}\frac{1}{F_L^*(\w')}\int_{s_2}^{\infty}ds\;e^{2i\w s-\epsilon s},\\
\big(B^{H^+}_{\w \w'}\big)_u&=& \frac{e^{-4i\w'}}{\pi}\sqrt{\frac{\w}{\w'}}\frac{1}{F_L(\w')}\int_{s_2}^{\infty}ds\;e^{2i\w s-\epsilon s}.
\eea \ees
Note that an integrating factor $0< \epsilon \ll 1$ has been introduced in these integrals to make them converge. Then, $\big(A^{H^+}_{\w \w'}\big)_u$ and $\big(B^{H^+}_{\w \w'}\big)_u$ become 
\bes \bea
\big(A^{H^+}_{\w \w'}\big)_u&=& \frac{e^{4i\w'}}{\pi}\sqrt{\frac{\w}{\w'}}\frac{1}{F_L^*(\w')}\frac{e^{2i\w s}}{2i\w-\epsilon},\\
\big(B^{H^+}_{\w \w'}\big)_u&=& \frac{e^{-4i\w'}}{\pi}\sqrt{\frac{\w}{\w'}}\frac{1}{F_L(\w')}\frac{e^{2i\w s}}{2i\w-\epsilon}.
\eea \ees
To find the large $s$ contribution to $\big(A^{\mathscr{I}^+}_{\w \w'}\big)_u$, we substitute ~\eqref{chiRL-def} into the third integral in ~\eqref{mat-general-s} and find
\bea
(A_{\w\w'0}^{\mathscr{I}^+})_u=-\frac{\w'e^{-i\w'(2-\log 2)}}{\pi\sqrt{\w\w'}}e^{i\gamma_A}\int_{0}^{\infty}ds \; e^{2i(\w-\w')s-\epsilon s}\Big(s^{-4i\w'-1}+s^{-4i\w'}\Big),\label{ch5-AILarges}
\eea
where again an integrating factor $0<\epsilon\ll 1$ has been introduced to make the integral converge at large $s$. Changing variables to $t=2i(\w'-\w)s+\epsilon s $, we find
\bea
(A_{\w\w'0}^{\mathscr{I}^+})_u &=&-\frac{\w'e^{-i\w'(2-\log 2)}}{\pi\sqrt{\w\w'}}e^{i\gamma_A}\int_{t_3}^{\infty}\frac{dt}{2i(\w'-\w)+\epsilon} \; e^{-t}\big(\frac{t}{2i(\w'-\w)+\epsilon}\big)^{-4i\w'-1}\nonumber \\ &&
-\frac{\w'e^{-i\w'(2-\log 2)}}{\pi\sqrt{\w\w'}}e^{i\gamma_A}\int_{t_3}^{\infty}\frac{dt}{2i(\w'-\w)+\epsilon} \; e^{-t}\big(\frac{t}{2i(\w'-\w)+\epsilon}\big)^{-4i\w'}\nonumber \\ &&
=-\frac{\w'e^{-i\w'(2-\log 2)}}{\pi\sqrt{\w\w'}}\frac{e^{i\gamma_A}}{(2i(\w'-\w)+\epsilon)^{-4i\w'}}\int_{t_3}^{\infty}dt \; e^{-t}t^{-4i\w'-1}\nonumber \\ &&
-\frac{\w'e^{-i\w'(2-\log 2)}}{\pi\sqrt{\w\w'}}\frac{e^{i\gamma_A}}{(2i(\w'-\w)+\epsilon)^{-4i\w'+1}}\int_{t_3}^{\infty}dt \; e^{-t}t^{-4i\w'}.\label{AIlargeS}
\eea
where $t_3=2i(\w'-\w)s_2$.

The integrals in ~\eqref{AIlargeS} can be written in terms of upper incomplete gamma functions as follows
 \bea
(A_{\w\w'0}^{\mathscr{I}^+})_u&=&
-\frac{\w'e^{-i\w'(2-\log 2)}}{\pi\sqrt{\w\w'}}\frac{e^{i\gamma_A}}{(2i(\w'-\w)+\epsilon)^{-4i\w'}}\Gamma(-4i\w',t_3)\nonumber \\ &&
-\frac{\w'e^{-i\w'(2-\log 2)}}{\pi\sqrt{\w\w'}}\frac{e^{i\gamma_A}}{(2i(\w'-\w)+\epsilon)^{-4i\w'+1}}\Gamma(-4i\w'+1,t_3).
\eea

The matching coefficient $(B_{\w\w'0}^{\mathscr{I}^+})_u$ is found in a similar way with the result
 \bea
(B_{\w\w'0}^{\mathscr{I}^+})_u&=&
\frac{\w'e^{i\w'(2-\log 2)}}{\pi\sqrt{\w\w'}}\frac{e^{i\gamma_B}}{(-2i(\w'+\w)+\epsilon)^{4i\w'}}\Gamma(4i\w',t_4)\nonumber \\ &&
+\frac{\w'e^{i\w'(2-\log 2)}}{\pi\sqrt{\w\w'}}\frac{e^{i\gamma_B}}{(-2i(\w'+\w)+\epsilon)^{4i\w'+1}}\Gamma(4i\w'+1,t_4),
\eea
where $t_4=-2i(\w'+\w)s_2+\epsilon s_2$.

In the next sections, we use the results obtained in this chapter so far to investigate the behaviors of the $in$ modes in several limits. We first check the validity of our result by looking at the modes as the null coordinate $v$ approaches $v_0$, which is the trajectory of the null shell. In this way, one can check the continuity of the $in$ modes at the null-shell surface.
We next look at the $in$ modes with large frequencies. Finally, we investigate the late-time behaviors of the $in$ modes.
\section{Continuity of  $f^{in}_{\w}$ at $v=v_0$}
We numerically computed $\big(f^{in}_{\w}\big)_v$ and $\big(f^{in}_{\w}\big)_u$ for various frequencies outside but near the null shell for a fixed radial coordinate $r$ and as a function of time $t_s$. The computation has been done for modes with different frequencies. The numerical results presented in Fig.~{fig:fv-continuity} and Fig.~{fig:fu-continuity} show that the $f^{in}_{\w}=\big(f^{in}_{\w}\big)_v+\big(f^{in}_{\w}\big)_u$ modes are continuous across the null-shell surface, as expected.
\section{Large Frequency Limit of $f^{in}_{\w}$}
The $f^{in}_{\w}$ modes with large $\w$ are computed in this section. An expression for $\big(f^{in}_{\w}\big)_v$ is given in ~\eqref{ch5-fin-v-second-2}. However, for $\big(f^{in}_{\w}\big)_u$ in ~\eqref{ch4-fin-upart}, a full numerical calculation is needed. The numerical results depicted in Fig.~{fig:fuv-large-w}, show that for a fixed spacetime point, at large enough frequency, $\w^2\gg \big|V(r)\big|_{\text{max}}$, where $V(r)$ is the effective potential in ~\eqref{chi-eq-Sch}, the $\big(f^{in}_{\w}\big)_v$ and $\big(f^{in}_{\w}\big)_u$ parts of the modes approach their no-scattering counter-terms $\big(V(r)=0\big)$, which are  $\big(f^{in}_{\w}\big)_v=Y_{00}\frac{e^{-i\w v}}{r\sqrt{4\pi \w}}$ and $\big(f^{in}_{\w}\big)_v=-Y_{00}\frac{e^{-i\w u(u_s)}}{r\sqrt{4\pi \w}}$ respectively.
\newpage
\begin{figure}[h]
\centering
\includegraphics[trim=0cm 0cm 0cm 0cm,clip=true,totalheight=0.23\textheight]{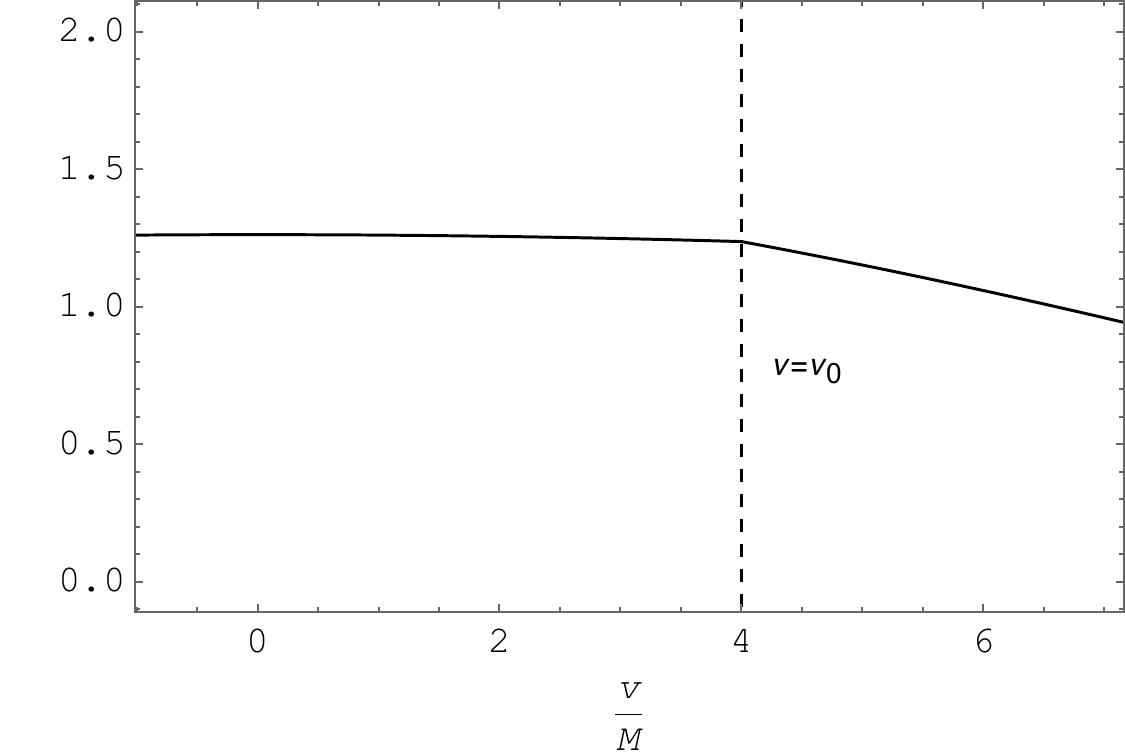}
\includegraphics[trim=0cm 0cm 0cm 0cm,clip=true,totalheight=0.23\textheight]{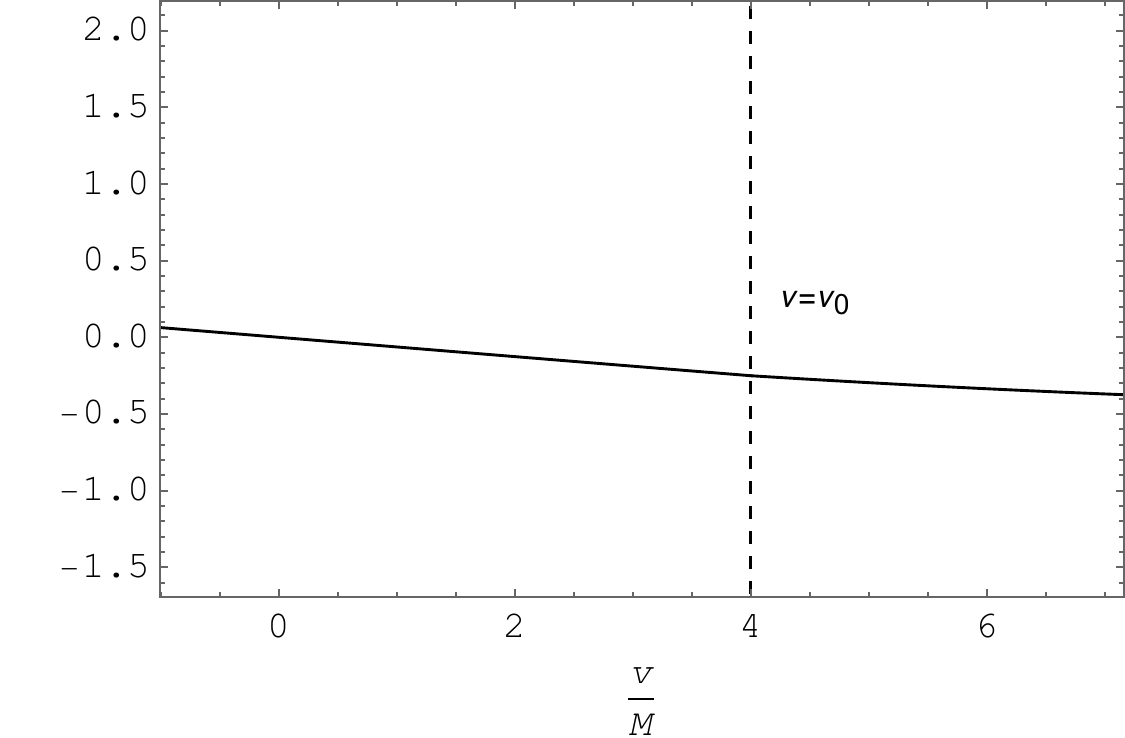}\\[10pt]
\includegraphics[trim=0cm 0cm 0cm 0cm,clip=true,totalheight=0.23\textheight]{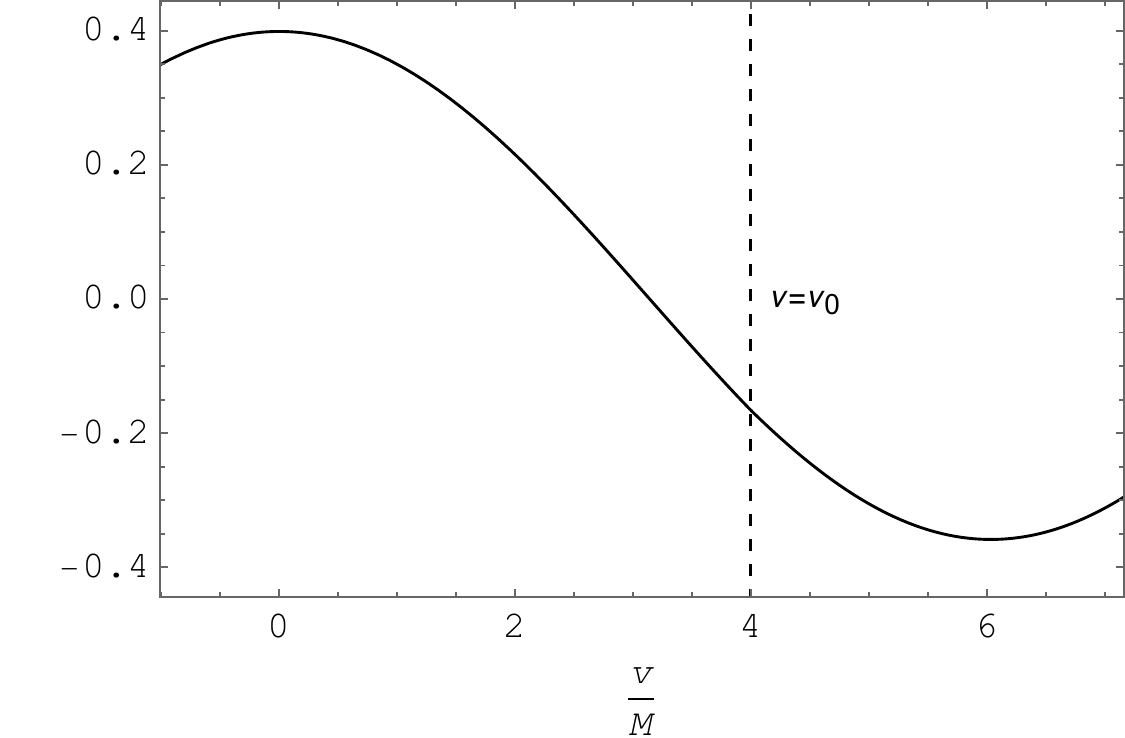}
\includegraphics[trim=0cm 0cm 0cm 0cm,clip=true,totalheight=0.23\textheight]{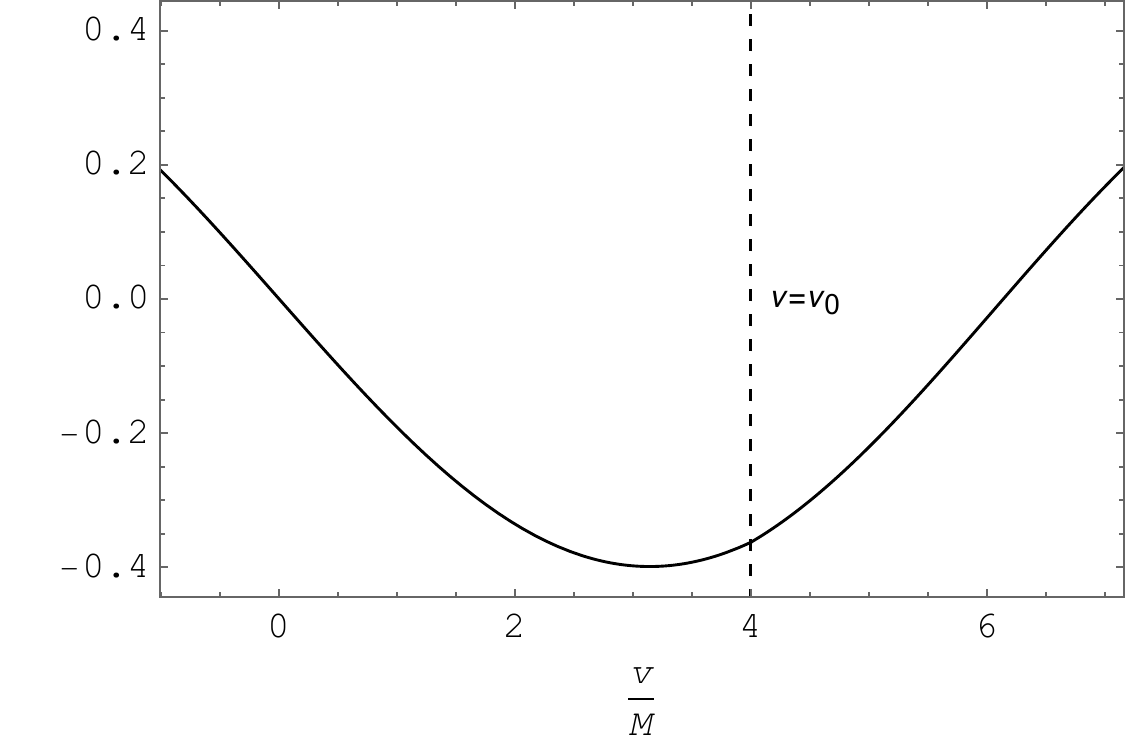}\\[10pt]
\includegraphics[trim=0cm 0cm 0cm 0cm,clip=true,totalheight=0.23\textheight]{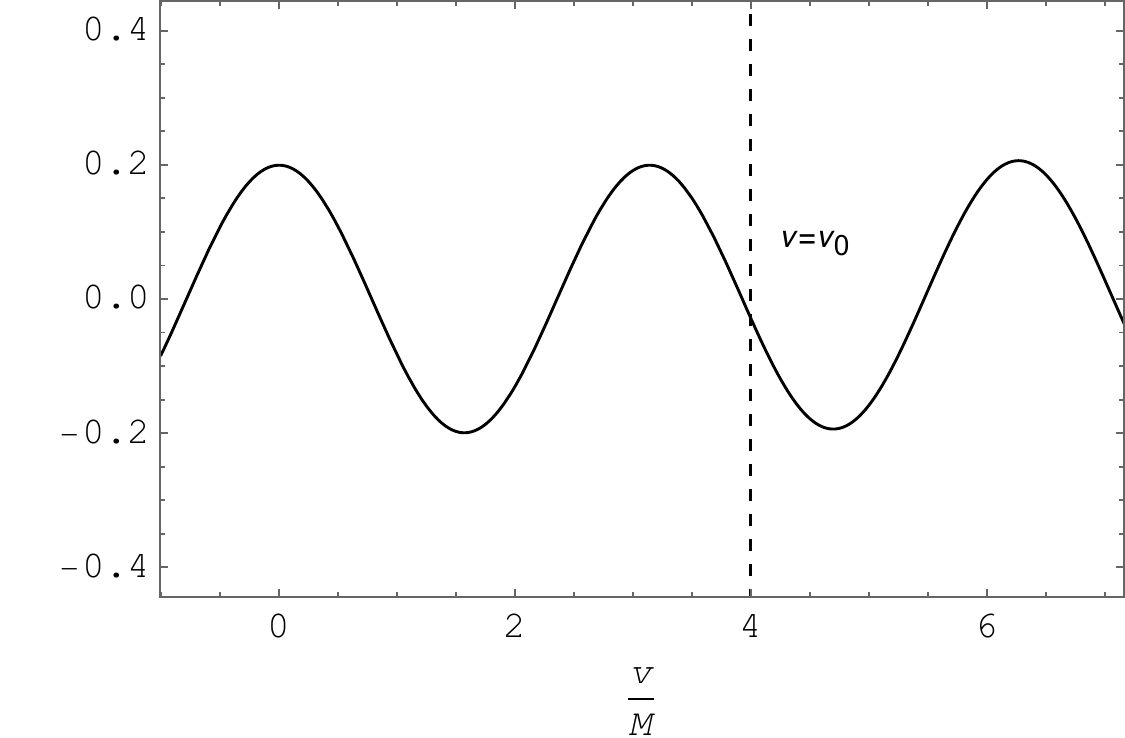}
\includegraphics[trim=0cm 0cm 0cm 0cm,clip=true,totalheight=0.23\textheight]{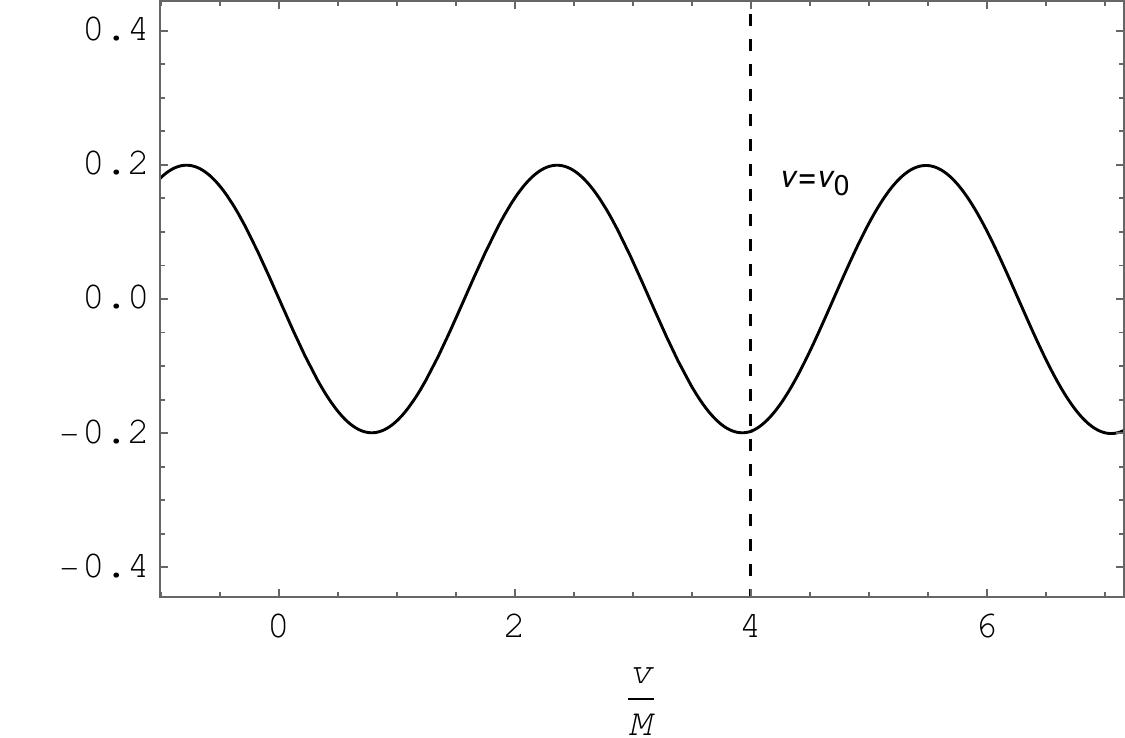}\\[10pt]

\label{fig:fv-continuity}
\end{figure}
\newpage
\begin{figure}[h]
\centering
\includegraphics[trim=0cm 0cm 0cm 0cm,clip=true,totalheight=0.23\textheight]{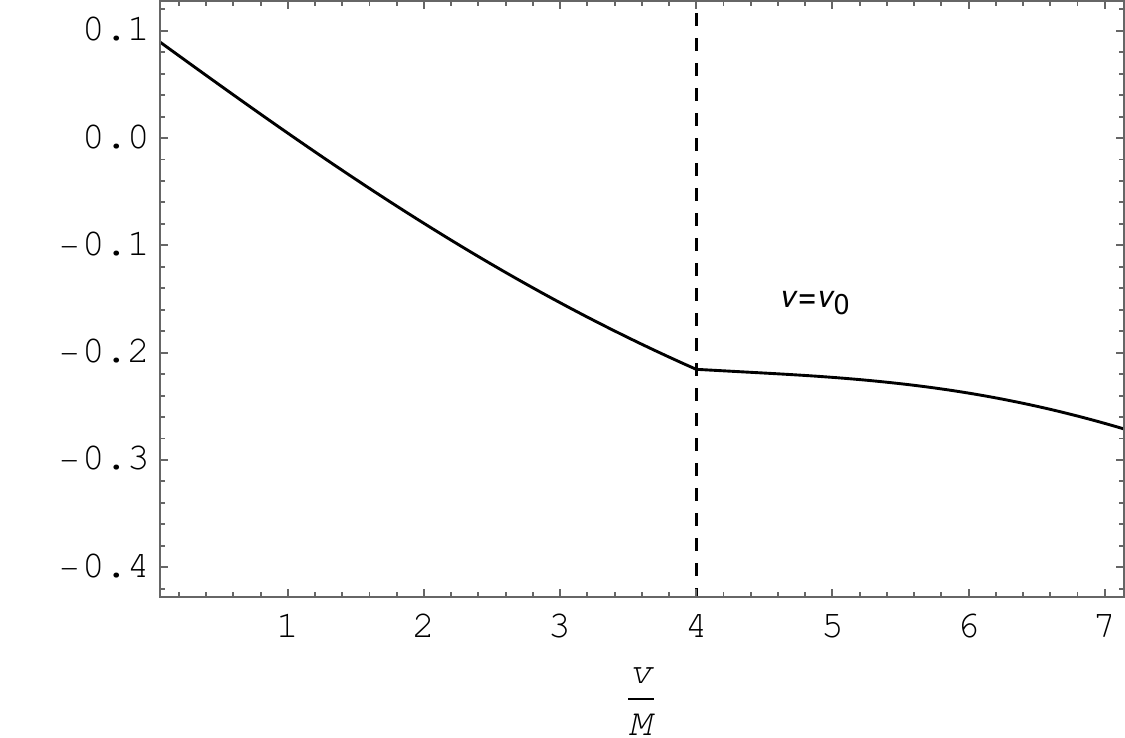}
\includegraphics[trim=0cm 0cm 0cm 0cm,clip=true,totalheight=0.23\textheight]{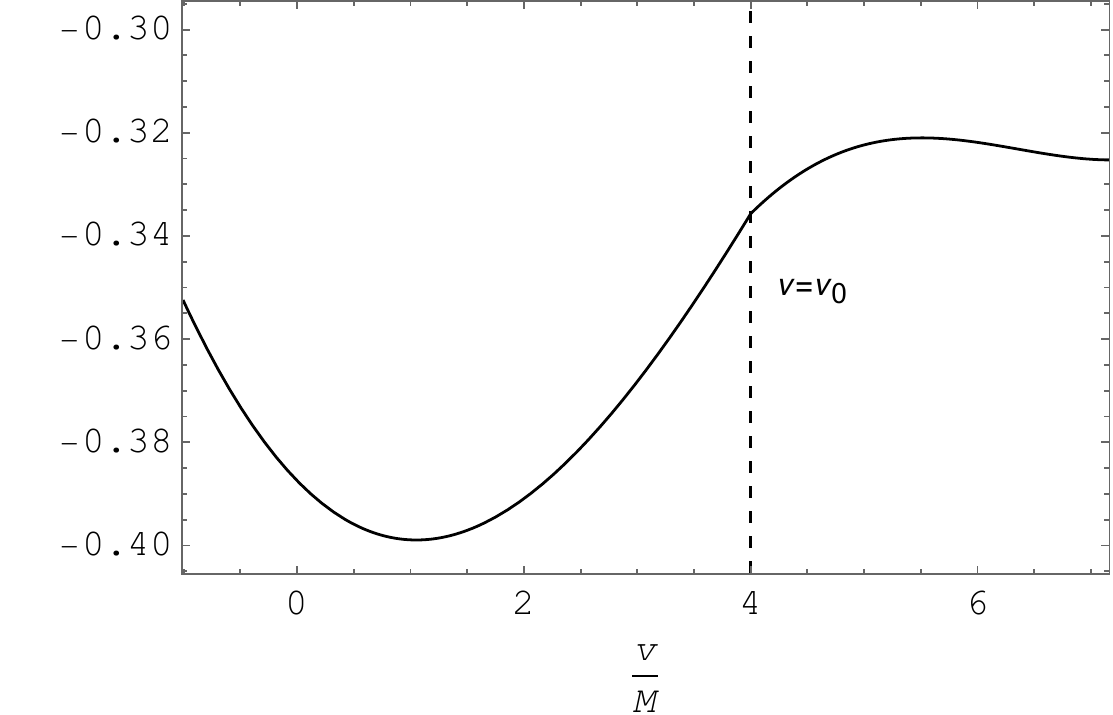}\\[10pt]
\includegraphics[trim=0cm 0cm 0cm 0cm,clip=true,totalheight=0.23\textheight]{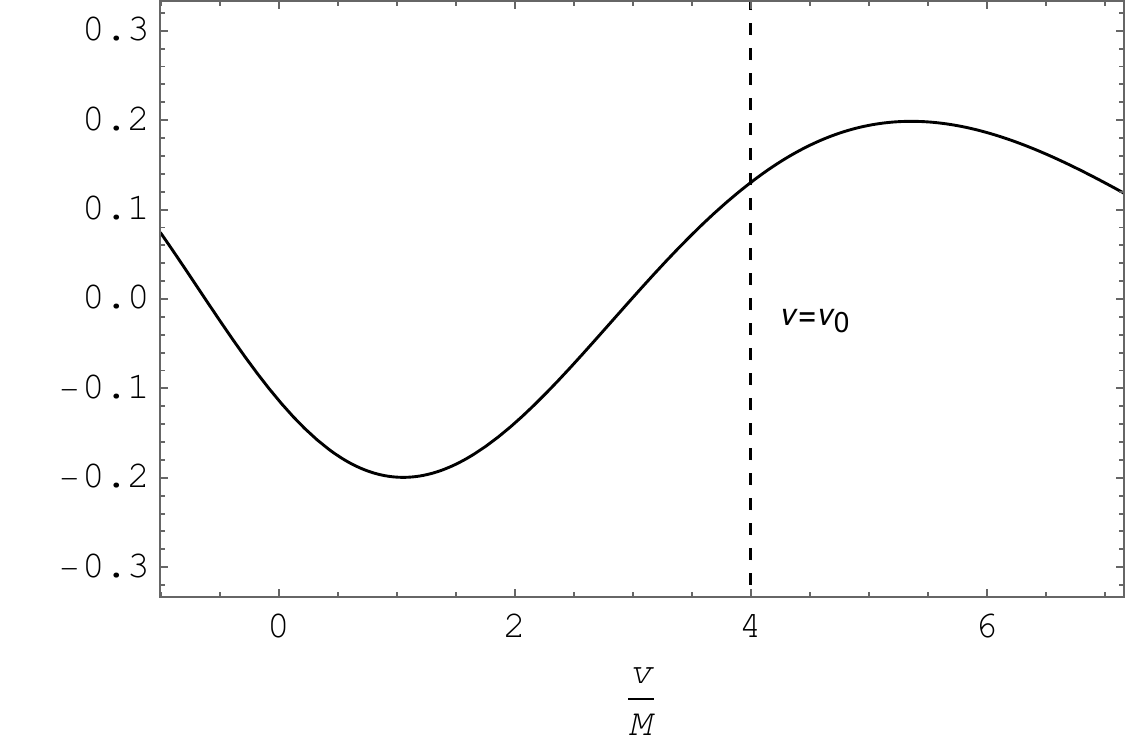}
\includegraphics[trim=0cm 0cm 0cm 0cm,clip=true,totalheight=0.23\textheight]{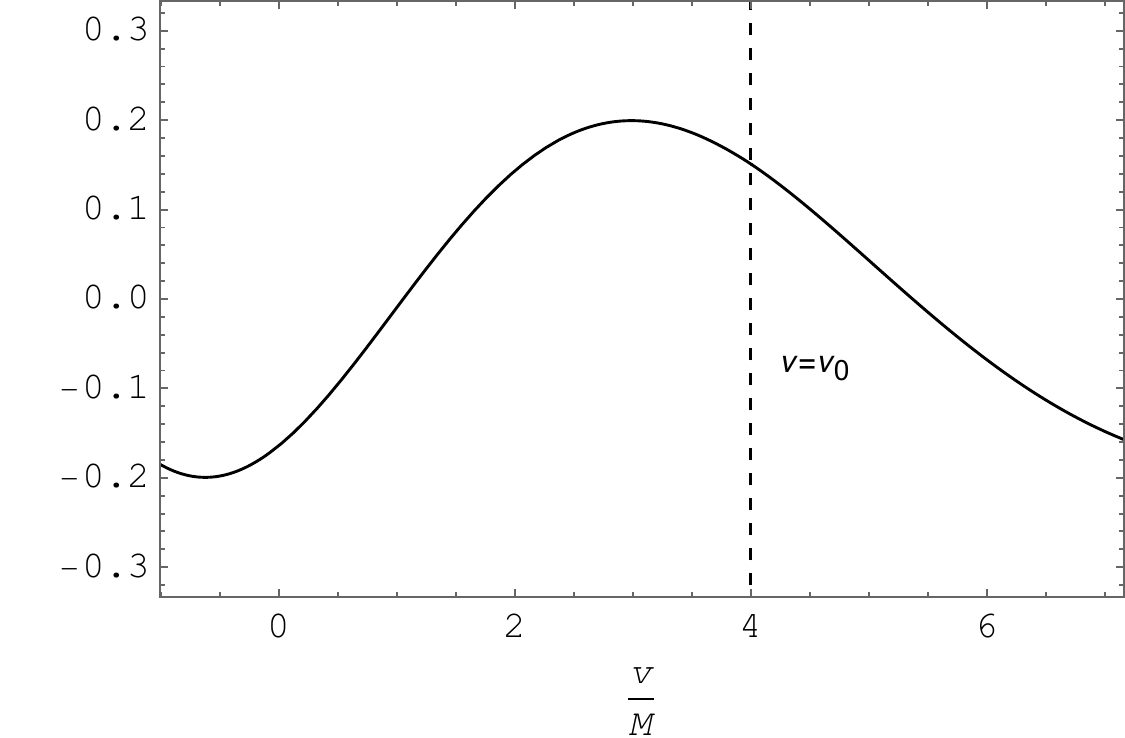}\\[10pt]

\caption{The real (left plot) and the imaginary (right plot) parts of the complex quantity $\sqrt{\frac{4\pi}{M}}(f^{in}_{\w})_u$ are plotted as a function of $v$ for  $M\w =0.5$ (top plots) and $M\w =2$ (bottom plots). The curves on the right-hand side of $v=v_0$ (outside the null shell), have been computed numerically.   }
\label{fig:fu-continuity}
\end{figure}
\newpage
\begin{figure}[h]
\centering
\includegraphics[trim=1cm 0cm 0cm 0cm,clip=true,totalheight=0.23\textheight]{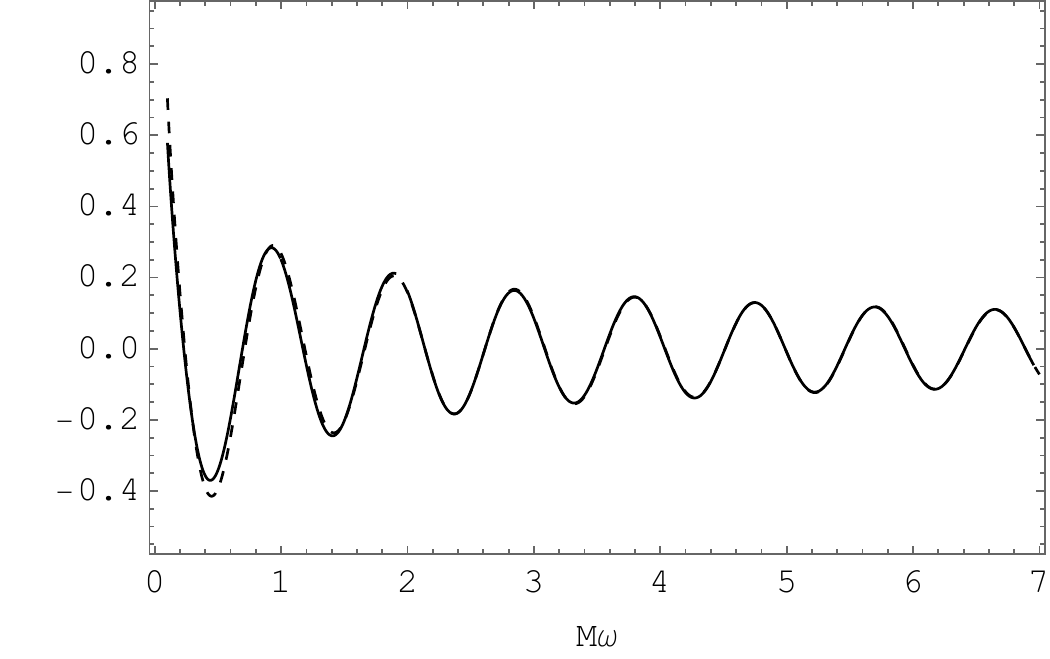}
\includegraphics[trim=0cm 0cm 0cm 0cm,clip=true,totalheight=0.23\textheight]{fvReal.pdf}\\[10pt]
\includegraphics[trim=1cm 0cm 0cm 0cm,clip=true,totalheight=0.23\textheight]{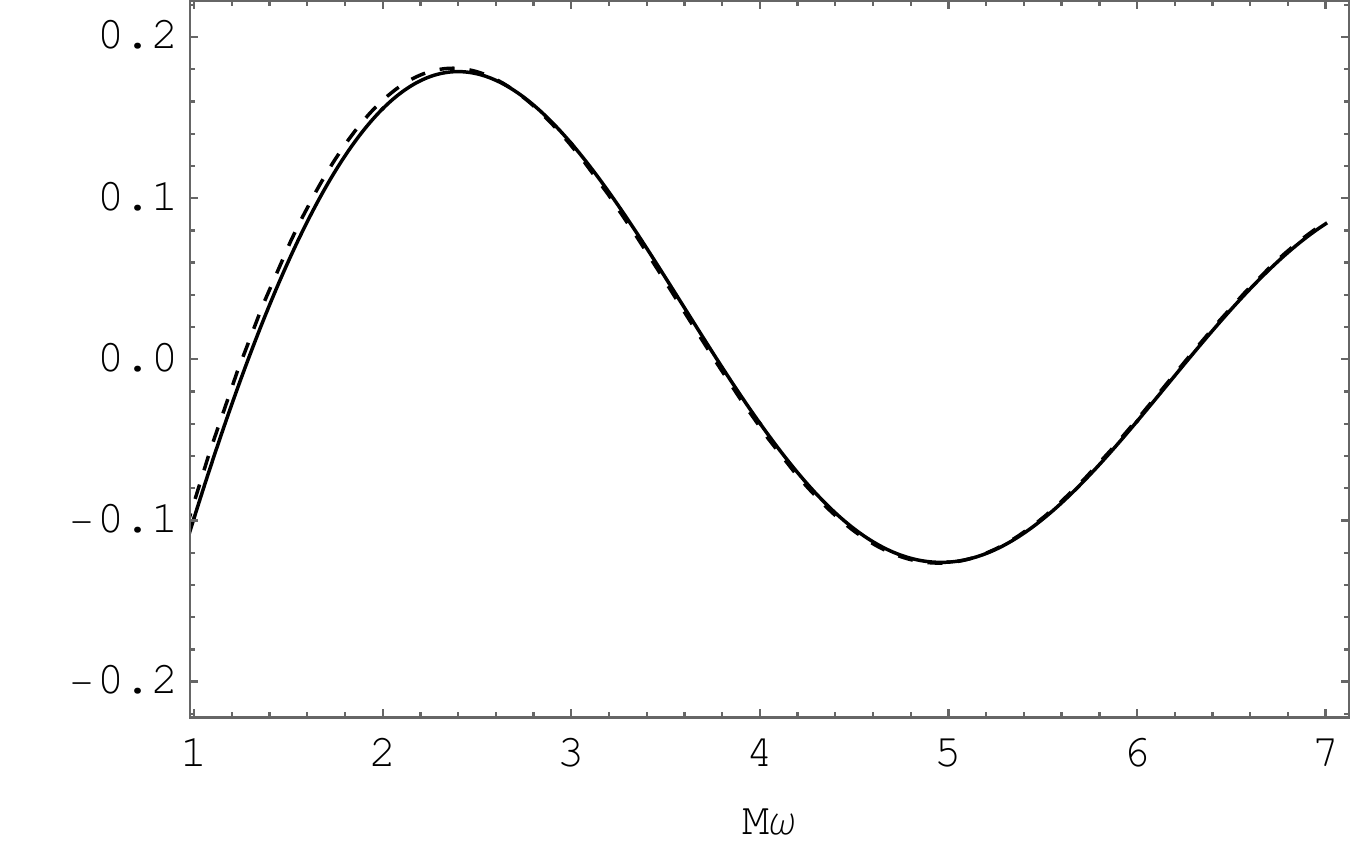}
\includegraphics[trim=0cm 0cm 0cm 0cm,clip=true,totalheight=0.23\textheight]{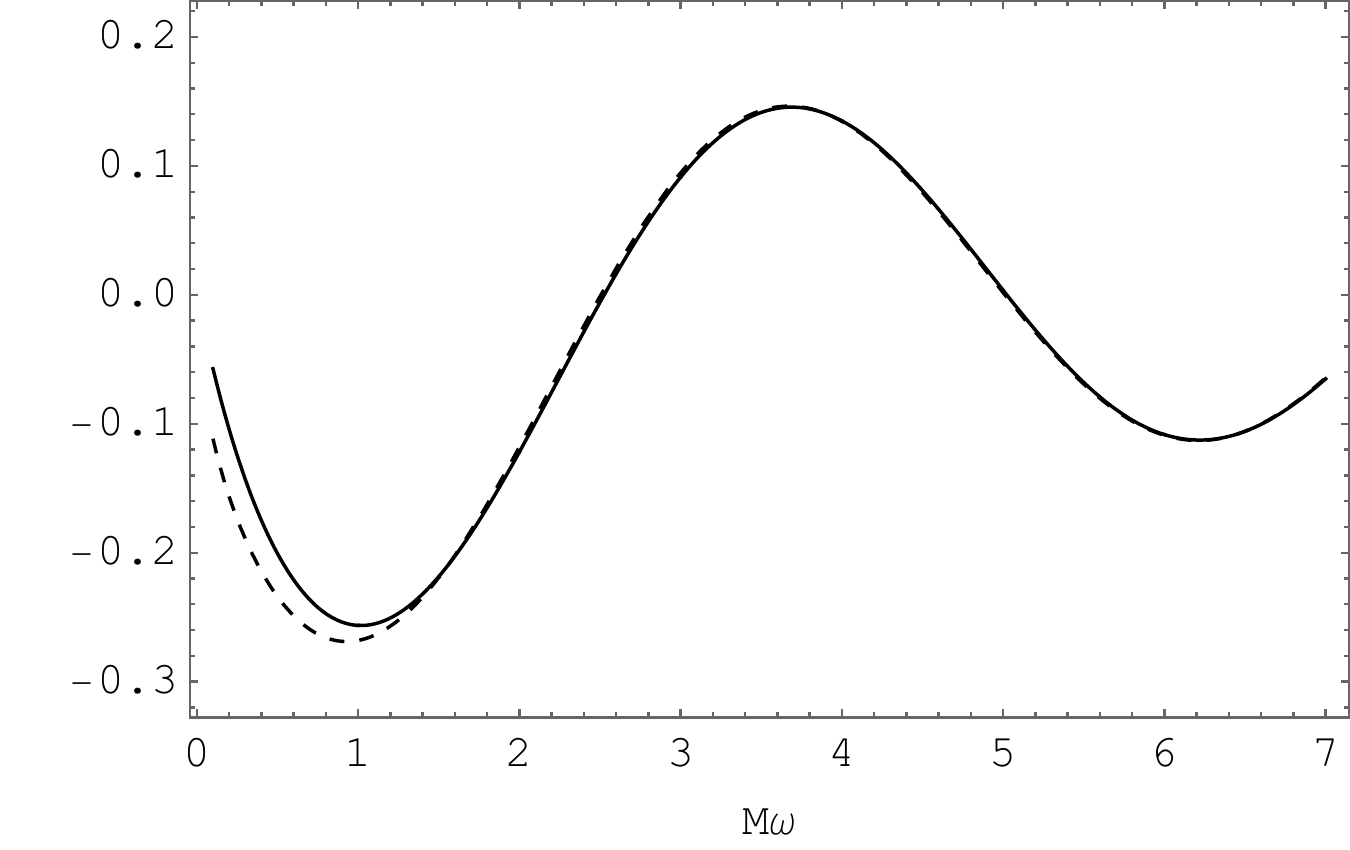}\\[10pt]
\caption{The solid curves in the top plots are the real (left) and the imaginary (right) parts of the complex quantity $\sqrt{\frac{4\pi}{M}}(f^{in}_{\w})_v$  as a function of $\w$ for $\frac{t_s}{M}=5$ and $\frac{r}{M}=3$. The solid curves in the bottom plots represent the real (left) and the imaginary (right) parts of $\sqrt{\frac{4\pi}{M}}(f^{in}_{\w})_u$  as a function of $\w$ for $\frac{t_s}{M}=5$ and $\frac{r}{M}=3$. The dashed curves in the plots depict the corresponding no-scattering terms. }
\label{fig:fuv-large-w}
\end{figure}
\newpage
\section{Late-time Behaviors of Modes:}
We next investigate the late-time behaviors of the $(f^{in}_{\w})_v$, $(f^{in}_{\w})_u$, $f^{\mathscr{I}^-}_{\w}$, and $f^K_{\w}$ modes. In ~\eqref{ch5-fin-v-second-2}, we showed an explicit expression for $(f^{in}_{\w}\big)_v$. One can use  ~\eqref{f-def-1}, \eqref{chi-def-2}, and ~\eqref{chi-scri-minus-def} to find the following expression for the $f^{\mathscr{I}^-}_{\w}$ modes with $\ell=0$,
\bea
f^{\mathscr{I}^-}_{\w}= \frac{Y_{00}}{r\sqrt{4\pi\w}}\Big\{\chi_L^{\infty}(r,t)-\frac{F_R^{*}(\w)}{F_L^*(\w)}\chi_R^{\infty}\Big\}e^{-i\w t_s}.\label{ch5-scri-minus}
\eea
One can see that the first terms in ~\eqref{ch5-fin-v-second-2} and ~\eqref{ch5-scri-minus} are equal. Furthermore, the amplitude of the contribution of the two integrals in ~\eqref{ch5-fin-v-second-2} decreases in time, but it approaches a constant value that is equal to the amplitude of the second term in ~\eqref{ch5-scri-minus}.
Therefore, $\big(f^{in}_{\w}\big)_v$ approaches $f^{\mathscr{I}^-}_{\w}$ at late times. Our numerical results (bottom plots in Fig.\;~{fig:fvdec} show that $\big(f^{in}_{\w}\big)_v-f^{\mathscr{I}^-}_{\w}$ decays to zero as power-law. The rate of the decay, to a good approximation, is $t_s^{-3}$.

To find $\big(f^{in}_{\w}\big)_u$, ~\eqref{ch4-fin-upart} has to be evaluated numerically. This task requires a numerical computation for the matching coefficients and the Boulware modes. The numerical computation of the matching coefficients is discussed in Sections 5.3 and 5.4. Using those results, we computed  $\big(f^{in}_{\w}\big)_u$ as a function of the time coordinate $t_s$ for fixed $\w$ and the spatial coordinate $r$. The results are depicted in Fig.\;{fig:fvdec}. It can be seen that at late times, $\big(f^{in}_{\w}\big)_u$ decays to zero. In Chapter 3, we found that the scattering due to a delta-function potential removes the infrared divergences in the Boulware modes. This probably causes the Kruskal modes $f^K_{\w}$ to decay at late times. In the 4D Schwarzschild spacetime, work is in progress to find the behaviors of these modes at late times. We also studied the late-time behavior of the Kruskal modes $f^K_{\w}$. We found that in the late-time limit, these modes decay as $t_s^{-3}$. This has been depicted in Fig.\;{fig:fK}.

In Fig.\;{fig:fvdec}, one can see that the real and imaginary parts of $\big\{(f^{in}_{\w})_v-f^{\mathscr{I}^-}_{\w}\big\}$ first undergo a series of damped oscillations and after that the decay is a power-law. A similar type of late-time behavior can be seen for the Kruskal modes $f^{K}_{\w}$ in Fig.\;{fig:fK}. We found the decay rate of $t_s^{-3}$ for both  $\big\{(f^{in}_{\w})_v-f^{\mathscr{I}^-}_{\w}\big\}$  and $f^{K}_{\w}$. This late-time behavior is similar to the late-time behavior found for the classical massless minimally-coupled scalar field in ~\cite{gcpp, burko-ori, barack}.

Our results also show that before the power-law tail becomes dominant, there is a period of damped oscillations. This is similar to the behavior of gravitational radiation emitted by a black hole after the "merger" phase. In this case, the exponential damping of the perturbation is due to the quasinormal modes of the black hole with complex frequencies. A study of quasinormal modes and the power-law tail was first done in \cite{Leaver}.
\newpage
\begin{figure}[h]
\centering
\includegraphics[trim=0cm 0cm 0cm 0cm,clip=true,totalheight=0.23\textheight]{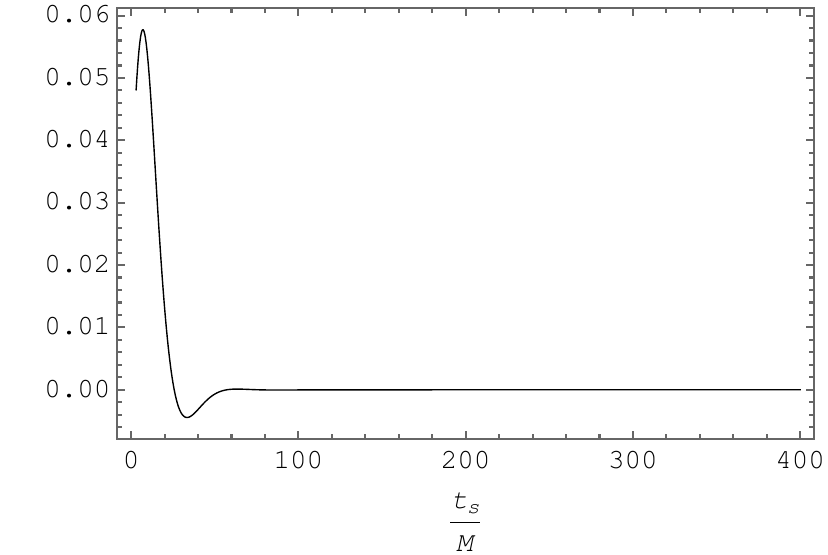}
\includegraphics[trim=0cm 0cm 0cm 0cm,clip=true,totalheight=0.23\textheight]{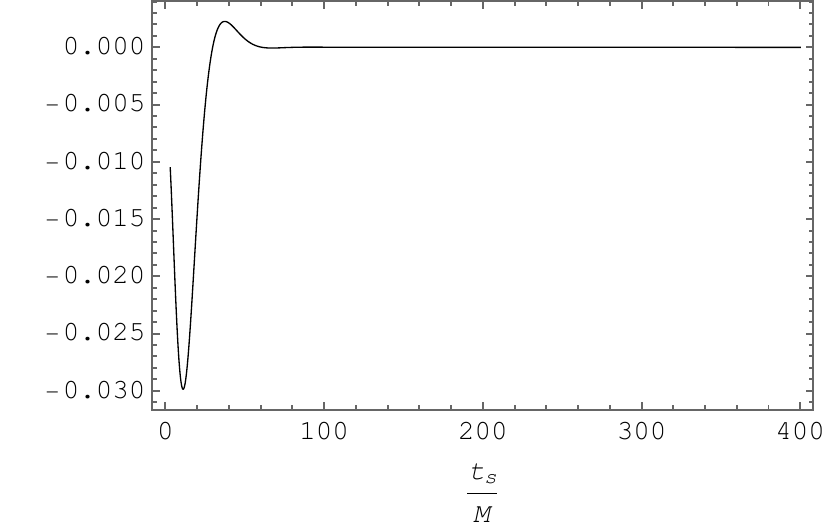}\\[10pt]
\includegraphics[trim=0cm 0cm 0cm 0cm,clip=true,totalheight=0.23\textheight]{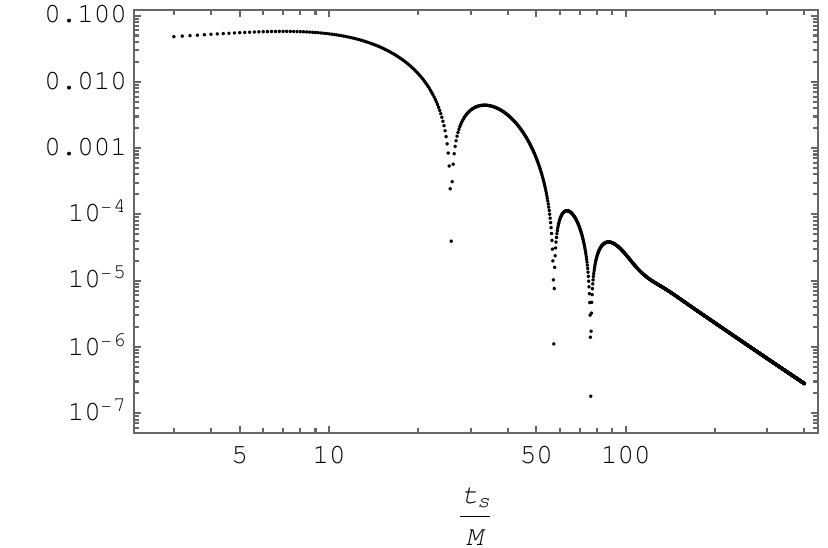}
\includegraphics[trim=0cm 0cm 0cm 0cm,clip=true,totalheight=0.23\textheight]{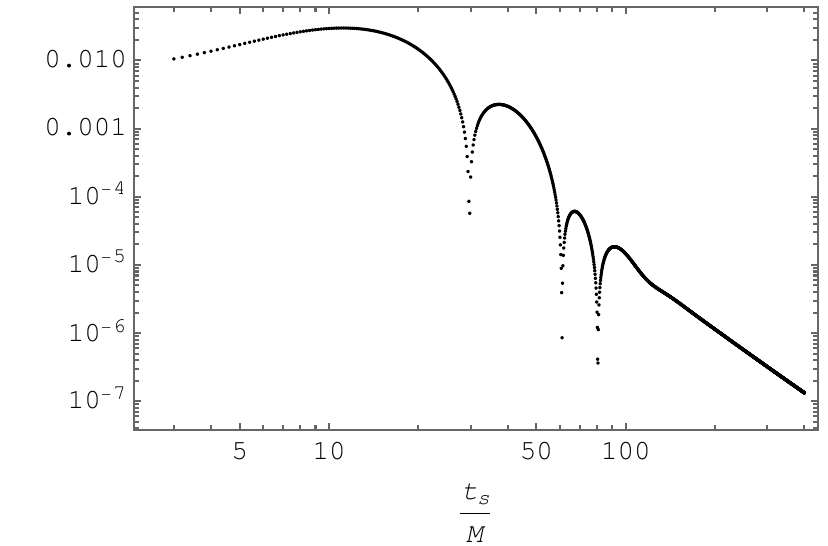}\\[10pt]
\caption{The top figures represent the real (left) and the imaginary (right) parts of the complex quantity $\sqrt{\frac{4\pi}{M}}\big\{(f^{in}_{\w})_v-f^{\mathscr{I}^-}_{\w}\big\}$ as a function of time for  $M\w =0.5$ and for $r=3M$. The two bottom figures are the log-log plots of the absolute values of the real (left) and imaginary (right) parts of $\sqrt{\frac{4\pi}{M}}\big\{(f^{in}_{\w})_v-f^{\mathscr{I}^-}_{\w}\big\}$. The bottom figures show that at late time $\big(f^{in}_{\w}\big)_v$ approaches $f^{\mathscr{I}^-}_{\w}$ as a power-law.}
\label{fig:fvdec}
\end{figure}
\newpage
\begin{figure}[h]
\centering
\includegraphics[trim=0cm 0cm 0cm 0cm,clip=true,totalheight=0.23\textheight]{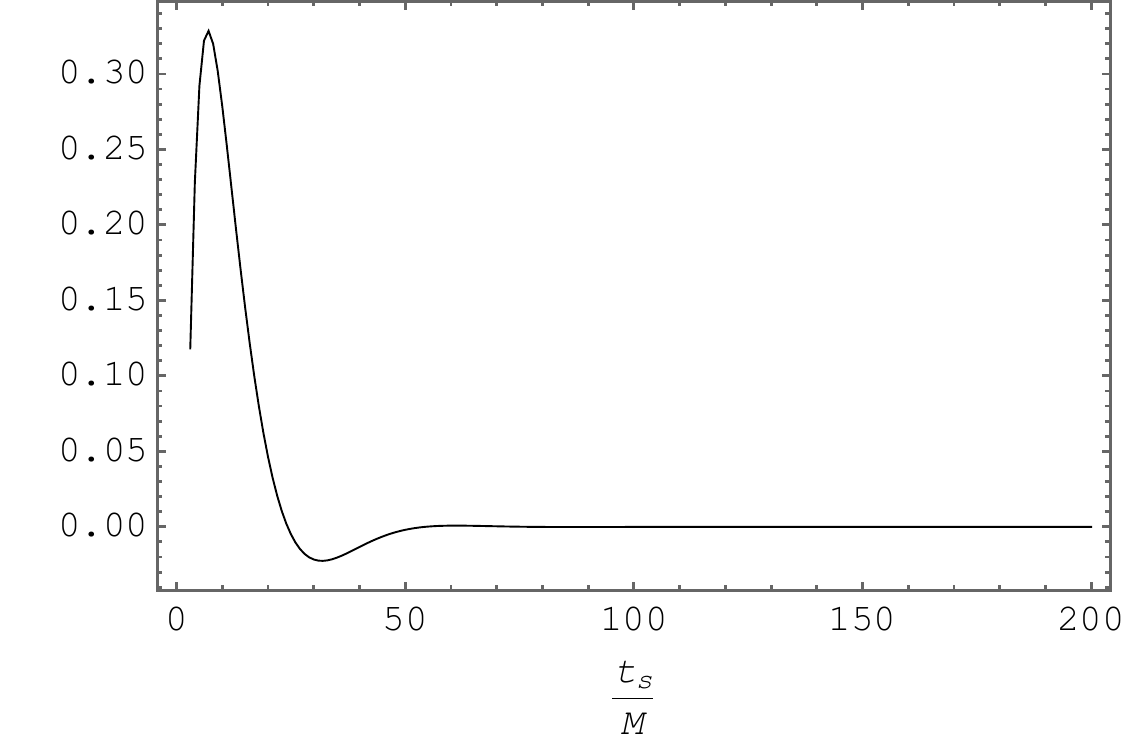}
\includegraphics[trim=0cm 0cm 0cm 0cm,clip=true,totalheight=0.23\textheight]{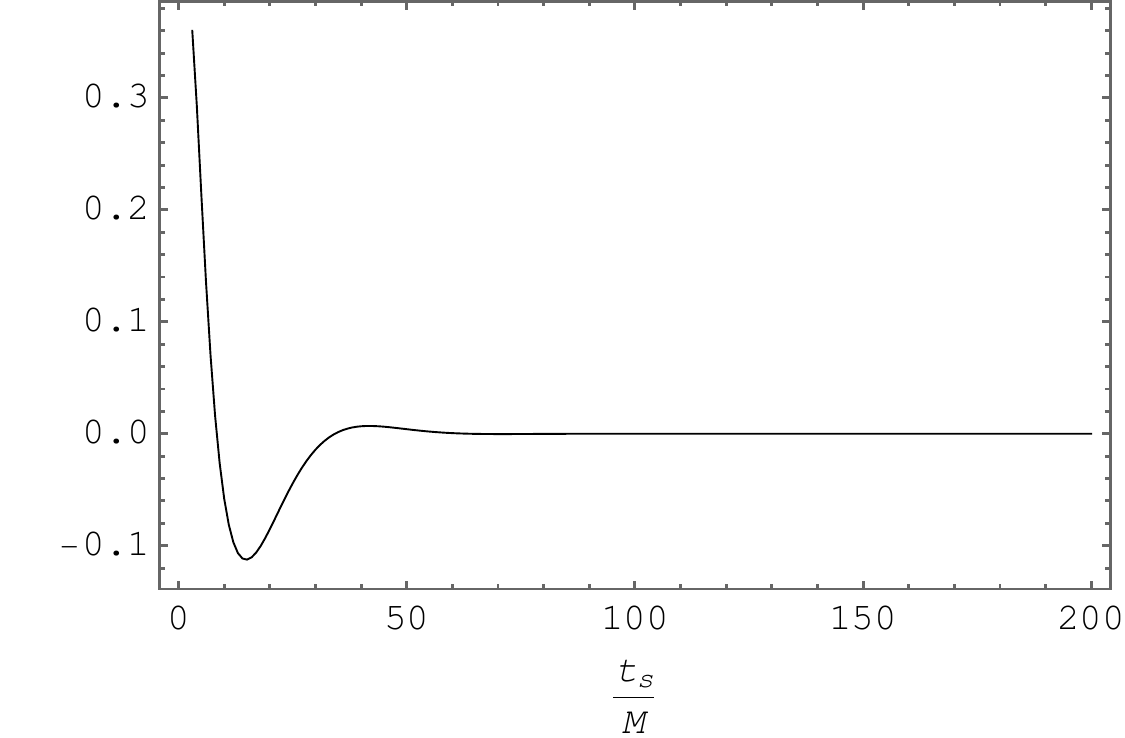}\\[10pt]
\includegraphics[trim=0cm 0cm 0cm 0cm,clip=true,totalheight=0.23\textheight]{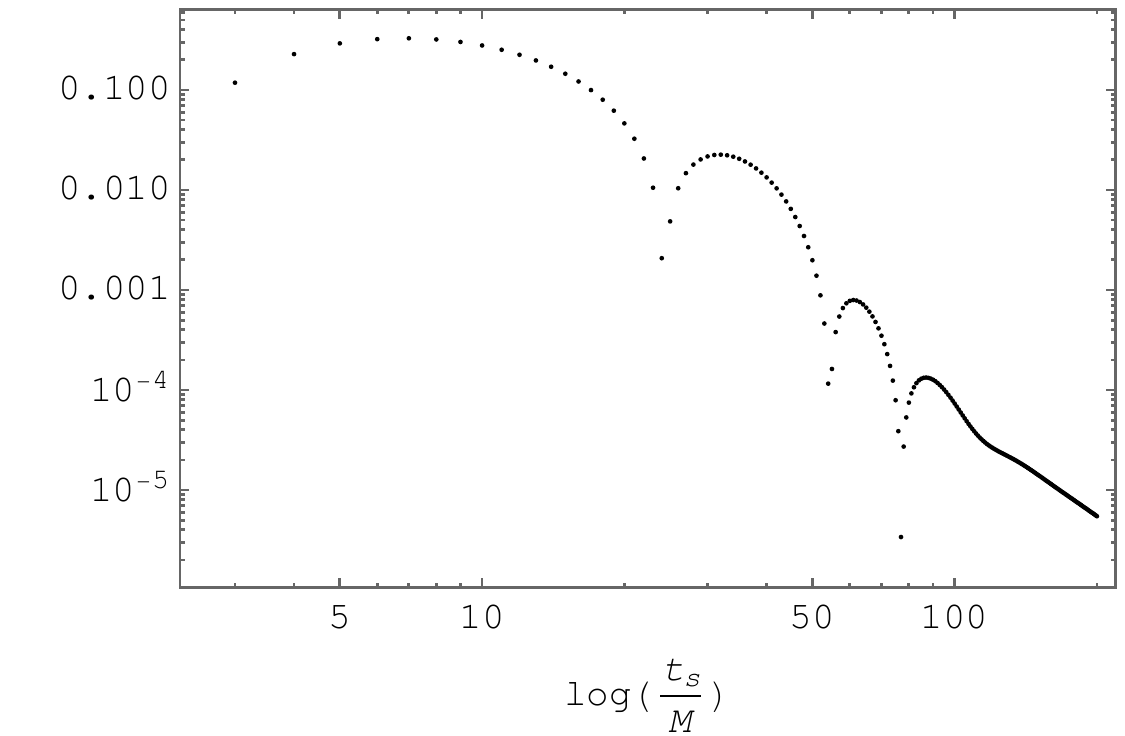}
\includegraphics[trim=0cm 0cm 0cm 0cm,clip=true,totalheight=0.23\textheight]{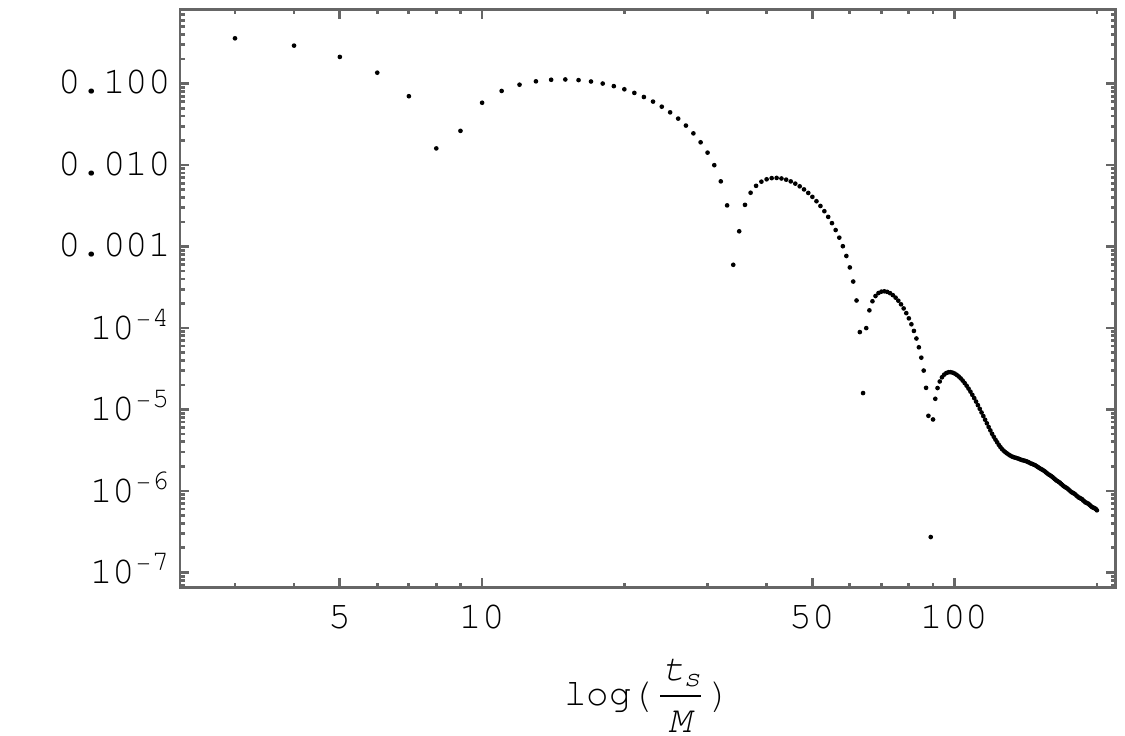}\\[10pt]
\caption{The top figures represent the real (left) and the imaginary (right) parts of the complex quantity $\sqrt{\frac{4\pi}{M}}f^K_{\w}$ as a function of time for  $M\w =0.5$ and for $r=3M$. The two bottom figures are the log-log plots of the absolute values of the real (left) and imaginary (right) parts of the same quantity. The bottom figures show that at late times, the Kruskal modes $f^{K}_{\w}$ decay as a power-law.}
\label{fig:fK}
\end{figure}
\newpage
\begin{figure}[h]
\centering
\includegraphics[trim=0.5cm 0cm 0cm 0cm,clip=true,totalheight=0.35\textheight]{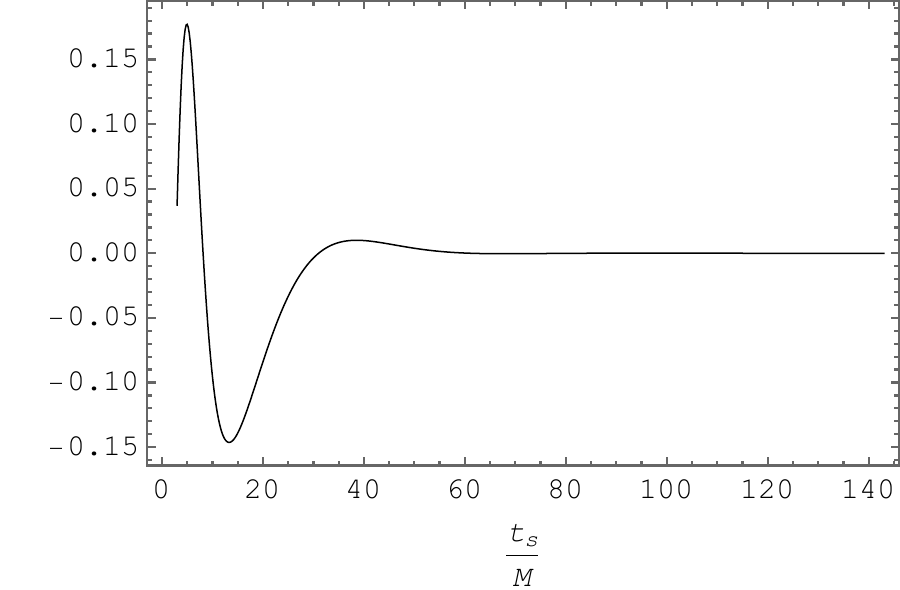}
\includegraphics[trim=0.5cm 0cm 0cm 0cm,clip=true,totalheight=0.35\textheight]{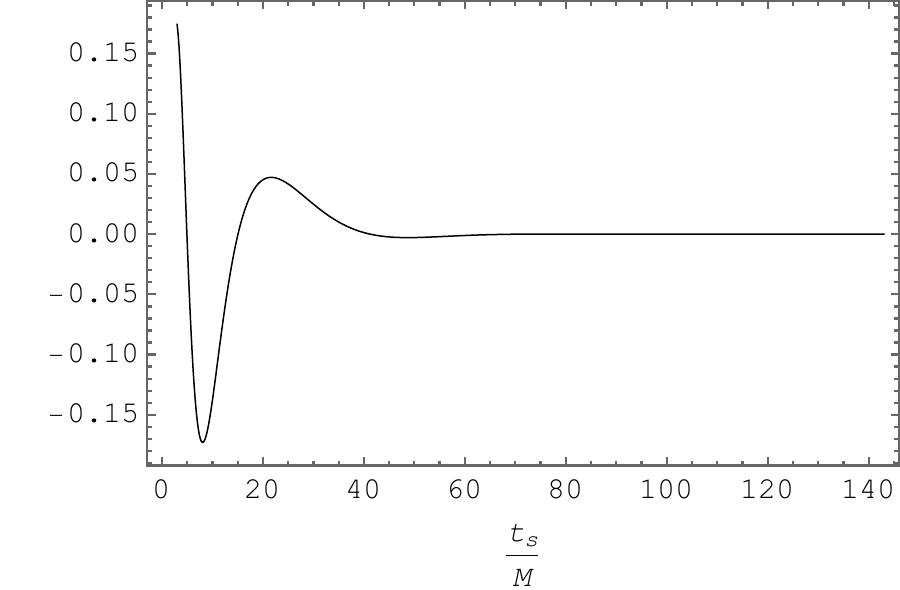}\\[10pt]
\caption{The top plot represents the real part and the bottom plot is the imaginary part of of the complex quantity $\sqrt{\frac{4\pi}{M}}(f^{in}_{\w})_u$ as a function of time for  $M\w =2.5$ and for $r=3M$. At large times, imaginary and real parts decay to zero.}
\label{fig:fudec}
\end{figure}
\chapter{Stress-energy Tensor in the collapsing null-shell spacetime}
\section{Introduction}
In this chapter, we first derive a general expression for the renormalized stress-energy tensor in terms of various multiple integrals for arbitrary space and time coordinates. We next discuss the numerical computation of these integrals. In Chapter 4, we showed that to renormalize the stress-energy tensor one can subtract the stress-energy tensor in the Unruh state, $\langle U|T_{\mu\nu}| U\rangle$, from the stress-energy tensor in the $in$ vacuum state, $\langle in|T_{\mu \nu}| in\rangle$. Then, to find the renormalized stress-energy tensor, one can add back the renormalized stress-energy tensor $\langle U|T_{\mu\nu}| U\rangle_{\text{ren}}$ to $\langle \Delta T_{\mu \nu}\rangle=\langle in|T_{\mu \nu}| in\rangle-\langle U|T_{\mu\nu}| U\rangle$. Therefore, this chapter focuses on the computation of $\langle \Delta T_{\mu \nu}\rangle$, since, as mentioned in Chapter 4, the renormalized expected value of stress-energy tensor for a massless minimally coupled scalar field in the Unruh state for a 4D Schwarzschild spacetime is computed in \cite{levi-ori} and \cite{levi}. Throughout this chapter, we only consider the contributions of the spherically-symmetric $in$ modes to the symmetric two-point function and the stress-energy tensor which are denoted by $G^{(1)}(x,x')$ and
$\langle T_{\mu \nu} \rangle$ in this chapter.
In a similar way to Chapter 5, we drop the subscripts $\ell$ and $m$ in the matching coefficients and the $in$ modes.
\section{Stress-energy Tensor}
To find $\langle T_{\mu \nu}\rangle$, we need first to construct the two-point function for $f^{in}_{\w}$ and then substitute that into ~\eqref{ch2-T}. The symmetric two-point function $G^{(1)}(x,x')$ for the modes $f^{in}_{\w}$ is
\bea
G^{(1)}(x,x')=\int_0^{\infty}\Big\{f^{in}_{\w}(x)f^{in *}_{\w}(x')+f^{in}_{\w}(x')f^{in *}_{\w}(x)\Big\}d\w=2\Re{\int_0^{\infty}f^{in}_{\w}(x)f^{in *}_{\w}(x')\;d\w}. \label{ch6-G(x,x')}
\eea
In the previous chapter, we showed that for the spherically-symmetric contribution to the $in$ modes, one has $f^{in}_{\w}=\big(f^{in}_{\w}\big)_v+\big(f^{in}_{\w}\big)_u$, where $\big(f^{in}_{\w}\big)_v$ is the contribution of the modes that are purely $v-$dependent inside the null shell and $\big(f^{in}_{\w}\big)_u$ are the contribution of the modes that only depend on $u$ inside the null shell. For the numerical calculations, before writing $f^{in}_{\w}$ in terms of $\big(f^{in}_{\w}\big)_v$ and $\big(f^{in}_{\w}\big)_u$, we write ~\eqref{ch6-G(x,x')} as follows
\bea
G^{(1)}(x,x')=2\Re{\int_0^{\lambda}f^{in}_{\w}(x)f^{in *}_{\w}(x')\;d\w+\int_{\lambda}^{\infty}f^{in}_{\w}(x)f^{in *}_{\w}(x')\;d\w}.\label{ch6-G(x,x')2}
\eea
If $\lambda$ is chosen to be small enough, then the contribution of the first integral is negligible because in the limit $\w\to 0$, the modes $f^{in}_{\w}$ approach zero. Therefore, we focus on computing the second integral in ~\eqref{ch6-G(x,x')2}.  It is worth mentioning that $\big(f^{in}_{\w}\big)_v$ and $\big(f^{in}_{\w}\big)_u$ are infrared divergent, but their sum is finite in the limit $\w\to 0$.  These properties are shown in Fig. {fig:fin-IR}.
\\[10pt]
\begin{figure}[h]
\centering
\includegraphics[trim=0cm 0cm 0cm 0cm,clip=true,totalheight=0.23\textheight]{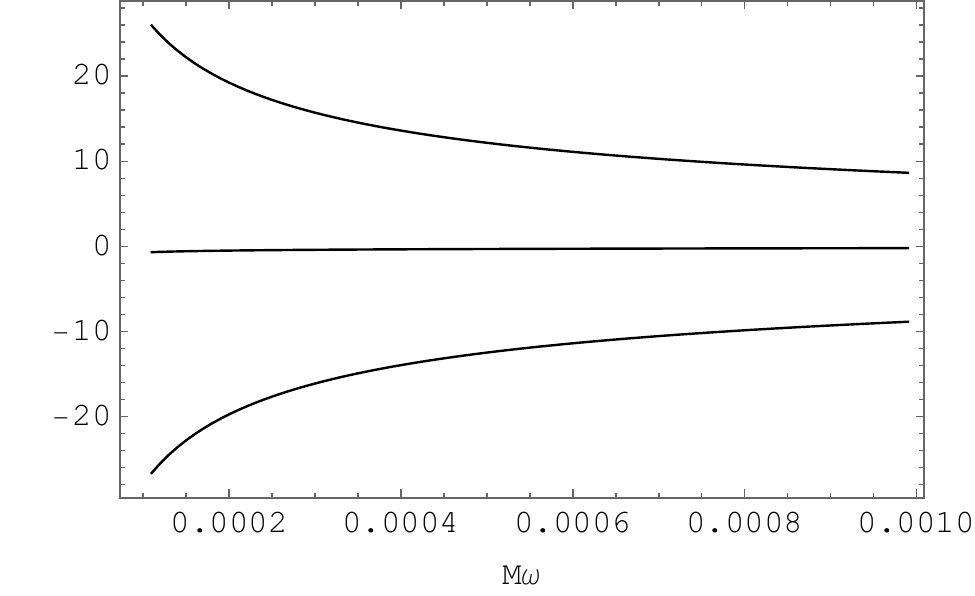}
\\[10pt]

\caption{The top curve is the real part of $\sqrt{\frac{4\pi}{M}}(f^{in}_{\w})_v$ as a function of frequency $\w$ for a fixed radial coordinate $r=3M$ and fixed time coordinate $t_s=3M$. The bottom curve is the real part of $\sqrt{\frac{4\pi}{M}}(f^{in}_{\w})_u$ for the same spacetime coordinates. The middle curve is the sum of these two quantities, which approaches zero for small frequencies. The imaginary part }
\label{fig:fin-IR}
\end{figure}

Replacing $f^{in}_{\w}$ with $\big(f^{in}_{\w}\big)_v+\big(f^{in}_{\w}\big)_u$, we find three types of integrals. 
\bes \bea
\big\{G^{(1)}_A(x,x')\big\}^{in}_v&=&2\Re{\int_{\lambda}^{\infty}\big(f^{in}_{\w} \big)_v(x)\big(f^{in *}_{\w}\big)_v(x')\;d\w},\\
\big\{G^{(1)}_B(x,x')\big\}^{in}_v&=&2\Re{\int_{\lambda}^{\infty}\big(f^{in}_{\w} \big)_u(x)\big(f^{in *}_{\w}\big)_u(x')\;d\w},\\
\big\{G^{(1)}_C(x,x')\big\}^{in}_v&=&2\Re{\int_{\lambda}^{\infty}\big(f^{in}_{\w} \big)_v(x)\big(f^{in *}_{\w}\big)_u(x')\;d\w}.
\eea \ees
In the next three sections, we discuss the computation of these integrals.

\section{Stress-energy Tensor: Contribution from $\big(f^{in}_{\w}\big)_v$}
In the previous chapter, we showed that $\big(f^{in}_{\w}\big)_v$ approaches $f^{\mathscr{I}^-}_{\w}$ as $t_s^{-3}$ at late times for fixed values of the spatial coordinate $r$. As mentioned previously, the $f^{\mathscr{I}^-}_{\w}$ modes are a set of modes that are positive frequency on past null infinity and are part of the Unruh state. To find the contribution to  $\langle \Delta T_{\mu \nu}\rangle$ from $\big(f^{in}\big)_v$, we only need to subtract the contribution of the $f^{\mathscr{I}^-}_{\w}$ modes to $\langle U|T_{\mu\nu}| U\rangle$ from the contribution to $\langle in|T_{\mu\nu}|  in\rangle$ from $\big(f^{in}\big)_v$. The general expression for the stress-energy tensor for a particular state is given in  ~\eqref{Delta-Tab-2}. Note that to find the contribution to $\langle \Delta T_{\mu \nu} \rangle$ from $\big(f^{in}\big)_v$, one needs first to find the difference between the contribution to the symmetric two-point correlation function from $(f^{in}_{\w})_v$ and the contribution from $f^{\mathscr{I}^-}_{\w}$, which is denoted by $(\Delta G)_v$. Here we first derive a general expression for $(\Delta G)_v$ in terms of some multiple integrals for arbitrary points in the spacetime and then later in this chapter numerically evaluate its contribution to the stress-energy tensor for particular spacetime coordinates.

The difference between the two-point functions is
\bea
\big\{\Delta G^{(1)}(x,x')\big\}_v=\big\{G^{(1)}_A(x,x')\big\}^{in}_v-\big\{G^{(1)}(x,x')\big\}^{\mathscr{I}^-}.
\eea
with
\bea
\Big\{G^{(1)}(x,x')\Big\}^{\mathscr{I}^-}&=&
\int_{0}^{\infty}\Big\{f^{\mathscr{I}^-}_{\w}(x)f^{\mathscr{I}^{-*}}_{\w}(x')+f^{\mathscr{I}^-}_{\w}(x')f^{\mathscr{I}^{-}*}_{\w}(x)\Big\}d\w
\nonumber \\
&=& 2 \operatorname{Re}\Big\{\int_{0}^{\infty}f^{\mathscr{I}^-}_{\w}(x)f^{\mathscr{I}^{-*}}_{\w}(x') d\w \Big\}.\label{ch6-GUnruh}
\eea
In\;~\eqref{f-def-1}, we showed the general form of solutions that can be obtained using separation of variables in ~\eqref{Box-phi}.  Therefore, $f^{\mathscr{I}^-}_{\w}$ can be written as
\bea
f^{\mathscr{I}^-}_{\w}=\frac{Y_{00}}{\sqrt{4\pi \w}}\psi^{\mathscr{I}^-}_{\w}(r,t_s),\label{ch6-fIminus}
\eea
where $\psi^{\mathscr{I}^-}_{\w}(r,t_s)=\chi^{\mathscr{I}^-}_{\w}(r)e^{-i\w t_s}$. Substituting ~\eqref{ch6-fIminus} into ~\eqref{ch6-GUnruh}, we find the contribution to the two-point function from the s-wave sector of the $f^{\mathscr{I}^-}_{\w}$ modes is
\bea
\Big(G^{(1)}(x,x')\Big)^{\mathscr{I}^-}&=&
2  |Y_{00}|^2 \operatorname{Re}\Big\{\int_{\lambda}^{\infty}d\w \frac{1}{4\pi r^2\w}  \chi_L^{\infty}(r)\chi_R^{\infty}(r)\Big\}\nonumber \\&&
-4  |Y_{00}|^2 \operatorname{Re}\Big\{\int_{\lambda}^{\infty}d\w \frac{1}{4\pi r^2\w}\frac{F_R(\w)}{F_L(\w)} \chi_L^{\infty}(r)\chi_L^{\infty}(r)\Big\}\nonumber \\&&
+2  |Y_{00}|^2 \operatorname{Re}\Big\{\int_{\lambda}^{\infty}d\w \frac{1}{4\pi r^2\w}  \frac{F_R^{*}(\w)}{F_L^{*}(\w)}\frac{F_R(\w)}{F_L(\w)}\chi_R^{\infty}(r)\chi_L^{\infty}(r)\Big\}.\label{gUnruhtt}
\eea
The two-point correlation function for $\big(f^{in})_v$ is given by 
\bea
\Big\{G^{(1)}(x,x')\Big\}^{in}_v&=&\int_{\lambda}^{\infty}\Big\{\big(f^{in}_{\w}(x)\big)_v\big(f^{in*}_{\w}(x')\big)_v+\big(f^{in}_{\w}(x')\big)_v\big(f^{in*}_{\w}(x)\big)_v\Big\}d\w
\nonumber \\
&=& 2 \operatorname{Re}\Big\{ \int_{\lambda}^{\infty}\big(f^{in}_{\w}(x)\big)_v\big(f^{in*}_{\w}(x')\big)_v d\w \Big\}.\label{ch6-DeltaG}
\eea
Substituting ~\eqref{ch4-fin-vpart2} into ~\eqref{ch6-DeltaG}, we find 
\bea
\Big\{G^{(1)}(x,x')\Big\}^{in}_v&=&2 |Y_{00}|^2  \operatorname{Re}\Big\{\int_{\lambda}^{\infty}d\w \Bigg(\frac{\chi_L^{\infty}(\w,r)e^{-i\w t_s}} {r\sqrt{4\pi \w}}\nonumber \\ &&
-\frac{i}{2\pi}\frac{1}{r\sqrt{4\pi\w}}\int_{0}^{\infty}\frac{F_R^*(\w')}{F_L^*(\w')}\frac{e^{-i(\w-\w')v_0}}{\w'-\w+i\epsilon}\chi_R^{\infty}(\w',r)e^{-i\w' t_s}d\w'\nonumber \\ &&
+\frac{i}{2\pi}\frac{1}{r\sqrt{4\pi\w}}\int_0^{\infty}\frac{F_R(\w')}{F_L(\w')}\frac{e^{-i(\w'+\w)v_0}}{\w'+\w-i\epsilon}\chi_L^{\infty}(\w',r)e^{i\w' t_s}d\w'\Bigg)\nonumber \\ &&
\times\Bigg(\frac{\chi_R^{\infty}(\w,r')e^{i\w t'_s}} {r'\sqrt{4\pi \w}}
+\frac{i}{2\pi}\frac{1}{r'\sqrt{4\pi\w}}\int_{0}^{\infty}\frac{F_R(\w'')}{F_L(\w'')}\frac{e^{i(\w-\w'')v_0}}{\w''-\w-i\epsilon}\chi_L^{\infty}(\w'',r')e^{i\w'' t'_s}d\w''\nonumber \\ &&
-\frac{i}{2\pi}\frac{1}{r'\sqrt{4\pi\w}}\int_0^{\infty}\frac{F^*_R(\w'')}{F^*_L(\w'')}\frac{e^{i(\w''+\w)v_0}}{\w''+\w+i\epsilon}\chi_R^{\infty}(\w'',r')e^{-i\w'' t'_s}d\w''\Bigg)\Big\}.\label{ch4-Gin-vpart}
\eea

By expanding ~\eqref{ch4-Gin-vpart}, we get nine terms. One of them is a one-dimensional integral which we call $I_0$ and is given by
\bea
I_0=2 |Y_{00}|^2  \operatorname{Re}\Big\{\int_{\lambda}^{\infty}d\w \frac{\chi_L^{\infty}(\w,r)\chi_R^{\infty}(\w,r)} {4\pi r^2\w}\Big\}.
\eea
Note that this integral will be canceled by the first term on the right side of ~\eqref{gUnruhtt} once we subtract $\Big\{G^{(1)}(x,x')\Big\}^{\mathscr{I}^-}$ from $\Big\{G^{(1)}(x,x')\Big\}^{\text{in}}_v$. We also get four two-dimensional integrals that are given by
\bes \bea
I_1&=&\operatorname{Re}\Big\{2 |Y_{00}|^2  \int_{\lambda}^{\infty}d\w \frac{i\chi_L^{\infty}(\w,r)e^{-i\w t_s}} {8 \pi^2 r^2\w}\nonumber \\ &&
\times\int_{0}^{\infty}\frac{F_R(\w'')}{F_L(\w'')}\frac{e^{i(\w-\w'')v_0}}{\w''-\w-i\epsilon}\chi_L^{\infty}(\w'',r)e^{i\w'' t_s}d\w''\Big\},\\[6pt]
I_2&=&\operatorname{Re}\Big\{-2 |Y_{00}|^2  \int_{\lambda}^{\infty}d\w \frac{i\chi_L^{\infty}(\w,r)e^{-i\w t_s}} {8 \pi^2 r^2\w}\nonumber \\ &&
\times\int_{0}^{\infty}\frac{F^*_R(\w'')}{F^*_L(\w'')}\frac{e^{i(\w+\w'')v_0}}{\w''+\w+i\epsilon}\chi_R^{\infty}(\w'',r)e^{-i\w'' t_s}d\w''\Big\},\\[6pt]
I_3&=&\operatorname{Re}\Big\{-2 |Y_{00}|^2  \int_{\lambda}^{\infty}d\w \frac{i\chi_R^{\infty}(\w,r)e^{i\w t_s}} {8 \pi^2 r^2\w}\nonumber \\ &&
\times\int_{0}^{\infty}\frac{F^*_R(\w')}{F^*_L(\w')}\frac{e^{-i(\w-\w')v_0}}{\w'-\w+i\epsilon}\chi_R^{\infty}(\w',r)e^{-i\w' t_s}d\w'\Big\},\\[6pt]
I_4&=&\operatorname{Re}\Big\{2 |Y_{00}|^2  \int_{\lambda}^{\infty}d\w \frac{i\chi_R^{\infty}(\w,r)e^{i\w t_s}} {8 \pi^2 r^2\w}\nonumber \\ &&
\times\int_{0}^{\infty}\frac{F_R(\w')}{F_L(\w')}\frac{e^{-i(\w+\w')v_0}}{\w'+\w-i\epsilon}\chi_L^{\infty}(\w',r)e^{i\w' t_s}d\w'\Big\}.
\eea \label{double-integrals}\ees \\[3pt]
We begin with computing the integrals $I_1$ and $I_2$. We computed the scattering coefficients $F_R(\w)$ and $F_L(\w)$ as well as the radial functions $\chi_L^{\infty}(\w,r)$ and $\chi_R^{\infty}(\w,r)$  numerically for various values of $\w$ and $r$. We will then use these numerical results to compute the above integrals. 

One can see that there is a singularity at $\w''=\w$ in the second integrand in $I_1$. To handle this singularity and for simplicity in the numerical calculations, we write integrals $I_1$ and $I_2$ as follows

\bea
I_1&=&2 |Y_{00}|^2  \operatorname{Re}\Bigg\{\int_0^{\infty}d\w \frac{i\chi_L^{\infty}(\w,r)e^{-i\w t_s}} {8 \pi^2 r^2\w}
\Bigg[\int_{0}^{\infty}\frac{F_R(\w'')}{F_L(\w'')}\frac{e^{i(\w-\w'')v_0}}{\w''-\w-i\epsilon}\chi_L^{\infty}(\w'',r)e^{i\w'' t_s}d\w''\nonumber \\ &&
-\int_{0}^{\infty}\frac{F_R(\w)}{F_L(\w)}\frac{1}{\w''-\w-i\epsilon}\chi_L^{\infty}(\w,r)e^{i\w'' t_s}d\w''
\nonumber \\ &&
+\int_{0}^{\infty}\frac{F_R(\w)}{F_L(\w)}\frac{1}{\w''-\w-i\epsilon}\chi_L^{\infty}(\w,r)e^{i\w'' t_s}d\w''\Bigg]
\Bigg\},\label{I1}
\eea 
\bea
I_2&=&-2 |Y_{00}|^2  \operatorname{Re}\Bigg\{\int_0^{\infty}d\w \frac{i\chi_L^{\infty}(\w,r)e^{-i\w t_s}} {8 \pi^2 r^2\w}
\Bigg[\int_{0}^{\infty}\frac{F_R^*(\w'')}{F_L^*(\w'')}\frac{e^{i(\w-\w'')v_0}}{\w''+\w+i\epsilon}\chi_R^{\infty}(\w'',r)e^{-i\w'' t_s}d\w''\nonumber \\ &&
-\int_{0}^{\infty}\frac{F_R(\w)}{F_L(\w)}\frac{1}{\w''+\w+i\epsilon}\chi_L^{\infty}(\w,r)e^{-i\w'' t_s}d\w''
\nonumber \\ &&
+\int_{0}^{\infty}\frac{F_R(\w)}{F_L(\w)}\frac{1}{\w''+\w+i\epsilon}\chi_L^{\infty}(\w,r)e^{-i\w'' t_s}d\w''\Bigg]
\Bigg\}.\label{I2}
\eea 
Note that the subtraction of 
\[\int_{0}^{\infty}\frac{F_R(\w)}{F_L(\w)}\frac{1}{\w''-\w-i\epsilon}\chi_L^{\infty}(\w,r)e^{i\w'' t_s}d\w''\] 
from $I_1$ removes the singularity at $\w=\w''$ in the first term inside the square bracket in $I_1$. Then, we can numerically compute these terms. Note that the last terms on the right in ~\eqref{I1} and ~\eqref{I2} are given by
\bea
(I_1)_{add}+(I_2)_{add}&=&2 |Y_{00}|^2  \operatorname{Re}\Bigg\{\int_0^{\infty}d\w \frac{i\chi_L^{\infty}(\w,r)e^{-i\w t_s}} {8 \pi^2 r^2\w}
\int_{0}^{\infty}\frac{F_R(\w)}{F_L(\w)}\frac{1}{\w''-\w-i\epsilon}\chi_L^{\infty}(\w,r)e^{i\w'' t_s}d\w''\Bigg\}\nonumber \\ &&
-2 |Y_{00}|^2  \operatorname{Re}\Bigg\{\int_0^{\infty}d\w \frac{i\chi_L^{\infty}(\w,r)e^{-i\w t_s}} {8 \pi^2 r^2\w}\nonumber \\ &&
\times \int_{0}^{\infty}\frac{F_R(\w)}{F_L(\w)}\frac{1}{\w''+\w+i\epsilon}\chi_L^{\infty}(\w,r)e^{-i\w'' t_s}d\w''\Bigg\}.
\eea
By changing the variable of integration over $\w''$ in the second curly bracket such that $\w''\to -\w''$, one finds
\bea
(I_1)_{add}+(I_2)_{add}&=&2 |Y_{00}|^2  \operatorname{Re}\Bigg\{\int_0^{\infty}d\w \frac{i\chi_L^{\infty}(\w,r)e^{-i\w t_s}} {8 \pi^2 r^2\w}\nonumber \\ &&
\times\int_{-\infty}^{\infty}\frac{F_R(\w)}{F_L(\w)}\frac{1}{\w''-\w-i\epsilon}\chi_L^{\infty}(\w,r)e^{i\w'' t_s}d\w''\Bigg\}.
\eea
Contour integration can now be used for the integral over $\w''$. Note that the integral over $\w''$ has a pole at $\w''=\w+i\epsilon$. For $t_s>0$, we close the contour in the upper half-plane and find
\bea
(I_1)_{add}+(I_2)_{add}&=&-2 |Y_{00}|^2  \operatorname{Re}\Bigg\{\int_0^{\infty}d\w \frac{\chi_L^{\infty}(\w,r)e^{-i\w t_s}} {4 \pi r^2\w}\frac{F_R(\w)}{F_L(\w)}\chi_L^{\infty}(\w,r)e^{i\w t_s}\Bigg\}\nonumber \\ 
&=&-2 |Y_{00}|^2  \operatorname{Re}\Bigg\{\int_0^{\infty}d\w \frac{\chi_L^{\infty}(\w,r)} {4 \pi r^2\w}\frac{F_R(\w)}{F_L(\w)}\chi_L^{\infty}(\w,r)\Bigg\}.
\eea
This is equal to half of the second term on the right-hand side of ~\eqref{gUnruhtt} and will go away when we subtract ~\eqref{gUnruhtt} from ~\eqref{ch4-Gin-vpart}. The sum of the integrals $I_3$ and $I_4$ can be computed similarly. It turns out the sum of these two terms is equal to the sum of the integrals $I_1$ and $I_2$. For $t_s<0$, we close in the lower half plane and find a contribution of zero.

We find four triple integrals by expanding ~\eqref{ch4-Gin-vpart}.
\bes \bea
I_5&=&\frac{|Y_{00}|^2}{8\pi^3rr'}  \operatorname{Re}\Bigg\{\int_0^{\infty}d\w' \int_0^{\infty}d\w'' \int_0^{\infty}\frac{d\w}{\w}\Bigg(
\frac{F_R^*(\w')}{F_L^*(\w')}\frac{e^{-i(\w-\w')v_0}}{\w'-\w+i\epsilon}\chi_R^{\infty}(r)e^{-i\w' t_s}\nonumber \\ &&
\frac{F_R(\w')}{F_L(\w')}\frac{e^{i(\w-\w')v_0}}{\w'-\w-i\epsilon}\chi_L^{\infty}(r')e^{i\w' t'_s}\Bigg)
\Bigg\},\\
I_6&=&-\frac{|Y_{00}|^2}{8\pi^3rr'}  \operatorname{Re}\Bigg\{\int_0^{\infty}d\w' \int_0^{\infty}d\w'' \int_0^{\infty}\frac{d\w}{\w}\Bigg(
\frac{F_R^*(\w')}{F_L^*(\w')}\frac{e^{-i(\w-\w')v_0}}{\w'-\w+i\epsilon}\chi_R^{\infty}(r)e^{-i\w' t_s}\nonumber \\ &&
\frac{F^*_R(\w')}{F^*_L(\w')}\frac{e^{i(\w+\w')v_0}}{\w'+\w+i\epsilon}\chi_R^{\infty}(r')e^{-i\w' t'_s}\Bigg)
\Bigg\},\\
I_7&=&-\frac{|Y_{00}|^2}{8\pi^3rr'}  \operatorname{Re}\Bigg\{\int_0^{\infty}d\w' \int_0^{\infty}d\w'' \int_0^{\infty}\frac{d\w}{\w}\Bigg(
\frac{F_R(\w')}{F_L(\w')}\frac{e^{-i(\w+\w')v_0}}{\w'+\w-i\epsilon}\chi_L^{\infty}(r)e^{i\w' t_s}\nonumber \\ &&
\frac{F_R(\w')}{F_L(\w')}\frac{e^{i(\w-\w')v_0}}{\w'-\w-i\epsilon}\chi_L^{\infty}(r')e^{i\w' t'_s}\Bigg)
\Bigg\},\\
I_8&=&\frac{|Y_{00}|^2}{8\pi^3rr'}  \operatorname{Re}\Bigg\{\int_0^{\infty}d\w' \int_0^{\infty}d\w'' \int_0^{\infty}\frac{d\w}{\w}\Bigg(
\frac{F_R(\w')}{F_L(\w')}\frac{e^{-i(\w+\w')v_0}}{\w'+\w-i\epsilon}\chi_L^{\infty}(r)e^{i\w' t_s}\nonumber \\ &&
\frac{F^*_R(\w')}{F^*_L(\w')}\frac{e^{i(\w+\w')v_0}}{\w'+\w+i\epsilon}\chi_R^{\infty}(r')e^{-i\w' t'_s}\Bigg)
\Bigg\}.
\eea \ees
These integrals have to be computed directly since there is no obvious way to remove the singularity in these integrals by subtracting and adding the Unruh counter terms. However, later in this chapter,  it is shown that at late times, the sum of these integrals approaches the third integral in ~\eqref{gUnruhtt}.

To find the contribution to $\langle \Delta T_{ab}\rangle$ from $(f^{in}_{\w})_v$ and $f^{\mathscr{I}^-}_{\w}$, one needs to compute various derivatives of $\Delta G^{(1)}_v(x,x')$ as follows 
\bea
\Delta \langle T_{ab} \rangle=\frac{1}{4}\big[\lim_{x' \to x} \big(\Delta G_{v; a' ;b}^{(1)}(x,x')+ \Delta G_{v; a ;b'}^{(1)}(x,x')\big)-g_{ab}g^{cd}\lim_{x' \to x} \Delta G_{v; c ;d'}^{(1)}(x,x')\big]\label{mainequation};
\eea
It is worth mentioning that for modes with $\ell=0$, the angular part of the modes is a constant since $Y_{00}=\frac{1}{2}\sqrt{\frac{1}{\pi}}$. Therefore, one only needs to compute the derivatives of $\Delta G^{(1)}(x,x')$ with respect to the radial coordinate $r$ and the time coordinate $t_s$. 
\section{Stress-energy Tensor: Contribution of $\big(f^{in}_{\w}\big)_u$}
In this section, we derive the contribution to the stress-energy tensor from $\big(f^{in}_{\w}\big)_u$ by computing the difference between the contribution to the symmetric two-point functions from $\big(f^{in}_{\w}\big)_u$ and from the Kruskal modes $f^{K}_{\w}$.
\bea
\Delta G^{(1)}_u(x,x')=\big\{G^{(1)}(x,x')\big\}^{in}_u-\big\{G^{(1)}(x,x')\big\}^{K}.
\eea

As a first step, we construct the contribution to the two-point function from $\big(f^{in}_{\w}\big)_u$.
\bea
\Big\{G^{(1)}(x,x')\Big\}^{in}_u=2 \operatorname{Re}\Bigg\{ \int_{0}^{\infty}\big(f^{in}_{\w}(x)\big)_u\big(f^{in*}_{\w}(x')\big)_u d\w\Bigg\}\label{TwoPoint}.
\eea
By substitution of ~\eqref{ch4-fin-upart} into 
~\eqref{TwoPoint}, we find 
\bea
\Big\{G^{(1)}(x,x')\Big\}^{in}_u&=&2 \operatorname{Re}\Bigg\{ \int_{0}^{\infty}d\w\int_{0}^{\infty}d\w_1\Big\{\big(A^{H^+}_{\w\w_1}\big)_uf^{H^+}_{\w_1}(x)+\big(B^{H^+}_{\w\w_1}\big)_uf^{H^{+*}}_{\w_1}(x)\nonumber \\ &&
+\big(A^{\mathscr{I}^+}_{\w\w_1}\big)_uf^{\mathscr{I}^+}_{\w_1}(x)+\big(B^{\mathscr{I}^+}_{\w\w_1}\big)_uf^{\mathscr{I}^{+*}}_{\w_1}(x)\Big\}\nonumber \\ &&
\times \int_0^{\infty}d\w_2\Big\{\big(A^{H^{+*}}_{\w\w_2}\big)_uf^{H^{+*}}_{\w_2}(x')+\big(B^{H^{+*}}_{\w\w_2}\big)_uf^{H^{+}}_{\w_2}(x')\nonumber \\ &&
+\big(A^{\mathscr{I}^{+*}}_{\w\w_2}\big)_uf^{\mathscr{I}^{+*}}_{\w_2}(x')+\big(B^{\mathscr{I}^{+*}}_{\w\w_2}\big)_uf^{\mathscr{I}^{+}}_{\w_2}(x')\Big\}
\Bigg\}.\label{ch5-upart}
\eea
The matching coefficients in ~\eqref{ch5-upart} are the non-analytical terms in ~\eqref{gen-mat} whose numerical computation is discussed in the previous chapter. For simplicity in the numerical calculations and to avoid working with too many integrals, we write the contribution of the $H^+$ modes in a way similar way to the 2D case, which was presented in ~\eqref{fin-A-B}; i.e.,
\bea
\Big\{G^{(1)}(x,x')\Big\}^{in}_u&=&2 \operatorname{Re}\Bigg\{ \int_{0}^{\infty}d\w \Big\{\big(f^{in }_{\w}\big)_{H^+}+\int_{0}^{\infty}d\w_1\big[A^{\mathscr{I}^+}_{\w\w_1}f^{\mathscr{I}^+}_{\w_1}(x)+B^{\mathscr{I}^+}_{\w\w_1}f^{\mathscr{I}^{+*}}_{\w_1}(x)\big]\Big\}\nonumber \\ &&
\times \Big\{\big(f^{in\; *}_{\w}\big)_{H^+}+\int_0^{\infty}d\w_2\big[A^{\mathscr{I}^{+*}}_{\w\w_2}f^{\mathscr{I}^{+*}}_{\w_2}(x')+B^{\mathscr{I}^{+*}}_{\w\w_2}f^{\mathscr{I}^{+}}_{\w_2}(x')\big]\Big\}
\Bigg\}.\label{expansion}
\eea
To find $\Delta G^{(1)}_u$, we subtract the two-point function for the $f^{K}_{\w}$ modes from ~\eqref{expansion} and substitute the result into ~\eqref{mainequation}. In ~\eqref{fK}, the modes $f^K_{\w}$ are represented in terms of the Boulware modes $f^{H^-}_{\w}$ using a Bogolubov transformation. Then the contribution to the two-point function 
from the $f^K_{\w}$ modes is 
\bea
\Big(G^{(1)}(x,x')\Big)^K&=&2 \operatorname{Re}\Bigg\{ \int_{0}^{\infty}d\w\int_{0}^{\infty}d\w_1 \Big[\alpha_{\w\w_1}\;f^{H^-}_{\w_1}(x)+\beta_{\w\w_1}\;f^{H^{-\;*}}_{\w_1}(x)\big]\nonumber \\&&
\times\int_0^{\infty}d\w_2\big[\alpha^*_{\w\w_2}\;f^{H^{-*}}_{\w_2}(x')+\beta^{*}_{\w\w_2}\;f^{H^{-}}_{\w_2}(x')\Big]\Bigg\},\label{ch5-GK}
\eea
where $\alpha^K_{\w\w'}$ and $\beta^K_{\w\w'}$ are given in ~\eqref{alphaK} and ~\eqref{betaK}. As mentioned above, ~\eqref{ch5-GK} has to be subtracted from ~\eqref{expansion}. To do this, we first expand ~\eqref{expansion} and find the following four integrals
\bes \bea
I_1&=&2 \operatorname{Re}\Bigg\{ \int_{0}^{\infty}d\w \big(f^{in}_{\w}(x)\big)_{H^+}\;\big(f^{in *}_{\w}(x')\big)_{H^+}\Bigg\}.
\\
I_2&=&2 \operatorname{Re}\Bigg\{ \int_{0}^{\infty}d\w \big(f^{in}_{\w}(x)\big)_{H^+}\int_{0}^{\infty}d\w_2\big[A^{\mathscr{I}^{+*}}_{\w\w_2}f^{\mathscr{I}^{+*}}_{\w_2}(x')+B^{\mathscr{I}^{+*}}_{\w\w_2}f^{\mathscr{I}^{+}}_{\w_2}(x')\big]\Bigg\}.\\
I_3&=&2 \operatorname{Re}\Bigg\{ \int_{0}^{\infty}d\w \big(f^{in \;*}_{\w}(x')\big)_{H^+}\int_{0}^{\infty}d\w_1\big[A^{\mathscr{I}^{+}}_{\w\w_1}f^{\mathscr{I}^{+}}_{\w_1}(x)+B^{\mathscr{I}^{+}}_{\w\w_1}f^{\mathscr{I}^{+\;*}}_{\w_1}(x)\big]\Bigg\}.
\\
I_4&=&2 \operatorname{Re}\Bigg\{ \int_{0}^{\infty}d\w\int_{0}^{\infty}d\w_1 \big[A^{\mathscr{I}^{+}}_{\w\w_1}f^{\mathscr{I}^{+}}_{\w_1}(x)+B^{\mathscr{I}^{+}}_{\w\w_1}f^{\mathscr{I}^{+\;*}}_{\w_1}(x)\big]\nonumber \\&&
\times\int_0^{\infty}d\w_2\big[A^{\mathscr{I}^{+*}}_{\w\w_2}f^{\mathscr{I}^{+*}}_{\w_2}(x')+B^{\mathscr{I}^{+*}}_{\w\w_2}f^{\mathscr{I}^{+}}_{\w_2}(x')\big]\Bigg\}.
\eea \ees
We next show that subtracting the two-point function for the Kruskal modes $f^K_{\w}$ from $I_4$ removes any ultraviolet divergences in the two-point function from $\big(f^{in}_{\w}\big)_u$. To subtract the two-point function for the Kruskal modes from $I_4$, we first write the matching coefficients $A^{\mathscr{I}^+}_{\w\w_1}$ and $B^{\mathscr{I}^+}_{\w\w_2}$ as follows
\bes \bea
\Big(A^{\mathscr{I}^+}_{\w\w_1}\Big)_{4D}&=&\Big(A^{\mathscr{I}^+}_{\w\w_1}\Big)_{4D}-\Big(A^{\mathscr{I}^+}_{\w\w_1}\Big)_{2D}+\Big(A^{\mathscr{I}^+}_{\w\w_1}\Big)_{2D}=\Delta A^{\mathscr{I}^+}_{\w\w_1}+\Big(A^{\mathscr{I}^+}_{\w\w_1}\Big)_{2D},\label{AI-subtraction}\\[6pt]
\Big(B^{\mathscr{I}^+}_{\w\w_1}\Big)_{4D}&=&\Big(B^{\mathscr{I}^+}_{\w\w_1}\Big)_{4D}-\Big(B^{\mathscr{I}^+}_{\w\w_1}\Big)_{2D}+\Big(B^{\mathscr{I}^+}_{\w\w_1}\Big)_{2D}=\Delta B^{\mathscr{I}^+}_{\w\w_1}+\Big(B^{\mathscr{I}^+}_{\w\w_1}\Big)_{2D}.\label{BI-subtraction}
\eea \ees
Substituting ~\eqref{AI-subtraction} and ~\eqref{BI-subtraction} into $I_4$, isolating the terms with only $2D$ matching coefficients, and subtracting ~\eqref{ch5-GK} from the result, we find the four triple integrals
\bes \bea
\Delta G_1&=&2 \operatorname{Re}\Bigg\{ \int_{0}^{\infty}d\w\int_{0}^{\infty}d\w_1 \int_0^{\infty}d\w_2\Big[\big(A^{\mathscr{I}^{+}}_{\w\w_1}\big)_{2D}\big(A^{\mathscr{I}^{+*}}_{\w\w_2}\big)_{2D}-\alpha_{\w\w_1}\alpha^*_{\w\w_2}\Big]f^{\mathscr{I}^+}_{\w_1}(x)f^{\mathscr{I}^{+*}}_{\w_2}(x')\Bigg\}\nonumber \\ &&
- 2\operatorname{Re}\Bigg\{ \int_{0}^{\infty}d\w\int_{0}^{\infty}d\w_1 \int_0^{\infty}d\w_2 \;\alpha_{\w\w_1}\alpha^*_{\w\w_2}\nonumber \\&&
\times\Big[f^{H^-}_{\w_1}(x)f^{H^{-*}}_{\w_2}(x')-f^{\mathscr{I}^+}_{\w_1}(x)f^{\mathscr{I}^{+*}}_{\w_2}(x')\Big]\Bigg\},\label{ch6-DeltaG1}\\
\Delta G_2&=&2 \operatorname{Re}\Bigg\{ \int_{0}^{\infty}d\w\int_{0}^{\infty}d\w_1 \int_0^{\infty}d\w_2\Big[\big(B^{\mathscr{I}^{+}}_{\w\w_1}\big)_{2D}\big(B^{\mathscr{I}^{+*}}_{\w\w_2}\big)_{2D}-\beta_{\w\w_1}\beta^*_{\w\w_2}\Big]f^{\mathscr{I}^{+*}}_{\w_1}(x)f^{\mathscr{I}^+}_{\w_2}(x')\Bigg\}\nonumber \\ &&
- 2\operatorname{Re}\Bigg\{ \int_{0}^{\infty}d\w\int_{0}^{\infty}d\w_1 \int_0^{\infty}d\w_2 \;\beta_{\w\w_1}\beta^*_{\w\w_2}
\nonumber \\ &&
\times\Big[f^{H^{-*}}_{\w_1}(x)f^{H^{-}}_{\w_2}(x')-f^{\mathscr{I}^{+*}}_{\w_1}(x)f^{\mathscr{I}^{+}}_{\w_2}(x')\Big]\Bigg\},\label{ch6-DeltaG2}\\
\Delta G_3&=&2 \operatorname{Re}\Bigg\{ \int_{0}^{\infty}d\w\int_{0}^{\infty}d\w_1 \int_0^{\infty}d\w_2\Big[\big(A^{\mathscr{I}^{+}}_{\w\w_1}\big)_{2D}\big(B^{\mathscr{I}^{+*}}_{\w\w_2}\big)_{2D}-\alpha_{\w\w_1}\beta^*_{\w\w_2}\Big]f^{\mathscr{I}^+}_{\w_1}(x)f^{\mathscr{I}^+}_{\w_2}(x')\Bigg\}\nonumber \\ &&
- 2\operatorname{Re}\Bigg\{ \int_{0}^{\infty}d\w\int_{0}^{\infty}d\w_1 \int_0^{\infty}d\w_2 \;\alpha_{\w\w_1}\beta^*_{\w\w_2}\nonumber \\ &&
\times\Big[f^{H^{-}}_{\w_1}(x)f^{H^{-}}_{\w_2}(x')-f^{\mathscr{I}^{+}}_{\w_1}(x)f^{\mathscr{I}^{+}}_{\w_2}(x')\Big]\Bigg\},\label{ch6-DeltaG3}\\
\Delta G_4&=&2 \operatorname{Re}\Bigg\{ \int_{0}^{\infty}d\w\int_{0}^{\infty}d\w_1 \int_0^{\infty}d\w_2\Big[\big(B^{\mathscr{I}^{+}}_{\w\w_1}\big)_{2D}\big(A^{\mathscr{I}^{+*}}_{\w\w_2}\big)_{2D}-\beta_{\w\w_1}\alpha^*_{\w\w_2}\Big]f^{\mathscr{I}^{+*}}_{\w_1}(x)f^{\mathscr{I}{+*}}_{\w_2}(x')\Bigg\}\nonumber \\ &&
- 2\operatorname{Re}\Bigg\{ \int_{0}^{\infty}d\w\int_{0}^{\infty}d\w_1 \int_0^{\infty}d\w_2 \;\beta_{\w\w_1}\alpha^*_{\w\w_2}\nonumber \\ &&
\times\Big[f^{H^{-*}}_{\w_1}(x)f^{H^{-*}}_{\w_2}(x')-f^{\mathscr{I}^{+*}}_{\w_1}(x)f^{\mathscr{I}^{+*}}_{\w_2}(x')\Big]\Bigg\}.\label{ch6-DeltaG4}
\eea \ees
One can see that the sum of the first terms on the right in ~\eqref{ch6-DeltaG1}, ~\eqref{ch6-DeltaG2}, ~\eqref{ch6-DeltaG3}, and ~\eqref{ch6-DeltaG4},
\bea
(\Delta G_C)_{\text{4D}}&=&2 \operatorname{Re}\Bigg\{ \int_{0}^{\infty}d\w\int_{0}^{\infty}d\w_1 \int_0^{\infty}d\w_2\Big[\big(A^{\mathscr{I}^{+}}_{\w\w_1}\big)_{2D}\big(A^{\mathscr{I}^{+*}}_{\w\w_2}\big)_{2D}-\alpha_{\w\w_1}\alpha^*_{\w\w_2}\Big]f^{\mathscr{I}^+}_{\w_1}(x)f^{\mathscr{I}^{+*}}_{\w_2}(x')\Bigg\}\nonumber \\ &&
+2 \operatorname{Re}\Bigg\{ \int_{0}^{\infty}d\w\int_{0}^{\infty}d\w_1 \int_0^{\infty}d\w_2\Big[\big(B^{\mathscr{I}^{+}}_{\w\w_1}\big)_{2D}\big(B^{\mathscr{I}^{+*}}_{\w\w_2}\big)_{2D}-\beta_{\w\w_1}\beta^*_{\w\w_2}\Big]f^{\mathscr{I}{+*}}_{\w_1}(x)f^{\mathscr{I}^+}_{\w_2}(x')\Bigg\}\nonumber \\ &&
+2 \operatorname{Re}\Bigg\{ \int_{0}^{\infty}d\w\int_{0}^{\infty}d\w_1 \int_0^{\infty}d\w_2\Big[\big(A^{\mathscr{I}^{+}}_{\w\w_1}\big)_{2D}\big(B^{\mathscr{I}^{+*}}_{\w\w_2}\big)_{2D}-\alpha_{\w\w_1}\beta^*_{\w\w_2}\Big]f^{\mathscr{I}^+}_{\w_1}(x)f^{\mathscr{I}^+}_{\w_2}(x')\Bigg\}\nonumber \\ &&
+2 \operatorname{Re}\Bigg\{ \int_{0}^{\infty}d\w\int_{0}^{\infty}d\w_1 \int_0^{\infty}d\w_2\Big[\big(B^{\mathscr{I}^{+}}_{\w\w_1}\big)_{2D}\big(A^{\mathscr{I}^{+*}}_{\w\w_2}\big)_{2D}-\beta_{\w\w_1}\alpha^*_{\w\w_2}\Big]\nonumber \\&&
\times f^{\mathscr{I}^{+*}}_{\w_1}(x)f^{\mathscr{I}^{+*}}_{\w_2}(x')\Bigg\},
\eea
has a similar form to  ~\eqref{Del-GC}. The only difference is due to the different forms of $f^{\mathscr{I}^+}_{\w}$ in the 2D and 4D cases. As we saw in Chapter 4, the contribution of ~\eqref{Del-GC} to the stress-energy tensor approaches zero at late times in the 2D case. However, as is shown later in this chapter, $(\Delta G_c)_{4D}$ and its contribution to the components of the stress-energy tensor approach non-zero constants at late times. We find numerically show that for $(\Delta G_c)_{4D}$, this constant is equal to the sum of the second terms in ~\eqref{ch6-DeltaG1}, ~\eqref{ch6-DeltaG2}, ~\eqref{ch6-DeltaG3}, and ~\eqref{ch6-DeltaG4}.
\section{Stress-energy Tensor: $uv-$dependent part}
 Given that $f^{in}_{\w}=\big(f^{in}_{\w}\big)_v+\big(f^{in}_{\w}\big)_u$, in addition to the contributions to the two-point function $\Big\{G^{(1)}(x,x')\Big\}^{in}$ from the parts which only depend on $\big(f^{in}_{\w}\big)_v$ and the parts that only depend on $\big(f^{in}_{\w}\big)_u$ , there is a contribution from the cross terms, which is
\bea
\Big\{G^{(1)}(x,x')\Big\}^{in}_{uv}=4 \operatorname{Re}\Bigg\{ \int_{0}^{\infty}\big(f^{in}_{\w}(x)\big)_v\big(f^{in*}_{\w}(x') \big)_ud\w\Bigg\}\label{ch6-G-uv}.
\eea
The above term is the only contribution to the stress-energy tensor that is not ultraviolet divergent. Also, there is no counter-term from the Unruh state that has to be subtracted from ~\eqref{ch6-G-uv}.

For numerical purposes, we divide ~\eqref{ch6-G-uv} into two parts as follows
\bea
\Big\{G^{(1)}(x,x')\Big\}^{in}_{uv}=4 \operatorname{Re}\Bigg\{ \int_{\lambda}^{\Lambda}\big(f^{in}_{\w}(x)\big)_v\big(f^{in*}_{\w}(x') \big)_ud\w+\int_{\Lambda}^{\infty}\big(f^{in}_{\w}(x)\big)_v\big(f^{in*}_{\w}(x') \big)_ud\w\Bigg\}\label{ch6-G-uv-decomp}.
\eea

The first integral on the right-hand side has to be computed numerically. In the second integral, $\Lambda$ is chosen to be large enough that $\big(f^{in}_{\w}\big)_v$ and 
$\big(f^{in}_{\w}\big)_u$ can be approximated by their no scattering forms $\frac{Y_{00}e^{-i\w v}}{r\sqrt{4\pi \w}}$ and $-\frac{Y_{00}e^{-i\w u}}{r\sqrt{4\pi \w}}$ respectively. Substituting these approximations into the second integral on the right side, we find
\bea
\int_{\Lambda}^{\infty}\frac{e^{-i\w(v-u)}}{4\pi \w}d\w=-\frac{1}{4\pi}\text{Ei}(i\Lambda(u-v))\pm i\pi.
\eea
The plus and minus signs above correspond to $u-v>0$ and $u-v<0$ respectively. 
\section{Numerical Results for $\langle \Delta T_{tt}\rangle$}
In this section, we provide preliminary results for the numerical computation of $\langle \Delta T_{tt}\rangle$. The contribution of $\big(f^{in}_{\w}\big)_v$ to $\langle \Delta T_{tt}\rangle$ is shown in Fig. {fig:Ttt(vpart)}. The results show that at late times, this contribution approaches the contribution of the $f^{\mathscr{I}^-}_{\w}$ modes. We also found that after a series of damped oscillations, the corresponding contribution to $\langle \Delta T_{tt}\rangle$ decays as a power-law.  

For the contribution of $\big(f^{in}_{\w}\big)_u$ to  $\langle \Delta T_{tt}\rangle$, we use the approximation given in (6.24) for $(\Delta G_C)_{4D}$. Work is in progress to numerically compute the exact contribution from $\big(f^{in}_{\w}\big)_u$ and the cross terms. 

However, the contribution to $\langle \Delta T_{tt}\rangle$ from $\big(f^{in}_{\w}\big)_u$ using the approximation discussed earlier has been numerically computed with the result presented in Fig.{fig:Ttt(upart)}. The results show that at late times, the contribution of $\big(f^{in}_{\w}\big)_u$  to the renormalized stress-energy tensor in the $in$ vacuum state approaches the contribution of the $f^K_{\w}$ modes to the stress-energy tensor in the Unruh state.
\newpage
\begin{figure}[h]
\centering
\includegraphics[trim=0.5cm 0cm 0cm 0cm,clip=true,totalheight=0.35\textheight]{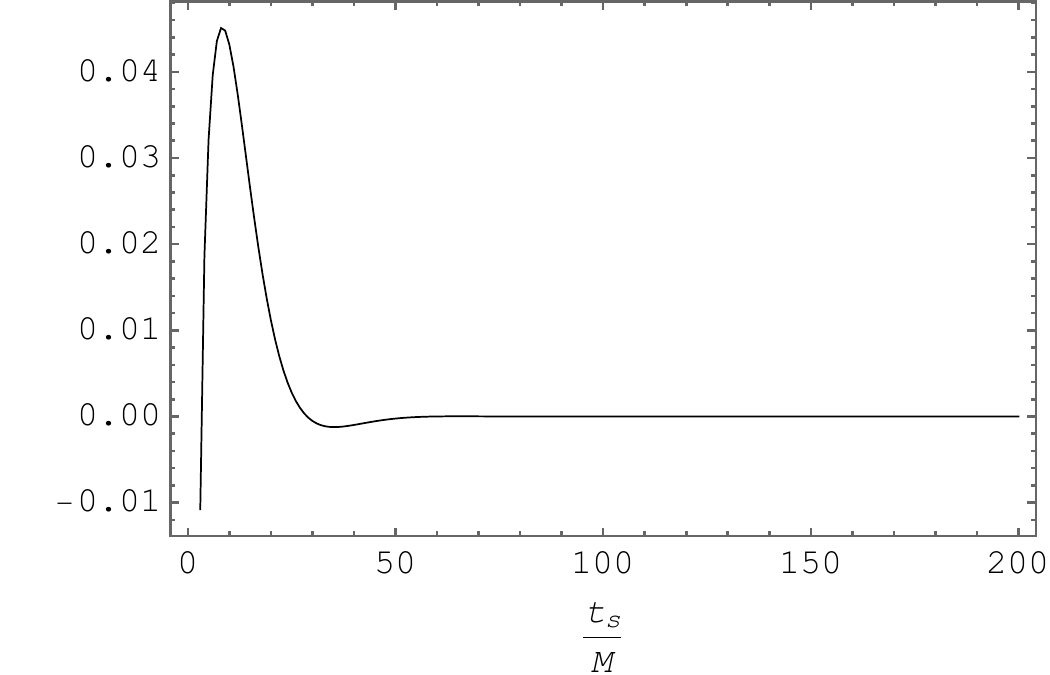}
\includegraphics[trim=0.5cm 0cm 0cm 0cm,clip=true,totalheight=0.35\textheight]{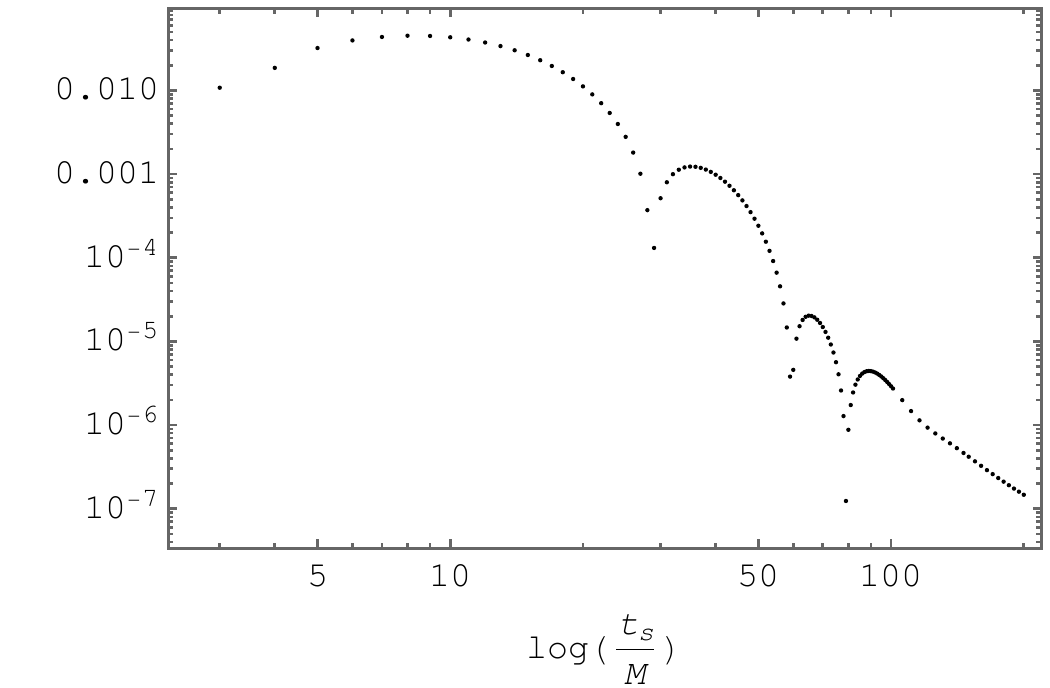}\\[10pt]
\caption{The top plot represents the contribution of $\big(f^{in}_{\w}\big)_v$ to the quantity $M^2\langle \Delta T_{tt} \rangle $ as a function of the time coordinate $t_s$ and for $r=3M$. At late times, this quantity approaches zero. The bottom plot shows the log-log plot for the same quantity. One can see a power-law decay at late times.}
\label{fig:Ttt(vpart)}
\end{figure}
\begin{figure}[h]
\centering
\includegraphics[trim=0cm 0cm 0cm 0cm,clip=true,totalheight=0.35\textheight]{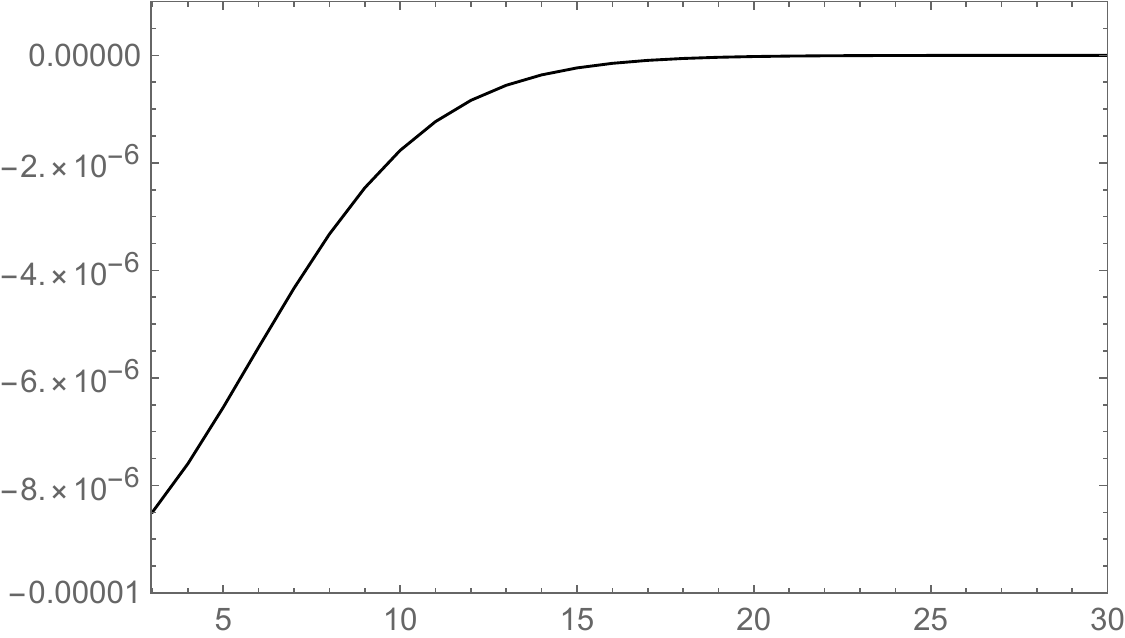}
\caption{This plot represents the contribution of $\big(f^{in}_{\w}\big)_u$ to the quantity $M^2\langle \Delta T_{tt} \rangle $ as a function of the time coordinate $t_s$ and for fixed spatial coordinate $r=3M$. At late times, this quantity approaches zero.}
\label{fig:Ttt(upart)}
\end{figure}
\chapter{Conclusions}
In Chapter 3, the behaviors of some of the mode solutions for a scalar field in the Unruh state, and the corresponding symmetric two-point correlation functions in various 2D eternal black hole spacetimes were studied. The study has been done outside the future horizon of each black hole and inside the cosmological horizon if it exists. I particularly studied the properties of the Kruskal modes of a massless minimally-coupled scalar field in an arbitrary asymptotically-flat static black-hole spacetime in which the mode equation has an effective potential proportional to a Dirac delta function.

In such a spacetime, it is convenient to expand the Kruskal modes in terms of the Boulware modes. We argued that the absence of infrared divergences in the Boulware modes, which is caused by scattering due to the Dirac delta function, strongly affects the late-time behavior of the Kruskal modes and makes them vanish at late times. The same type of behavior was observed for the Kruskal modes of a massive minimally-coupled scalar field in a 2D Schwarzschild-de Sitter spacetime. If the effective potential is zero, then there is no
scattering, the Boulware modes have infrared divergences, and the Kruskal modes
approach constant values at late times.

It has been shown in ~\cite{a-t} that for a massless minimally-coupled scalar field in the case of a 2D spacetime with a static patch and a black-hole and/or cosmological horizon, the two-point function for the Kruskal modes grows linearly in time for a pair of points with the same time coordinates and different fixed radial coordinates for a natural set of time and space coordinates for the static patch. This happens when one uses the first Unruh formulation defined in \cite{unruh:1976} and Chapter 3. However, the linear growth in time is not observed if the second Unruh formulation is used. We showed that when there is scattering due to a Dirac delta function potential in the mode equation, the two-point function for the Kruskal modes does not grow linearly in time and the two Unruh formulations are in agreement.

In Chapter 4, we presented a method for computing the stress-energy tensor for a massless minimally-coupled scalar field outside the future horizon of a  4D black hole that has formed from the collapse of a spherically-symmetric null shell. This method involves two parts. The first is the computation of the modes of the scalar field outside the null-shell spacetime, where the geometry is the Schwarzschild geometry. This has been accomplished by expanding the $in$ modes in terms of a complete set of modes in an eternal Schwarzschild spacetime. We found partially analytic expressions for the matching coefficients. 

The second part of the method we proposed involves renormalizing the stress-energy tensor in the region outside the null shell by subtracting the unrenormalized stress-energy tensor for the field in the Unruh state. Since the ultraviolet divergent behavior of the stress-energy tensor does not depend on the state of the quantum field, this difference is expected to be finite. Then, the renormalized stress-energy tensor can be found by adding the renormalized stress-energy tensor in the Unruh state \cite{levi-ori,levi} to this difference.

We did some tests to check the validity of this method. For one, we numerically computed the spherically symmetric contribution to the $in$ modes on the future horizon and showed that it approaches the known value on the null shell. This is expected since the $in$ modes are continuous across the null shell. We next applied our method to the case of a mode equation with a Dirac delta function as an effective potential. We found analytic expressions for part of the matching coefficients in this case and used them to check the continuity of the part of the $in$ modes that go like $e^{-i\w v}$ inside the null shell, on the surface of the null shell. 

Furthermore, the modes in a 2D null-shell spacetime were used to construct the stress-energy tensor and renormalize it in the 2D case, where the exact result is known  \cite{Fabbri:2005mw, hiscock, Davies-Fulling-Unruh}. We found that the numerical result for the renormalized stress-energy tensor found using our method is in agreement with the known analytical result.

In Chapter 5, we applied our method of computing the $in$ modes to the computation of the spherically-symmetric $in$ modes in the null-shell spacetime in 4D. In this case, the $in$ modes can be decomposed into two parts. The first part consists of the solutions to the mode equation which initially (on the Cauchy surface we are using) depend only on the null coordinate $v$ and the second part corresponds to the solutions which inside the null shell depend only on the null coordinate $u$. We computed these solutions and studied their behaviors in different limits. We first verified that the $in$ modes are continuous across the null shell. Furthermore, we found that, for a fixed spacetime point, in the large frequency limit where the effective potential is negligible, the $in$ modes approach their no-scattering counterparts. 

We also showed that for a fixed frequency and a fixed space point, at late times, the difference between the part of the modes which goes like $e^{-i\w v}$ inside the null shell and the $f^{\mathscr{I}^-}_{\w}$ modes approaches zero at late times as a power-law. This may be an important result since it could provide useful information regarding the late-time behavior of the stress-energy tensor. We numerically computed the part of the $in$ modes which goes like $e^{-i\w u}$ inside the shell and the Kruskal modes $f^{K}_{\w}$ and found that at late times, it also decays as power-law.

In Chapter 6, we have numerically computed the contribution of the modes that go like $e^{-i\w v}$ inside the shell and $f^{\mathscr{I}^-}_{\w}$ for the Unruh state to the
difference between the $t_st_s$ component of the null shell and Unruh stress-energy
tensors. We found that this difference approaches zero as a power law in time.

For the part of the $in$ modes that behave like $e^{-i\w u}$ inside the null shell, and the Kruskal modes, the contribution to the difference between the null-shell and Unruh stress-energy tensors is approximated. Work is in progress to compute the exact contributions of these modes to the stress-energy tensor. Work is also in progress to compute the cross terms that depend on both the part of the $in$ modes that go like $e^{-i\w v}$ and the part of the in modes that go like $e^{-i\w u}$ inside the null shell.

There are still questions that have to be addressed regarding the stress-energy tensor in the collapsing null-shell spacetime. One of these is finding the contribution of the non-spherically-symmetric modes. As a plan for future research, we are interested in finding these modes and their contribution to the stress-energy tensor. The other topic for future work is the behavior of the modes and corresponding stress-energy tensor inside the future horizon of the black hole. Finding the stress-energy tensor, in this case, might shed some light on where in the interior quantum field theory in curved spacetime is valid.

\addcontentsline{toc}{chapter}{Curriculum Vitae}  
\end{document}